\documentclass[twoside,gimmsy2e]{article} %
\usepackage{amsmath}    
\usepackage{amsthm}     
\usepackage{amscd}      
\usepackage{amssymb}    
\usepackage{eucal}      
\usepackage{latexsym}   
\usepackage{graphicx}   
\usepackage{verbatim}   
\usepackage{gimmsy2e}   

\makeatletter
\@addtoreset{equation}{subsection}

\makeatother

\def\intprod{\mathbin{\hbox to 6pt{%
                 \vrule height0.4pt width5pt depth0pt
                 \kern-.4pt
                 \vrule height6pt width0.4pt depth0pt\hss}}}

\let\hook\intprod

\newcommand{\Aut}{\operatorname{Aut}}
\newcommand{\Diff}{\operatorname{Diff}}

\newcommand{\ns}{\mspace{-1.5mu}}             
\newcommand{\ps}{\mspace{1.5mu}}              

\newcommand{\cc}{\mathcal C}
\newcommand{\cd}{\mathcal D}
\newcommand{\ce}{\mathcal E}
\newcommand{\cf}{\mathcal F}
\newcommand{\cg}{\mathcal G}

\newcommand{\cj}{\mathcal J}

\newcommand{\cl}{\mathcal L}

\newcommand{\cp}{\mathcal P}

\newcommand{\cy}{\mathcal Y}
\newcommand{\cz}{\mathcal Z}

\newcommand{\eps}{\epsilon}
\newcommand{\varep}{\varepsilon}
\newcommand{\sig}{\sigma} 	
\newcommand{\Sig}{\Sigma}

	\newcommand{\Ff}{\mathfrak{F}}
\newcommand{\fg}{\mathfrak{g}}	
	\newcommand{\Fh}{\mathfrak{H}}
	
	\newcommand{\Fj}{\mathfrak{J}}

	\newcommand{\Ft}{\mathfrak{T}}

	\newcommand{\Fx}{\mathfrak{X}}

\begin{document}

\pagenumbering{roman}
\thispagestyle{empty}


\title
{\huge \bf Momentum Maps \\[.5ex] and \\[.5ex] Classical Fields
\\[1ex] {\Large\it Part I: Covariant Field Theory}\\[1.5ex]}
\author{%
{\bf Mark~J.~Gotay} 
\thanks{Research partially supported as a Ford Foundation Fellow, 
by NSF grants DMS 88-05699, 92-22241, 96-23083, and 00-72434, and
grants from ONR/NARC.} 
\\[-2pt]
Department of Mathematics\\[-2pt]
University of Hawai`i\\[-2pt]
Honolulu, Hawai`i 96822, USA\\[-2pt]
gotay@math.hawaii.edu\\
\and 
{\bf James~Isenberg} 
\thanks{Partially supported by NSF grant PHY 00-99373.}
\\[-2pt]
Department of Mathematics\\[-2pt]
University of Oregon\\[-2pt]
Eugene, Oregon 97403, USA \\[-2pt]
jim@newton.uoregon.edu\\
\and
{\bf Jerrold~E.~Marsden} 
\thanks{Partially supported by NSF grant DMS 96-33161.}
\\[-2pt]
Control and Dynamical Systems 107-81\\[-2pt] 
California Institute of Technology\\[-2pt]
Pasadena, California 91125, USA\\[-2pt]
marsden@cds.caltech.edu\\
\and
{\bf Richard~Montgomery}
\thanks{Partially supported by NSF grant DMS 87-02502.} 
\\[-2pt]
Department of Mathematics\\[-2pt] 
University of California at Santa Cruz\\[-2pt] 
California 95064, USA \\[-2pt]
rmont@math.ucsc.edu\\ [12pt]
\and 
\centerline{\rm With the collaboration of}
\and
\and 
{\bf J\c edrzej  \'Sniatycki}
\and
{\bf Philip B.Yasskin}
}
\date{November 18, 1997 \\  (Revised November 22, 2003)}

\thispagestyle{empty}

\maketitle

\newpage
\thispagestyle{empty}
\tableofcontents

\addtocontents{toc}{\protect\vspace{5ex}}


\newpage
\setcounter{page}{0}
\pagenumbering{arabic}
\pagestyle{myheadings}

\setcounter{section}{0}
\setcounter{subsection}{0}

\section{Introduction}

This is the first of a five part work in which we study the
Lagrangian and Hamiltonian structure of classical field theories with
constraints.  Our goal is to explore some of the connections between
initial value constraints and gauge transformations in such theories
(either relativistic or not). To do this, in the course of this book,
we develop and use a number of tools from symplectic and
multisymplectic geometry. Of central importance in our analysis is the
notion of the ``energy-momentum map'' associated to the gauge group of
a given classical field theory. We hope to demonstrate that, as
illustrated and explained below, many different and apparently
unrelated facets of field theories can be thereby tied together and
understood in an essentially new way.  

Our research is motivated by previous exhaustive studies of several
standard examples following the methods of Choquet--Bruhat,
Lichnerowicz, Dirac--Bergmann, and  Arnowitt--Deser--Misner. In
particular, if one performs an initial value (or ``canonical'')
analysis of Einstein's theory of gravity (either alone or coupled to
Dirac, electromagnetic, or other fields), Yang--Mills theory,
relativistic fluid theories, bosonic string theory, topological field
theories, and so forth, one finds that they share the following
striking features. (See Fischer and Marsden [1979a,b], Isenberg and
Nester [1980],  K\"unzle and Nester [1984], Bao, Isenberg, and Yasskin
[1985], Batlle, Gomis, and Pons [1986], and Horowitz [1989] and 
references therein for discussions of these and other field theories.)

\begin{myproclaim}
The Euler--Lagrange equations are
underdetermined; that is, there are not enough evolution equations to
propagate all the field components.  This is because the theory has
gauge freedom. The corresponding gauge group is known at the outset.
\end{myproclaim}

As a consequence, some of the fields do not have their evolution
determined by the Euler--Lagrange equations. These ``kinematic fields''
have no physical significance. Amongst the remaining fields are the
``dynamic fields'' $\psi$ (with conjugate momenta $\rho$); these are
the fields that have physical meaning.

The situation is further complicated in that the initial data
$(\psi(0),\rho(0))$ cannot be freely specified.

\begin{myproclaim}
The Euler--Lagrange equations are overdetermined in that
they include constraints
\begin{equation}\label{eqn1:1} 
\Phi^i(\psi, \rho) = 0  
\end{equation}
on the choice of Cauchy data.
\end{myproclaim}

Equations (\ref{eqn1:1}) typically comprise an elliptic system
whose component equations can be of variable order. For simplicity we
suppose that all constraints are ``first class'' in the terminology of
Dirac [1964]. (The case when second class constraints are present will 
be discussed as the need arises.)

It is remarkable that the Euler--Lagrange equations are simultaneously
underdetermined  and overdetermined. The next feature ties these two
properties together.

\begin{myproclaim}
The constraint functions $\Phi^i$ generate gauge
transformations of the dynamic fields via the canonical symplectic
structure on the space of Cauchy data.
\end{myproclaim}

Thus, the presence of (first class) initial value constraints is
intimately related to the gauge freedom of the theory.  Since gauge
transformations are part of the dynamics and since the constraints
generate gauge transformations, one expects  the functions $ \Phi^i
$ to appear in the Hamiltonian.  Indeed, in these examples, we find: 

\begin{myproclaim}
The Hamiltonian\/ \textup{(}with respect to a slicing of spacetime,
usually taken to consist of Cauchy surfaces\/\textup{)} has the form
\begin{equation}\label{eqn1:2}
 H = \int_\Sigma \sum_i \alpha_i \Phi^i (\psi, \rho)\, d\Sigma
\end{equation}
depending linearly on the ``atlas fields'' $\alpha_i$.  Here $\Sigma$ is
a hypersurface modeling a typical member of the given slicing.
\end{myproclaim}
The atlas fields, discussed in more detail later, are closely related
to the kinematic fields mentioned above in item {\bf 1}. (Being
sections of an associated bundle, they may be regarded as certain
combinations of the kinematic fields and  elements of the Lie algebra
of the gauge group.) The atlas fields are arbitrarily specifiable;
intuitively, one can view them as governing the propagation of
coordinates on a Cauchy surface (and on bundles of fields over a
Cauchy surface) in time---hence the name. Their  importance derives
from the fact that they ``drive'' the entire gauge ambiguity of the
theory (cf. the next item).

Introduce an $L_2$ inner product on the space of Cauchy data (with
corresponding adjoint denoted by $^*$), and let ${\mathbb J} $ be an
almost complex structure compatible with both the given inner product
and the symplectic form. Let $\lambda$  be a parameter (``time'')
labeling the slicing mentioned above. From (\ref{eqn1:2}), it follows
that

\begin{myproclaim}
The evolution equations for the dynamic fields
$(\psi,\rho)$ take the ``adjoint form''
\begin{equation}\label{eqn1:3}
\frac{d}{d\lambda} \binom{\psi}{\rho} = \mathbb J\cdot \sum_i \left[ D
\Phi^i (\psi(\lambda), \rho (\lambda))\right]^* \alpha_i(\lambda).
\end{equation}
\end{myproclaim}

Equations (\ref {eqn1:3}) are typically a {\it hyperbolic\/} system;
hence, for fixed $\alpha_i(\lambda )$, they uniquely determine the
evolution of the dynamic fields and their conjugate momenta from given
initial data. Here, the atlas fields $\alpha_i(\lambda)$ are to be
specified {\it ab initio\/};  this amounts to a gauge-fixing.
Equations (\ref {eqn1:3}) display, in the clearest and most concise
way, the interrelations between the dynamics, the initial value
constraints, and the gauge ambiguity of a theory.

The constraint equations (\ref {eqn1:1}) are preserved by the evolution
equations (\ref {eqn1:3}). Of course,

\begin{myproclaim}
The covariant Euler--Lagrange equations are equivalent,
modulo gauge  transformations, to the combined evolution and constraint
equations \textup{(\ref{eqn1:3})} and\/ \textup{(\ref{eqn1:1})}.
\end{myproclaim}

This result, together with items {\bf 4}  and {\bf 5}, constitute the
theoretical background necessary  to analyze the structure of the
solution space of the Euler--Lagrange equations. For the theories
mentioned above, we have:

\begin{myproclaim}
The space of solutions of the field equations is not
necessarily a smooth manifold.  It may have quadratic singularities
occurring exactly at those solutions that are symmetric  {\em(}in the
sense of being invariant under the action of a nondiscrete subgroup of
the gauge group{\em)}.
\end{myproclaim}

See Fischer, Marsden, and Moncrief [1980], Arms, Marsden, and Moncrief
[1981,\,1982], Marsden [1988], Arms [1977,\,1981], Arms, Gotay, and 
Jennings [1990], Arms, Cushman, and Gotay [1991], Sjamaar and Lerman
[1991], and Interlude~III for more information on the meaning of item
{\bf 7}. These results are an outgrowth of the extensive literature on 
linearization stability, which concerns the success or failure of the
linearization process as far as predicting the valid directions of
perturbation theory, as well as that on symplectic  reduction, which is
the means of eliminating those degrees of freedom of a system that are 
associated with symmetries.

Noether's theorem (aka the ``first Noether theorem'') and the formal
Dirac--Bergmann analysis of constraints do much to predict and explain
features {\bf 1}--{\bf 6}. However, we wish to go further and provide
(realistic) sufficient conditions which guarantee that they {\it
must\/} occur in a given field theory. In this work we provide such
criteria for the first six conditions and lay the groundwork needed
for the analysis of the seventh. One of the goals is thus to derive
the ``adjoint formalism'' for classical field theories.

\medskip 

A key tool throughout this work is the ``energy-momentum map,'' which
we define in Part II. A recurrent theme in this work is that {\em the
energy-momentum map encodes essentially all the dynamical information
carried by a given classical field theory\/}: its  Hamiltonian, its
initial value constraints, its gauge freedom, and even its
stress-energy-momentum tensor. Indeed, in Part~III we shall prove

\begin{quote}  {\bf The Energy-Momentum Theorem.}\enspace 
{\it The constraints\/ \textup{(\ref{eqn1:1})} are given by the
vanishing of the energy-momentum map associated to the gauge group of
the theory.}
\end{quote} 
Thus the constraint functions $\Phi^i$ are ``components'' of the
energy-momentum map. Item {\bf 4} then shows that the same is true for
the Hamiltonian.

Certain of these results are known  in specific instances and
have to some extent acquired ``folk theorem'' status. But we wish to
prove that they hold rather generally. In this regard we emphasize that
the energy-momentum map does more than merely provide reformulations
of known results; it leads to fundamental insights into the
foundations of physical theories. In view of their  importance---and
ubiquity---in classical field theory, one is tempted to elevate the
statement ``momentum maps are everything'' to a general principle!

To understand the role played by the energy-momentum map, recall the
two traditional approaches to classical field theory, which we may
characterize as ``group-theoretical'' and ``canonical.'' The
group-theoretical approach is concerned with the gauge covariance of a
given system, and is based upon Noether's theorem. It operates on the
(spacetime) covariant level, and is usually phrased in Lagrangian
terms. (See Trautman [1967], Olver [1986], and references contained
therein for relevant background.) In contrast, the canonical approach
is used to analyze the initial value problem in a Hamiltonian setting;
the basic tool here is the Dirac--Bergmann theory of constraints. Such
an analysis requires that the theory be space + time decomposed
relative to a fixed choice of Cauchy surface in spacetime. (See Dirac
[1964], Hanson, Regge, and Teitelboim [1976], Gotay, Nester, and Hinds
[1978], Isenberg and Nester [1980], as well as the review by
Sundermeyer [1982] for detailed accounts.)

The group-theoretical and canonical aspects of a mechanical system are
linked by the momentum map (see, for example, Abraham and Marsden
[1978] and Marsden and Ratiu [1999]). One would like to have an
analogous connection in field theory which, in particular, relates
gauge symmetries to initial value constraints. But---and this is a
crucial point---{\it the standard notion of a momentum map associated
to a symplectic group action usually  cannot be carried over to
spacetime covariant field theory\/}. The reason is that the gauge
group typically does not act in the space + time decomposed (or
``instantaneous'') framework, because spacetime diffeomorphisms move
Cauchy surfaces. Because the Hamiltonian formalism is only defined
relative to a fixed Cauchy surface, there is no possibility of
obtaining a momentum map for the gauge group in the usual sense on the
instantaneous phase space.\footnote{\ It is possible that a momentum
map in the sense of groupoids  may be appropriate for this setting. 
(See, for instance, Mikami and Weinstein [1988].) This is, however, a
subject for future research.}

The prime example where this situation arises is Einstein's theory 
 of vacuum gravity, in which case the gauge group is the spacetime
diffeomorphism group. The only remnants of this group on the
instantaneous level are, in the ADM (Arnowitt, Deser and Misner [1962])
language, the superhamiltonian $\mathfrak  H$ and supermomenta
$\mathfrak  J$ which are interpreted as the generators of temporal and
spatial  deformations of a Cauchy surface, respectively. But these
deformations do not form a group, nor are
$\mathfrak  H$ and $\mathfrak  J$ components of a momentum map. (See
Kucha\v r [1973] for further details.) More generally, this difficulty
appears for any theory which is ``param\-etr\-ized'' in that its
gauge group ``includes'' the spacetime diffeomorphism group (in an
appropriate sense).

This circumstance forces us to work on the covariant level. But it is
then necessary to construct a covariant counterpart to the
instantaneous Hamiltonian formalism. In the spacetime covariant (or
``multisymplectic'') framework we develop here---which is an  extension
and refinement of the formalism of Kijowski and Szczyrba [1976]---the
gauge group {\it does\/} act. This enables us to define the notion of a
covariant (or ``multi-'') momentum map on the corresponding covariant
(or ``multi-'') phase space. This covariant momentum map satisfies
Noether's theorem, and has all the properties one would expect. It
remains to correlate this object with the dynamics once a Cauchy
surface has been singled out. To this end, a careful study of the
mechanics of the space + time  decomposition shows that the covariant
momentum map induces an ``energy-momentum'' map $\Phi$ on the
instantaneous phase space. It is not a genuine momentum map, as there
is no group action in this context; nonetheless, $\Phi$ is the
instantaneous ``shadow'' of a covariant  momentum map, and this is
enough. The energy-momentum map is the crucial object that  reflects
the gauge transformation covariance of a classical field theory in the
instantaneous formalism. In the ADM formulation of gravity, $\Phi  =
-(\mathfrak  H, \mathfrak  J)$, so that the superhamiltonian and
supermomenta are the components of the energy-momentum map.

The energy-momentum map thus synthesizes the group-theoretical and
ca\-nonical approaches to classical field theory. Consonant with this
observation, we view the gauge group of a theory as being fundamental,
and suppose it to be known {\it ab initio\/}. This is  certainly the
case in the standard examples cited previously. Our plan is then to use
the  energy-momentum map associated to the gauge group of the theory
to study, and indeed derive, items {\bf 1}--{\bf 6} above. This shift
in viewpoint proves to be surprisingly fruitful, enabling us to obtain
deeper results than by the traditional methods alone.

\medskip

The  theoretical underpinnings of the adjoint formalism for classical
field theories consist of four components: 
\begin{enumerate} 
\renewcommand{\labelenumi}{\bf \Roman{enumi}.}
\item  
a covariant analysis of field theories; 
\item  
a space + time  decomposition of the covariant
formalism followed by an initial value analysis of field  theories;
\item  
a study of the relations between gauge symmetries and initial
value constraints; and 
\item  
the actual derivation of the adjoint
formalism. 
\end{enumerate} 
The first two components constitute the foundation for the last two,
where most of the hard work is concentrated. We present each of these
four components as a paper in this series, starting with the covariant
analysis. The fifth paper in this series applies our formalism to
Einstein's theory of gravity. Before beginning Part~I, we now comment
briefly on each of the five parts, in turn. (For the convenience of
the reader, we append a table of contents of Parts II--V at the end of
this paper.)

\medskip 

In Part I we develop some of the basic theory of classical fields
from a spacetime covariant viewpoint. Throughout we restrict attention
to first order theories; these are theories whose Lagrangians involve
no higher than first derivatives of the fields.

We begin in Chapter 2 of Part I with a study of the covariant
Lagrangian and Hamiltonian  formalisms, mirroring the approach usually
taken in classical mechanics. Let $X$ be an 
$(n+1)$-dimensional parameter space, which in applications is usually
spacetime. We suppose that we are given a bundle $Y$ over $X$ whose
sections are the fields of interest; $Y$ is the  {\bfi covariant
configuration bundle\/}. The field-theoretic analogue of the tangent
bundle of mechanics is the first jet bundle $J^{1}Y$  of $Y$. By
taking an appropriate dual of $J^{1}Y$  we obtain the space $Z$ which
is, roughly speaking, the covariant cotangent bundle of the bundle
$Y$. As such, it carries a canonical $(n+1)$-form $\Theta$.   The
{\bfi multisymplectic\/}
$(n+2)$-form $\Omega = - d\Theta$  is the covariant generalization of
the symplectic form in Hamiltonian mechanics. The pair $(Z, \Omega)$ is
called {\bfi multiphase space\/}.

Next, in Chapter 3, we introduce a Lagrangian density $\mathcal L$ on
$J^{1}Y$. The covariant Lagrangian and Hamiltonian formalisms are then
related by the {\bfi covariant Legendre transformation\/} $\mathbb
F\mathcal L: J^1 Y\to  Z$ which is defined, as in mechanics, by means
of the fiber derivative of $\mathcal L$. Pulling the form $\Omega$ on
$Z$ back by
$\mathbb F\mathcal L$ we obtain the {\bfi Cartan form\/} on $J^{1}Y$ 
(also variously known as the ``Hamilton--Cartan'' and
``Poincar\'{e}--Cartan'' form). We use it to construct an intrinsic
formulation of the variational calculus and the Euler--Lagrange
equations, cf. Goldschmidt and Sternberg [1973] and Garc\'{\i}a
[1974].  

Our approach to covariant field theory has a number of advantages over
other formulations.  For instance, unlike those of Kijowski and
Tulczyjew [1979], Ragionieri and Ricci [1981],  G\"{u}nther [1987],
Sardanashvily [1993], and Kanatchikov [1998], our multisymplectic
structure and corresponding covariant  Hamiltonian formalism are
intrinsically defined, independent of other constructs (such as a
background connection). Other formalisms such as those of Goldschmidt
and Sternberg [1973] and \'{S}niatycki [1984], although intrinsic, do
not carry multisymplectic structures at all. The approach we take here
is also promising for higher order field theories, which are plagued
by a number of ambiguities; see Gotay [1991a,b] and references
therein. Perhaps most importantly, our approach allows us to define
covariant momentum maps for groups that act nontrivially on~$X$. Such
group actions also arise naturally in fluid dynamics (Moreau [1982])
and continuum mechanics (Lew, Marsden, Ortiz, and West [2002]) in
relation to taking ``horizontal variations'' of the action.

In Chapter 4 we discuss symmetries and conservation laws of classical
field theories in terms of ``covariant momentum maps,''  generalizing
the concept of momentum map familiar from mechanics. If one
introduces  a Lie group $\mathcal G$ (possibly infinite-dimensional)
with Lie algebra
$\mathfrak  g$ that acts on
$(Z, \Omega)$ in an appropriate way, then one can define a {\bfi
multi\/}- or {\bfi covariant momentum map\/}
$J: Z\to \mathfrak  g^*\otimes \Lambda^{n} Z$ that intertwines the
group action with the  multisymplectic structure via the equation
$$
\xi_Z \hook\, \Omega = \mathbf{d} \langle J,\xi\rangle,
$$ 
where $\langle J,\xi\rangle$ denotes $J$ paired with $\xi \in \fg$,
$\xi_Z$ is the infinitesimal generator on $Z$ of the one-parameter
group generated by $\xi$, and $\mathbf{d}$ is the exterior
derivative. It often happens that such group actions are lifted from
actions on $Y$, in which case one has an explicit formula for
$J$; see \S4C. Then in \S4D we prove Noether's theorem, which
provides the basic relation between symmetries of the Lagrangian and
conserved quantities. Our multisymplectic formulation of Noether's
theorem  is specifically designed with later applications to  the
relationship between constraints and the momentum  map in mind. Here
we also state and prove a converse to Noether's theorem, subject to a
certain transitivity assumption on the group action.

\medskip 
\newpage
All the discussion in Part I is spacetime covariant. In Part II we use
{\bfi space $+$ time decompositions\/} to reformulate classical field
theories as infinite-dimensional dynamical systems. The transition from
the multisymplectic to the instantaneous formalism once a Cauchy
surface (or a slicing by Cauchy surfaces) has been chosen is essential
to define the energy-momentum map and thence to cast the field
dynamics into adjoint form.

Consider  an $n$-dimensional Cauchy surface $\Sigma \subset X$
(assumed compact and  boundaryless). In the instantaneous context, the
configuration space for field dynamics is the space $\cy_\Sig$ of
sections of $Y$ restricted to $\Sigma$,  and the phase space is
$T^*\cy_\Sigma$ with its canonical symplectic form. We emphasize that
this symplectic structure lives on the space of Cauchy data for the
evolution equations, and {\it not\/} on the space of solutions thereof
(as in, e.g., Zuckerman [1987] and Crnkovi\' c and Witten [1987]).
This is an important distinction, as the former space is typically
better behaved than the latter (cf. item {\bf 7} as well as the
discussion in Horowitz [1989]).  It also obviates the problem of
constructing a differentiable structure on the space of solutions, as
was attempted by Kijowski and Szczyrba [1976].

In Part~II, the main result of Chapter 5 is that the multisymplectic
structure on $Z$ induces the canonical  symplectic structure on
$T^*\cy_\Sigma$. To bridge the gap between the covariant and
instantaneous  formalisms, we introduce an intermediate space
$\cz_\Sigma$, the space of sections of $Z$  restricted to $\Sigma$.
The multisymplectic $(n+2)$-form $\Omega$ on $Z$ induces a  {\it
pre\/}symplectic $2$-form $\Omega_\Sigma$ on $\cz_\Sigma$ by
integration over $\Sigma$.  Reducing $\cz_\Sigma$ by the kernel of
$\Omega_\Sigma$ produces a symplectic manifold $\cz_\Sigma/(\ker
\Omega_\Sigma)$,  which we prove is canonically symplectomorphic to
$T^*\cy_\Sigma$.

This shows how to space + time decompose covariant multisymplectic
phase spaces into instantaneous symplectic phase spaces.  Next, in
\S\S6A--D, we perform a similar decomposition of dynamics using the
notion of slicings. An important result here  states that
the dynamics is compatible with the space + time decomposition in the
following sense: Solutions of the instantaneous Hamilton equations
correspond directly to solutions of the covariant Euler--Lagrange 
equations and vice versa. This fact---a prerequisite for establishing
item {\bf 6}---is  usually taken for granted in the literature. Here we
provide a proof, assuming only mild regularity  assumptions that are
satisfied in cases of interest.

An important tool throughout this work is the {\bfi initial value
analysis.\/} One of the hallmarks  of field theories is that not all of
the Euler--Lagrange equations necessarily describe the temporal
evolution of fields; some of them may impose constraints on the choice
of initial  data. The constraints that  are first class reflect the
gauge symmetry of the theory and eventually will be related to the
vanishing of various (energy-) momentum maps. Thus, it is important to
fully analyze the roles of constraints and gauge transformations in
dynamics and the relations between them. To this end, we utilize the
symplectic version of the  Dirac--Bergmann treatment of degenerate
Hamiltonian systems (patterned after Gotay, Nester, and Hinds [1978],
cf. also Sundermeyer [1982]). This formalism, which we summarize in
\S6E, yields an essentially complete understanding of items
{\bf 1--3}, and lays the groundwork for  correlating the first class
constraints with the gauge group of the theory via the 
energy-momentum map.

In Chapter 7 we space + time decompose multimomentum maps associated to
covariant group actions, thereby correlating these objects with
momentum mappings (in the usual sense) in the instantaneous formalism.
Thus, suppose that $\cg$ is a group of automorphisms of $Y$. As
emphasized earlier, $\cg$ usually does not act on $\cy_\Sigma$ or on 
$T^*\cy_\Sigma$ as it need not stabilize $\Sigma$. Nonetheless, by
integrating over $\Sigma$, the covariant momentum map $J$ gives rise to
a map $\cz_\Sigma\to\fg^*$, which can then be shown to project to a
well-defined map $\ce_\Sigma:\cp_\Sigma \to \fg^*$, where the
``primary constraint set'' $\cp_\Sigma$ is the image of $T\cy_\Sigma$
in  $T^*\cy_\Sigma$ under the instantaneous Legendre transform. This
map $\ce_\Sigma$ is the {\bfi energy-momentum map\/}.  It is
typically not a true momentum map. However, if one considers only the
subgroup $\cg_\Sigma$ of $\cg$ that  stabilizes the Cauchy surface 
(and that therefore acts in the instantaneous formalism), then
$\ce_\Sigma$ restricts to a genuine momentum map for the
$\cg_\Sigma$-action on $T^*\cy_\Sigma$ that corresponds in relativity
to the supermomenta, without the inclusion of the superhamiltonian. In
the general case $\ce_\Sigma$  corresponds to both $\Fh$ and $\Fj$.

\medskip 

Parts I and II provide the requisite background for the study of items
{\bf 4}--{\bf 6}, which we carry out in Parts~III and IV. Part III, in
particular, focuses on establishing a relationship between the initial
value constraints on the one hand, and the energy-momentum map
corresponding to the gauge transformations on the other. (See the
Energy-Momentum Theorem.) We begin the study of this relationship in
Chapter 8 of Part~III, in which we characterize the ``gauge group'' of
a given classical  field theory. Surprisingly, this notion, although
familiar and intuitive, has never been precisely defined in any
generality. It is---for the moment---to be distinguished from the
notion of ``instantaneous gauge transformation'' \`a la Dirac that was
alluded to above. Of course, one of the main goals of Part III is to
show that these two notions of ``gauge'' are in fact the same. This
will follow once item {\bf 3} and the Energy-Momentum  Theorem are
established.

The {\bfi gauge group\/} $\cg$ is taken to be a subgroup of Aut$(Y)$,
the group of automorphisms of the bundle $Y$. It must satisfy certain
properties, viz.: 
\begin{enumerate}
\renewcommand{\labelenumi}{$($\roman{enumi}$)$}
\item 
$\cg$ acts by {\bfi symmetries\/}, for example, the
Lagrangian density $\cl$  is $\cg$-equivariant. 
\item 
$\cg$ is {\bfi localizable}, which intuitively means that
elements of $\cg$ can be  ``turned off'' in the ``future'' or the
``past.'' This property is crucial; it is what distinguishes a gauge
group from a mere symmetry group. 
\item 
$\cg$ is the largest such
subgroup of Aut$(Y)$. 
\end{enumerate} 
In \S8A we give a principal bundle construction of
$\cg$.

One should bear in mind, however, that some field theories may have no
gauge freedom  (or initial value constraints) at all; the Klein--Gordon
field on a noninteracting background  spacetime is a simple example of
such a system. Theories like this can be subsumed into our formalism by
requiring  them to be {\bfi  param\-etr\-ized\/}, in the sense that 
$\cg$ maps onto $\Diff(X)$ (or at least a localizable subgroup
thereof) under the natural projection  $\Aut(Y) \to \Diff(X)$. Thus,
the gauge groups of param\-etr\-ized theories are {\it always\/}
nontrivial. Many field theories are ``already'' param\-etr\-ized
(e.g., the Polyakov string, topological field theories, gravity). In
any case, this requirement involves no loss of generality, since a
theory that  is not param\-etr\-ized can often be made so, for
example, by simply introducing a metric on $X$ and treating  it as an
auxiliary variable (which may or may not be {\bfi variational} in the
sense that it is to be varied in the action principle to obtain
Euler--Lagrange equations), or else by coupling it to gravity. 
Kucha\v{r} [1973] gives an alternate way of parametrizing a theory.
Henceforth we assume that the systems under consideration are
param\-etr\-ized. (Although our formalism is best suited to
param\-etr\-ized theories, much of it---with appropriate
modifications---will apply to ``background'' theories as well. These
points are discussed  further in the text; see also Interlude~I.)

We start building towards the main results of Part III by
establishing the following  result, which is basic to the entire
development. Fix a hypersurface $\Sigma$ and an element 
$\zeta \in \fg$ with the property that  its associated spacetime vector
field $\zeta_X$ is transverse to $\Sigma$.  (Such a vector field exists
since the theory is param\-etr\-ized.) The Lie algebra element $\zeta$
may be thought of as defining an ``evolution direction''; it is
analogous to the ``lapse'' and ``shift'' in relativity. In \S7F we
show that the instantaneous Hamiltonian density $H_{\Sigma, \zeta}$ on
$\cp_\Sigma$ corresponding to the evolution direction $\zeta \in \fg$
is given by
\begin{equation}\label{eqn1:4}
H_{\Sigma, \zeta} = - \langle\ce_\Sigma,\zeta\rangle.
\end{equation}
Thus the instantaneous Hamiltonian is a component of
$\ce_\Sigma$ (hence the appellation  {\it energy\/}-momentum map). It
follows that $\ce_\Sigma$ actually generates the dynamics of  the
system. This result is familiar in the ADM formulation of gravity:
using the fact that $\ce_\Sigma = -(\Fh,\Fj)$ and setting $\zeta  =
(N,{\mathbf M})  \in \Fx (X)$, (\ref{eqn1:4}) reduces to
$$  H_{\Sigma,  (N,{\mathbf M})} =  N\Fh +  {\mathbf M}\cdot \Fj.
$$  Equation (\ref{eqn1:4}) forms the basis for item {\bf 4}; in this
regard, the fact that $H_{\Sigma,\zeta}$ is manifestly linear in
$\zeta$ presages its linearity in the atlas fields, discussed later
for general field theories.

Next, in Chapter 9, we prove the {\bfi Vanishing Theorem\/} (aka the
``second Noether theorem''), which states that on any solution of the
Euler--Lagrange equations, the multimomentum  map integrated over a
hypersurface is zero. This result follows from Noether's theorem,
$\cg$-equivariance, and localizability. The Vanishing Theorem implies
that the energy-momentum  map must vanish on admissible Cauchy data
for the evolution equations. (The fact that $\ce_\Sigma$ must {\it
vanish\/}, as opposed to  merely being constant, is a direct
consequence of localizability.)

With these results in hand, in Chapter 10 we establish the
{\bfi Energy-Momen\-tum Theorem}, which asserts that for any field
theory which is first class and satisfies certain technical conditions,
the set $\cc_\Sigma$ of admissible Cauchy data for the evolution
equations is exactly the zero level set of the energy-momentum  map
associated to the gauge group:
\begin{equation}\label{eqn1:5}
\cc_\Sigma = \ce_\Sig^{-1} (0).
\end{equation}
The set $\cc_\Sig$ is known as the ``final constraint set'' in Dirac's
terminology. More succinctly, this theorem says that {\it the initial
value constraints are given by the vanishing of the energy-momentum
map\/}. Although often quoted in the literature, to our knowledge no
proof of this fact has ever been given. It may at first appear
that (\ref{eqn1:5}) is ``just'' the Vanishing Theorem, but this is not
the case. Certainly the Vanishing Theorem implies that the
vanishing of the energy-momentum map yields initial value constraints,
which can then be shown to be first class. But it is far from obvious
that in fact {\it all\/} such constraints arise in this fashion.
Moreover, the proper context for the second Noether theorem is the
multisymplectic formalism described above. Only in this setting is one
able to obtain constraints that are associated to gauge 
transformations that move Cauchy surfaces in spacetime, such as the
superhamiltonian constraint in general relativity. It should be
observed that other covariant Hamiltonian formalisms are deficient in
this regard, for example, that of Kijowski and Tulczyjew [1979].

One corollary of the Energy-Momentum  Theorem is that since $\cg
\ne \varnothing $,  first class constraints are {\it always\/} present
in param\-etr\-ized theories (in which all fields are variational).
Thus, with this proviso, items {\bf 1--3} are never vacuous. Another
corollary is that the covariant and instantaneous notions of gauge
coincide; more precisely, when space $+$ time  decomposed, the gauge
group generates exactly the instantaneous gauge transformations in the
sense of the Dirac--Bergmann constraint theory. This indicates that
the above definition of $\cg$ is ``correct.''

There are calculational advantages that accrue from this theorem as
well. For instance, one can always compute the initial value
constraints of a given field theory according to the Dirac--Bergmann
procedure. But the ensuing calculations can be quite involved, and the
results  are not always easy to interpret. An analysis based on the
gauge group and the energy-momentum map,  on the other hand, is often
substantially simpler---and has the advantage of attaching a 
group-theoretical interpretation to the constraints. A case in point is
again  provided by ADM gravity. If one merely carries out the
Dirac--Bergmann analysis, one obtains the superhamiltonian and
supermomentum constraints, but it is entirely unclear from this
analysis what their geometric significance is. Only by ``hindsight''
is one able to correlate these constraints with temporal and spatial
deformations of a Cauchy surface, reflecting the 4-dimensional
diffeomorphism gauge freedom of the theory. The problem  is that the
spacetime diffeomorphism group is ``hidden'' once one performs a space
$+$ time decomposition---a  necessary prerequisite to the initial value
analysis. But our procedure yields these constraints rather routinely,
and their geometric meaning is clear from the beginning. (See Part V
for the application of our techniques to gravity.) Thus, one has a
method for identifying the final constraint set that does not in
principle require one to invoke the entire constraint preservation
algorithm. The criteria which must be checked are not always easy to
verify but, in any case, the theorems of Chapter 10 do guarantee an
equivalence between the results of the Dirac--Bergmann algorithm and
the results of an analysis  based on the energy-momentum map of a
given gauge group action.

The Energy-Momentum Theorem is really a statement about (first class)
{\it secondary\/} constraints, as $\ce_\Sigma$ is only defined on the
primary constraint set $\cp_\Sigma \subset T^*\cy_\Sigma$. In Chapter
11 we show that (first class) {\it  primary\/} constraints can be
recovered in much the same way, but the proof is more difficult and
requires  different techniques. For the primaries the relevant object
is not
$\cg$ but rather a certain  foliation $\dot \cg_\Sigma$ of 
$\cy_\Sigma$ derived from $\cg$; intuitively, $\dot \cg_\Sigma$
consists of the ``time derivatives'' of elements of the subgroup
$\cg_\Sig$ which stabilizes $\Sig$. This foliation has a ``momentum
map'' $\dot \cj_\Sig$ and, provided zero is a regular value, the
primary constraints are obtained by equating $\dot \cj_\Sig$ to zero.
It is possible---with some additional assumptions on the nature of the
gauge group---to realize this foliation as the orbits of a genuine
subgroup of $\cg_\Sig$; this is the subject of \S12B.

We have thus far seen that both the Hamiltonian and the initial value
constraints are ``components'' of the energy-momentum map. In
Interlude~II we show, following Gotay and Marsden [1992], that the
stress-energy-momentum  tensor of a given collection of fields arises
in a similar manner, again demonstrating the capacity of the
multimomentum map to unify disparate concepts into a single geometric
entity. The stress-energy-momentum tensor ${\Ft^{ \mu}}_\nu$ so
obtained satisfies a generalized version of the classical
``Belinfante--Rosenfeld formula'' (Belinfante [1940], Rosenfeld
[1940]), and hence naturally incorporates the canonical
stress-energy-momentum tensor and the ``correction terms'' that are
necessary to make the latter gauge-covariant. Furthermore, in the
presence of a metric on
$X$, our $\Ft^{\mu\nu}$ coincides with the Hilbert 
stress-energy-momentum tensor and hence is automatically symmetric.

\medskip 

Part IV derives the adjoint formalism for field theories. This is the
capstone of our work, where all the various aspects of a classical
field theory meld into a coherent whole. The adjoint formalism is the
starting point for investigations into the structure of the space of
solutions of the Euler--Lagrange equations, linearization stability,
quantization, and related questions.

To accomplish this objective, in Chapter 12 we distinguish two types of
field variables that play an especially important role in the initial
value analysis. The {\bfi dynamic fields\/} are those whose evolution
from given initial data is determined by well-posed evolutionary
equations. These are the fields that carry the dynamical content of
the theory and with which the adjoint formalism is concerned. By way
of contrast, the evolution of the  {\bfi kinematic fields\/} is
determined not by the Euler--Lagrange equations,  but rather by the
way the  spacetime $X$ and the covariant configuration bundle $Y$ are
sliced. In fact, the kinematic fields are closely related to the first
class primary constraints and their existence can be  directly
attributed to the presence of gauge symmetries.

The space of dynamic fields is $\cd_\Sig =\cy_\Sig/\dot\cg_\Sig$. The 
kinematic fields are abstractly realized  as sections of this quotient;
this definition makes sense\, as the primary constraints are components
of the momentum map associated to the foliation $\dot\cg_\Sig$, so
their canonically conjugate fields are param\-etr\-ized by $\dot
\cg_\Sig$.  By a slight generalization of the cotangent bundle
reduction theorem (Abraham and Marsden [1978], Kummer [1981], Marsden
[1992]), the  instantaneous phase space $T^*\cy_\Sig$ can then be cut
down to the space $T^*\cd_\Sig$ of dynamic fields $\psi$ and their
conjugate momenta $\rho$. (Physically, this reduction consists of
imposing the primary constraints and then factoring out the kinematic
fields.)

The goal is to project the dynamics on the primary constraint set
$\cp_\Sig$ in $T^*\cy_\Sig$ to parametric dynamics on $T^*\cd_\Sig$.
The parameters will be certain combinations of the  kinematic fields
and elements of $\fg$ which we call ``atlas fields.''

Unfortunately, the energy-momentum map $\ce_\Sig$ on $\cp_\Sig$ does
not directly pro\-ject to $T^*\cd_\Sig$, as it involves the kinematic
fields. But $\ce_\Sig$, thought of as a map $\cp_\Sig
\times \fg \to \mathbb R$, also depends upon $\xi \in \fg$. By
algebraically lumping the kinematic fields together with elements of
the Lie algebra of the gauge group in a ``$\dot
\cg_\Sig$-equivariant'' fashion, one forms the nondynamic {\bfi atlas
fields\/} $\alpha_i$.  Their abstract definition turns out to be
rather delicate,  and necessitates some technical assumptions
regarding the action of the gauge group on
$Y$. In any case, the atlas fields are constructed in such a way that
$\ce_\Sig$ depends only upon $(\psi,\rho)$ and the $\alpha_i$. The
energy-momentum map then can be pushed down to a function
$\Phi_\Sig$ on $T^*\cd_\Sig$ depending upon the parameters $\alpha_i$. 
This function $\Phi_\Sig$ is the ``reduced'' energy-momentum map. We
observe that the Energy-Momentum Theorem remains valid in this
context, since the secondary constraints restrict only
$(\psi,\rho)$---the kinematic fields being correlated with the
primary  constraints. Combining these results with (\ref{eqn1:4}) then
leads to item {\bf 4}.

The essential content of item {\bf 4} is not so much that the
Hamiltonian can be expressed  in terms of the constraints, as
noteworthy as that is. (Indeed, this is already apparent from
(\ref{eqn1:4}).) Rather, it is the fact that the Hamiltonian is {\bfi 
linear\/} in the atlas fields.  This is vital for the {\bfi adjoint
form} (\ref{eqn1:3}) of the evolution equations, which now follows
readily from (\ref{eqn1:2}) by writing Hamilton's equations explicitly
on $T^*\cd_\Sig$.

The salient features of the adjoint form (\ref{eqn1:3}) are:
\begin{enumerate}
\item [$\bullet$]
 The constraints generate the dynamics, via the
appearance of the  energy-momentum map in (\ref{eqn1:3}).
\item [$\bullet$]
 The atlas fields ``drive'' the entire gauge
ambiguity of the theory. This is a remarkable result, especially when
viewed from the standpoint of Dirac constraint theory: no matter how
complex the cascade of constraints, the gauge freedom generated by the
totality of first class constraints is encapsulated in the fields
$\alpha_i$  (which are closely linked to the first class {\it  
primaries\/}).
\item [$\bullet$]
 The gauge freedom of the theory is completely
subsumed in the choice of slicing. This is convenient computationally,
as the form of the adjoint equations (\ref{eqn1:3}) is ideally  suited
for studying evolution relative to various slicings.          
\item [$\bullet$]
 The dynamics is independent of the choice of
slicing, which is a consequence of the preceding remark.
\end{enumerate}

These statements are familiar in the case of the ADM formulation of
gravity, but our results are more general and include theories which
have ``internal'' (i.e., not spacetime-based) gauge freedom, such as
Yang--Mills theory.

The independence of the adjoint formalism on the choice of
slicing together with the results of Chapter 6 give item {\bf 6}. The
adjoint formalism is now ready to be used for such purposes as the
singularity analysis of the solution space, as we have mentioned.

\medskip 

The results of this work apply to a large class of systems including
most field theories of current interest. To bolster this
claim, throughout the work we provide a number of illustrations of the
formalism as we develop it. These examples include particle mechanics,
Maxwell's theory of electromagnetism,  a topological field
theory, and bosonic string theory. In Part V we present (the Palatini
variant of) Einstein's theory of gravity as a stand-alone example.

All these examples have first order Lagrangians and, except for 
Palatini gravity and the topological field theory, all of them involve
only constraints which are first class in the sense of Dirac. Indeed,
many of the theorems as stated in this work are restricted to field
theories with these properties. However, we do believe that the
essential thrust of this work holds for more general classical field
theories. Second class constraints, for example, should not cause
trouble. The main effect of the presence of second class constraints is
the failure of equality in (\ref{eqn1:5}); we are only guaranteed that
$
\mathcal{C}_{\Sigma} \subset \ce^{-1}_{\Sigma} (0)$.  As a consequence,
one can no longer compute the final constraint set using the
energy-momentum map alone. However, the strong tie between first class
constraints and gauge transformations should remain, and one should
still be able to provide theoretical support for items {\bf 1}--{\bf
7}, including the adjoint form (\ref{eqn1:3}), for such theories. Note
that one way to deal with second class constraints is to eliminate
them via a Dirac bracket construction (\'Sniatycki [1974]); while this
presents no problem in principle, it is sometimes unsatisfactory from
a computational standpoint. Another way to handle second class
constraints might be based on the work of Lusanna [1991,\,1993], which
relates them to transformations that leave the Lagrangian {\it
quasi\/}-invariant. Note also that in well-known  field theories with
second class constraints---for example, the Einstein--Dirac theory
(see Bao, Isenberg, and Yasskin [1985], and Bao [1984]) and the KdV
equation (see Gotay [1988])---those constraints do not introduce undue
difficulties. 

We also expect---modulo some technical difficulties---that the
formalism developed here can  be straightforwardly applied to higher
order field theories. Such a generalization is necessary in particular
to directly treat Einstein's theory of gravity, since the Hilbert
Lagrangian is second order in the metric. In fact, the multisymplectic 
setup and its space + time decomposition have already been extended to
field theories of any order in Gotay [1988,\,1991a,b]; see also
Kouranbaeva and Shkoller [2000]. However, for third or higher order
systems the Cartan form is not uniquely determined by the Lagrangian
(Kol\'a\v{r} [1984]). This circumstance  bears on an important fact
which underlies our analysis; viz. for first (or second) order 
theories, a symmetry of the Lagrangian is also a symmetry of the
Cartan form. (This is essential for Noether's theorem among other
things.) Thus in the higher order case it is necessary to determine
whether there exist $\cg$-invariant Cartan forms. The work of Mu\~noz
[1985] may be helpful in this context.

We believe that one further generalization should hold. Namely, the
principal ideas of this work should apply to cases in which one is
naturally led to consider a Poisson, as opposed to a symplectic
structure, such as in certain formulations of general relativistic 
perfect fluids and plasmas (cf. Bao, Marsden, and Walton [1985] and
references therein). It could be useful to develop a covariant
analogue of Poisson manifolds (perhaps along the lines of Marsden,
Montgomery, Morrison, and Thompson [1986]) and to develop a version of
covariant reduction that could start with the multisymplectic ideas of
this work and produce Poisson structures as in the nonrelativistic
case. (See, for example, Marsden, Weinstein, Ratiu, Schmid, and
Spencer [1983] and Marsden [1992] for a summary of the nonrelativistic
reduction theory, and Holm [1985] for indications of how the
relativistic theory should proceed.) Recent results on reduction for
field theories can be found in Castrillon and Marsden [2002].

Finally, we remark that multisymplectic geometry may be used for other 
interesting purposes besides the investigation of the intrinsic
structure of classical field theories.  In particular, Marsden and
Shkoller [1999] and Marsden, Patrick, and Shkoller [1998] have shown
that Bridges' [1997] treatment of water waves and their instabilities
may be understood in multisymplectic terms and that one can build
numerical integrators using these ideas. One can also do continuum
mechanics from the multisymplectic point of view as in Marsden, 
Pekarsky, Shkoller, and West [2001]; this leads to a natural way of
incorporating `constraints' such as incompressibility. Covariant
techniques are also proving useful in understanding the dynamical
structure of the gravitational field; see Ashtekar, Bombelli, and Reula
[1991] and Anco and Tung [2002] for some results in this direction.

\section*{Acknowledgments} 

We thank Judy Arms, David Bao, Paul Dedecker, Christian G\"unther,
Marvin Ortel, Gunter Schwarz, Jim Stasheff, and Wiktor Szczyrba for
their helpful comments. We thank the University of California
Committee on Research and Cornell University which helped defray the
typing costs. We also thank MSRI at Berkeley, whose hospitality
allowed us to frequently collaborate  during the 1988--1989 symplectic
geometry year. 

This project began in Calgary in 1979 and continued in
Aix-en-Provence, Annapolis, Berkeley, Boston, Boulder, Brno, Canberra,
Champaign-Urbana, Clausthal-Zellerfeld, College Station, Eugene,
Hamburg, H\"amelschenburg, Honolulu, Houston, Ithaca, Logan, Los
Alamos, Minneapolis, Montr\'{e}al, Palo Alto, Paris, Pasadena,
Philadelphia, Potsdam, San Diego, San Luis Obispo, Santa Barbara,
Santa Cruz, Seattle, Stockholm, Sydney, Toronto, Waterloo, Washington,
and in the Friendly Skies. It has been a long journey, and over the
years it has been gratifying that a number of ideas that originated
here (such as the notions of the dual jet bundle and the covariant
momentum map) have found their way into the standard mathematical
physics literature. However, the main results of this work, such as
those pertaining to the link between the covariant and the
instantaneous formulations of dynamics, the Energy-Momentum Theorem,
and atlas fields and the adjoint formalism, as outlined in the
Introduction, have not previously been available.

We thank the people who have been patient with us during the writing
of this work, especially our long-suffering wives, children, and
relatives (who thought it would \emph{never} end). We appreciate both
our readers and the typesetters, Sue Knapp, Marnie McElhiney, Wendy
McKay, June Meyerman, Teresa Wild, Esther Zack, and Arlene Baxter, Sean
Brennan, Jo Butterworth, David Mostardi, and Margaret Pattison of
MSRI.  This work is dedicated to graduate students who may undertake 
to read it, and to the spirit of G. Gimmsy, may she rest in peace.

\newpage
\part*{\large\bfseries
{\LARGE I}---{\LARGE C}OVARIANT
{\LARGE F}IELD {\LARGE T}HEORY}

\addcontentsline{toc}{part}{I---Covariant Field Theory}{\null}
\bigskip\bigskip

In this part we develop some of the basic theory of classical
fields from a covariant  viewpoint. This is done in the framework of
multisymplectic geometry, which is analogous to the  geometry of the
cotangent bundle in classical nonrelativistic mechanics---which we
often call  particle mechanics (even though it also includes continuum
mechanics). We also develop  the geometry of the jet bundle
and the Euler--Lagrange equations, which is analogous to the  geometry
of the tangent bundle and the Lagrange equations of motion in the case
of particle  mechanics. The final chapter of this part discusses
conservation laws and Noether's theorem from the point of view of
covariant momentum maps, again generalizing the concept of momentum
map  familiar from particle mechanics.  

\section{
Multisymplectic Manifolds}

There are two major approaches to the symplectic description of
classical field theory, the covariant (or ``multisymplectic") formalism
and the instantaneous (or ``3+1'') framework. In the latter 
approach dynamics is described in terms of the {\it
infinite\/}-dimensional space of fields at a given  instant of time,
whereas in the former dynamics is phrased in the context of the {\it
finite\/}-dimensional space of fields at a given event in spacetime. 
Each formalism has its own advantages.  

In this and the next several sections, we summarize those aspects of
the multisymplectic  formalism and its relation to the 3+1 framework
that will be important for subsequent  developments.  There are many
recent references for the multisymplectic formalism, such as Dedecker 
[1953,\,1977], Gaw\c edzki [1972], Goldschmidt and Sternberg [1973]
(see also Guillemin and Sternberg [1977,\,1984]), Kijowski
[1973,\,1974], Garc\'\i a [1974], Kijowski and Szczyrba [1975,\,1976], 
Kijowski and Tulczyjew [1979], Aldaya and de Azc\'arraga [1980a], 
Kupershmidt [1980], Kosmann--Schwarzbach [1981], Ragionieri and Ricci
[1981], Kastrup \linebreak[4] [1983],
\'Sni\-atycki [1970a,\,1984], de~Le\'on and Rodrigues [1985], G\"unther
[1987], Gotay [1991a,b], Sardanashvily [1993], Kanatchikov [1998],
Marsden, Patrick, and Shkoller [1998], and Castrillon and Marsden
[2002]. Historically, much credit should also be given to Cartan
[1922], De~Donder [1930], Weyl [1935] and Lepage [1936]. Our
presentation here is self-contained, but not  exhaustive; topics not
needed for our main results as well as alternative approaches have
been  omitted.
\medskip

Throughout this work, all manifolds and maps are assumed to be of
class~$C^\infty$. Our conventions follow Misner, Thorne, and
Wheeler [1973] to a large extent.

\subsection{
The Jet Bundle}

	Let $X$ be an oriented manifold, which in many examples is spacetime,
and let $\pi_{XY}: Y \to  X$ be a finite-dimensional fiber bundle
called the {\bfi   covariant configuration bundle\/}.  The fiber 
$\pi^{-1}_{XY} (x)$ of $Y$ over $x \in X$ is denoted $Y_x$ and the
tangent space to $X$ at $x$ is denoted $T_xX$,  etc.   The physical
fields will be sections of this bundle, which is the covariant
analogue of the  configuration space in classical mechanics.

	We shall develop classical field theory in parallel to classical
mechanics.  First we discuss  the field-theoretic analogues of the
tangent and cotangent bundles.  The role of the tangent bundle is 
played by $J^{1}Y$, the {\bfi first jet bundle\/} of $Y$. We recall its
definition. We say that two local section $\phi_1$, $\phi_2$ of $Y$
{\bfi agree to first order\/} at $x \in X$ if each is defined in some
neighborhood of $x$ and if their first order Taylor expansions are
equal  at $x$ (in any coordinate system). In other words, they agree
to first order if and only they have the same target
$y = \phi_1 (x) = \phi_2 (x)$ and if their linearizations
$T_x \phi_1 $ and $T_x \phi_2 $ at $x$ coincide as linear maps 
$T_x X \rightarrow T_y Y$. Agreement to first order defines an
equivalence relation on the set of all local sections; the resulting
set of equivalence classes is the first jet bundle $J^1 Y$. Saunders
[1989] contains much information on jet bundles.

It is convenient to use the natural identification of $J^{1}Y$  with
the {\it affine\/} bundle over $Y$ whose fiber above $y \in Y_x$
consists of those linear mappings $\gamma : T_xX \to T_yY$ satisfying
\begin{equation} \label{eqn2A:1} 
T\pi_{XY} \circ \gamma = \text{Identity on } T_x X. 
\end{equation}
The map $\gamma$  corresponds to $T_x \phi$ for a  local section
$\phi$. The vector bundle underlying this affine bundle is the bundle
whose fiber over $y \in Y_x$ is the  space $L(T_xX, V_yY)$ of linear
maps of $T_xX$ to $V_yY$, where
\begin{equation} \label{eqn2A:2} 
 V_y Y = \{ v \in T_y Y \mid T\pi_{XY} \cdot v = 0\} 
\end{equation}
is the fiber above $y$ of the {\bfi vertical subbundle\/}
$VY \subset TY$.  Note that for $\gamma \in J^1_y Y$, we have the
splitting
\begin{equation} \label{eqn2A:3}
 T_yY  = \operatorname{im}  \gamma \oplus V_yY. 
\end{equation}

\begin{remark}[Remark \ 1.]
One can view sections of $J^{1}Y$  over $Y$ as
\emph{Ehresmann connections} on $Y$. (See Hermann [1975] or Marsden,
Montgomery, and Ratiu [1990] for a brief review of  Ehresmann
connections.) The horizontal-vertical splitting (\ref{eqn2A:3}) defines
the connection  associated to such a section.  If $Y$ is a principal
$G$-bundle, then $J^{1}Y \to Y$ is the bundle  whose equivariant
sections are connections in the usual sense of principal bundles.
\footnote{\ In this  case $J^{1}Y$, as a principal bundle over the
connection bundle $C Y  = J^1 Y  /G$, itself has a  canonical
connection. For the bundle $Q \times \mathbb R \to Q$ regarded  as  an
$\mathbb R$-bundle, $J^1(Q \times \mathbb R)$ is isomorphic  to the
bundle $T^*Q \times \mathbb R$ and the canonical connection is the
usual contact one-form $\theta$. The curvature of this connection is
the canonical symplectic form $\mathbf{d} \theta$,  up to a sign
depending on conventions.}
\end{remark}
	
In Chapter 3 we shall study Lagrangians on $J^{1}Y$  and the attendant
Euler--Lagrange equations,  etc., by a procedure parallel to that used
for Lagrangians on the tangent bundle of a configuration  manifold. 
Our choice of $J^{1}Y$  as the field-theoretic tangent bundle reflects
the fact that all of the theories we shall consider have
Lagrangians which depend at most on the fields and their {\it first\/}
derivatives.  For higher order field theories, the appropriate objects
to study are the higher order jet bundles; see, for example,  Aldaya
and Azc\'arraga [1980b], Shadwick [1982], Ferraris and Francaviglia
[1983], Krupka [1987], Hor\'ak and Kol\'a\v r [1983], de Le\'on and
Rodrigues [1985], Gotay [1991a,b], and references therein.
\medskip

We let $\dim X = n + 1$ and the fiber dimension of $Y$ be $N$. 
Coordinates on $X$ are  denoted $x^\mu, \;\mu = 0, 1, 2, \dots , n$,
and fiber coordinates on $Y$ are denoted by $y^A,\;  A = 1, \dots ,
N$.  These induce coordinates $v^A{}_\mu$ on the fibers of $J^{1}Y$. 
If $\phi : X \to Y$ is a section of $\pi_{XY}$, its  tangent map
$T_x\phi$ at $x \in X$ is an element of $J^{1}_{\phi(x)}Y$. Thus, the
map $x \mapsto T_x\phi$ is a section of $J^{1}Y$  regarded as a
bundle over $X$.  This section is denoted
$j^1\ns\phi$ and is called the {\bfi first jet prolongation\/} of
$\phi$.  In coordinates, $j^1\ns\phi$ is given by
\begin{equation}\label{eqn2A:4} 
 x^\mu \mapsto \left( x^\mu, \phi^A \!\left(x^\mu \right), \partial_\nu
\phi^A\!
\left(x^\mu \right) \right),
\end{equation}
where $\partial_\nu = {\partial}/{\partial x^\nu}$.  A section of the
bundle $J^{1}Y \to X$  which is the first jet of a section of $Y
\to X$  is said to be {\bfi holonomic\/}. In this work we shall use
global sections, although at certain points in the arguments one should
really work with local sections. It will be clear in each instance what
is meant.

\startrule
\vskip-12pt
            \addcontentsline{toc}{subsection}{Examples}
\begin{examples}
\mbox{}\\[-18pt]
\paragraph{\bf a\ \; Particle Mechanics.}\enspace 
For non-relativistic classical mechanics with configuration  space
$Q$,  let $X = \mathbb R$ (parameter time) and $Y = \mathbb R \times
Q$,  with $\pi_{XY}$ the projection onto the first factor.  The first
jet bundle $J^{1}Y$ is the bundle whose holonomic sections are
tangents of sections $\phi : X \to Y$ (for example, curves in $Q$), so
we can identify $J^1Y  = \mathbb R \times TQ$. 
(Note however that
$J^{1}Y$  is an {\it  affine\/} bundle, while $\mathbb R \times TQ$
has a {\it  vector\/} bundle structure). Using coordinates $(t, q^A)$
on $\mathbb R \times Q$, the induced coordinates on $J^{1}Y$  are the
usual tangent coordinates $(t, q^A, v^A)$.
\medskip

In the case of a relativistic particle we would take $Q$ to be the
physical spacetime equipped with a  Lorentzian metric $g$.  (In this
context, $A = 0,1, 2, 3$ is now a spacetime index.)

\paragraph{\bf b\ \; Electromagnetism.}\enspace 
For electromagnetism on a fixed background (so that the  spacetime
metric is ``frozen''), $X$ is four-dimensional spacetime and $Y$ is the
bundle $\Lambda^1 X$  of one-form potentials $A$.  Coordinates for
$J^{1}Y$  are $(x^\mu, A_\nu, v_{\nu\mu})$,  where the derivative index
$\mu$ is placed at the end:   
\begin{equation*}\label{2Aex:b1}
 v_{\nu\mu} \circ A = \partial_\mu A_\nu = A_{\nu,\mu}. 
\end{equation*}
The one-forms $A$ may be thought of as connections on a principal
$U(1)$-bundle. However, here we treat electromagnetism
as an ordinary one-form field theory.  Similarly, one could treat
Yang--Mills fields with all the results of the paper holding, by
taking $Y$ to be the connection bundle over spacetime (i.e., the bundle
whose sections are connections of a  given principal bundle). 
\medskip

We also consider the case when the spacetime metric $g$ is
variable (but still not variational), so that the theory is
parametrized. (See Interlude I. If we wanted the spacetime metric
to be variational, we could couple electromagnetism to gravity.) To
accomplish this, we merely append $g$ to the other field variables
as a {\em parameter\/}, replacing $Y$ and $J^1 Y$  by 
${\tilde Y} = Y \times_X S_2^{\ps 3,1}(X)$ and
$J^1 Y \times_X S_2^{\ps 3,1}(X)$, respectively, where
$S_2^{\ps 3,1}(X)$ is the bundle of symmetric covariant symmetric
two-tensors of Lorentz signature $(-,+,+,+).$ The coordinates on
$J^1 Y \times_X S_2^{\ps 3,1}(X)$ are now
$(x^{\mu},A_{\nu},v_{\nu\mu};g_{\sigma\ns\rho})$, where we use a
semicolon to separate variational fields from parametric ones. Note, in
particular, that there are no ``multivelocities''
$v_{\sigma\ns\rho\mu}$ corresponding to the $g_{\sigma\ns\rho}.$

\paragraph{\bf c\ \; A Topological Field Theory.}\enspace  Examples
{\bf b} and {\bf d} following are both ``metric'' field theories, that
is, theories which carry a metric (of some signature, usually
Lorentzian) on the parameter space $X$. Here we give a simple example
of a non-metric field theory: the {\bfi {\rm (}abelian{\rm )}
Chern--Simons theory.\/} Topological field theories---of which the
Chern--Simons theory is perhaps the best-known illustration---behave
rather differently than metric theories, and exhibit a variety of
interesting phenomena. We refer the reader to Horowitz [1989] for a
fuller discussion of such theories.

Superficially, the Chern--Simons theory resembles electromagnetism. We
take $X$ to be a 3-manifold and $Y =\Lambda^{1}X$. Coordinates on
$J^{1}Y$ are then $(x^{\mu},A_{\nu},v_{\nu\mu})$. As with
electromagnetism, one could view $A$ as a connection on some principal
$U(1)$-bundle over $X$. One could also consider non-abelian
generalizations, where $U(1)$ is replaced by an arbitrary Lie group.
But the key difference here is that we do not introduce a metric on
$X$.

\paragraph{\bf d\ \; Bosonic Strings.}\enspace   
Let $X$ be a 2-dimensional manifold with coordinates $x^\mu$  for 
$\mu = 1, 0$. Also, let $(M, g)$ be a $(d+1)$-dimensional spacetime 
manifold $M$ with coordinates $\phi^A$ for $A = 0, 1, \dots, d$ and a
fixed metric $g$ of Lorentz signature $(-,+, ..., +)$. A {\bfi bosonic
string\/} is a map $\phi : X \to  M$.  Equivalently, its graph is a
section of the product bundle $Y = X \times M$ over $X$.  To write down
the Lagrangian, we will also need a metric (or at least a conformal
metric) on $X$.  Thus, let $S^{1,1}_2(X)$ denote the bundle over $X$ of
symmetric covariant rank two tensors of Lorentz signature $(-,+)$. 
Then a section of $S^{1,1}_2(X)$ is a metric $h$ on $X$.

	There are (at least) three approaches to classical bosonic string
theory which differ in the  choice of independent variables and the
Lagrangian. In the first (due to Polyakov [1981]), $\phi$ and  
$h$ are taken as independent variables and the Lagrangian is taken as
the harmonic map Lagrangian.   In the second (due to Nambu [1970]),
$\phi$ is taken as the independent variable, while $h$ is taken to be
the induced metric
\begin{equation}\label{2Aex:e1}
 h_{\sigma\ns\rho} (x) = \left( \phi^*g \right)_{\sigma\ns\rho}
(x) = g_{AB}\! \left( \phi(x) \right) 
\partial_\sigma \phi^A \partial_\rho\phi^B. 
\end{equation}
Plugging these into the harmonic map Lagrangian, it reduces to the
minimal surface (i.e., area)  Lagrangian.  Between these two extremes
is the third version which takes $\phi$ and the conformal  part of $h$
as the independent variables.  Since the harmonic map Lagrangian is
conformally  invariant, it may be reexpressed in terms of these
variables.  For simplicity, we shall only consider  the Polyakov model
in this paper since it treats the metric $h$ variationally.  This
model has also  been treated by Batlle, Gomis and Pons [1986]; however
they assume that $g$ is flat whereas our $g$  is not necessarily
flat.  Our analysis of the bosonic string should also be compared with
that of  Beig [1989].  One can analyze the Nambu model in a similar
way; see, for example, Scherk  [1975].

With $\phi$ and $h$ chosen as the independent variables, the covariant
configuration bundle is  
$Y = \left (X \times M \right) \times_X S^{1,1}_2 \left( X \right)$ and 
\begin{equation*}\label{2Aex:e2}
 J^1  Y  = J^1 \left( X \times M \right) \times_X
J^1(S^{1,1}_2(X)).
\end{equation*}
The coordinates on $J^{1}Y$  are then
$(x^\mu, \phi^A, h_{\sigma\ns\rho}, v^A{}_\mu, w_{\sigma\ns\rho\mu})$.
\vskip -18pt
\end{examples}
\startrule
\vskip -24pt
\begin{remark}[Remark \ 2.]
Concerted efforts to  understand Einstein's theory of gravity have
provided much of the motivation and insight for this work. It is
therefore perhaps surprising that the (vacuum) gravitational field is
a somewhat exceptional case.  Here $X$ is again  spacetime and the
natural choice for $Y$ is $S^{3,1}_2(X)$. With this choice, however,
one is forced to work on the second jet bundle $J^2 Y$ since the 
Hilbert Lagrangian depends upon the second derivatives of the metric.
To avoid this, one may adopt the Palatini approach, using the metric
{\it and\/} the torsion-free  connection as the primary fields. In
terms of these variables, the Hilbert Lagrangian is first order,  and
so the analysis may be done on the first jet bundle. The price we must
pay for this simplification is that the Palatini formalism is plagued
by a plethora of second class constraints. \'Sniatycki [1970b],
Szczyrba [1976b], and  Binz,  \'Sniatycki,  and Fischer [1988] have
studied this approach and it is the one we shall follow. An
alternative approach has been developed by Ashtekar [1986,\,1987]
using another set of canonical variables. It would be of interest to
develop the  formalism herein for this formulation of gravity; first
steps in this direction have been taken by Esposito, Gionti, and
Stornaiolo [1995]. 

Because of its length and intricacy, we defer the analysis of
Palatini gravity until Part V. In the meantime, 
we observe that bosonic strings mimic gravity to a significant extent,
while having the virtue of being a first order theory with only first
class constraints.
\end{remark}

\newpage

\subsection{
The Dual Jet Bundle}
 
Next we introduce the field-theoretic analogue of the cotangent
bundle.  We define the {\bfi dual  jet bundle\/} $J^{1}Y^\star$ to be
the {\it  vector\/} bundle over $Y$ whose fiber at $y \in Y_x$ is the
set of {\it  affine\/} maps from $J^{1}_yY$ to $\Lambda^{n+1}_x X$,
where $\Lambda^{n+1} X$ denotes the bundle of $(n+1)$-forms on
$X$.\footnote{\ The space of affine maps from an affine space into a
linear space forms a vector space.  Hence $J^1 Y^\star$ is a vector
bundle, despite the fact that $J^1 Y$ is only an affine bundle.}  A
smooth section of $J^{1}Y^\star$ is therefore an affine bundle map of
$J^{1}Y$  to $\Lambda^{n+1} X$ covering $\pi_{XY}$. We choose affine
maps since $J^1Y$ is an affine bundle, and we map into
$\Lambda^{n+1} X$ since we are  ultimately thinking of integration as
providing the pairing on sections. 

Fiber coordinates on $J^{1}Y^\star$ are $\left( p, p_A{}^\mu \right) $,
which correspond to the affine map given in  coordinates by
\begin{equation} \label{eqn2B:1}
v^A{}_\mu \mapsto \left( p + p_A{}^\mu v^A{}_\mu \right) d^{\ps
n+1}\ns x
\end{equation}
where 
\[
d^{\ps n+1}\ns x =  dx^0  \wedge dx^1 \wedge \dots
\wedge dx^n.
\]

Analogous to the canonical one- and two-forms on a cotangent bundle,
there are canonical  forms on $J^{1}Y^\star$.  To define these, another
description of $J^{1}Y^\star$ will be convenient.  Namely, let $\Lambda
:= \Lambda^{n+1} Y$ denote the bundle of $(n+1)$-forms on $Y$,  with
fiber over $y \in Y$ denoted by $\Lambda_y$ and with projection
$\pi_{Y\ns \Lambda}: \Lambda \to Y$.  Let $Z \subset \Lambda$ be the
subbundle whose fiber is given by
\begin{equation} \label{eqn2B:2}
Z_y  =  \left\{z \in \Lambda_y \mid \mathbf{i}_v\mathbf{i}_w z = 0
\text{ for all } v, w \in V_yY\right\},
\end{equation}
where $\mathbf{i}_v$ denotes left interior multiplication by $v$.

Elements of $Z$ can be be written uniquely as
\begin{equation}\label{eqn2B:3}
 z  =  p\,d^{\ps n+1}\ns x + p_A{}^\mu dy^A \wedge d^{\ps n}\ns  x_\mu 
\end{equation}
where
\begin{equation*}
 d^{\ps n}\ns  x_\mu  =  \partial_\mu \hook \, d^{\ps n+1}\ns x .
\label{eqn2B:4}
\end{equation*}
Here $\hook$ also denotes the interior product. Hence fiber coordinates
for $Z$ are also $(p,p_A{}^\mu)$.

Equating the coordinates $(x^\mu, y^A, p, p_A{}^\mu)$ of $Z$ and of
$J^{1}Y^\star$ defines a vector bundle  isomorphism
\begin{equation*}\label{eqn2B:5}
\Phi : Z \to J^1 Y^\star. 
\end{equation*}
Intrinsically, $\Phi$ is defined by
\begin{equation}
\langle\Phi \left( z \right),\gamma\rangle = \gamma^*z \in 
\Lambda^{n+1}_x X 
\label{eqn2B:6}
\end{equation}
where $z \in Z_y$, $\gamma \in J^1_y Y$, $x = \pi_{XY}(y)$, and
$\langle\cdot,\cdot\rangle$ denotes the dual pairing.  To see this, 
note that if $\gamma$  has fiber coordinates $v^A{}_\mu$, then
\begin{equation}
\gamma^*dx^\mu  =  dx^\mu \quad \text{ and  } 
\quad \gamma^*dy^A  =  v^A{}_\mu
dx^\mu 
\label{eqn2B:7}
\end{equation}
and so
\begin{equation}
\gamma^* \!\left( p\,d^{\ps n+1}\ns x + p_A{}^\mu dy^A 
\wedge d^{\ps n}\ns x_\mu \right)  = 
\left( p + p_A{}^\mu v^A{}_\mu \right) d^{\ps n+1}\ns x, 
\label{eqn2B:8}
\end{equation}
where we have used 
\[
dx^\nu \wedge d^{\ps n}\ns  x_\mu =
\delta^\nu_\mu \,d^{\ps n+1} \ns x.
\]

The inverse of $\Phi$ can also be defined intrinsically, although it
is somewhat more  complicated.  Notice that {\it a connection is not
needed to define this isomorphism\/}.

	We summarize:

\begin{prop}\label{prop2.1}
The spaces $J^{1}Y^\star$ and $Z$ are
canonically isomorphic as vector bundles  over $Y$.  
\end{prop}
 
We shall construct canonical forms on $Z$ and then use the
isomorphism between $J^{1}Y^\star$  and $Z$ to transfer these to
$J^{1}Y^\star$.  We  first define the {\bfi canonical\/}
$(n+1)$-{\bfi form\/} $\Theta_\Lambda$ on $\Lambda$ by 
\begin{align} \label{eqn2B:9}
 \Theta_\Lambda \!\left( z \right) \left( u_1, \dots, u_{n+1}\right) 
       &= z\left( T\pi_{Y\ns\Lambda} \cdot u_1, \dots,
T\pi_{Y\ns\Lambda} \cdot u_{n+1}\right) \nonumber \\[1.5ex] 
       &= \left( \pi^*_{Y\ns\Lambda}z \right) \left( u_1, \dots,
u_{n+1}\right),
\end{align}
where $z \in \Lambda$ and $u_1, \dots, u_{n+1} \in T_z\Lambda$.  Define
the {\bfi canonical\/} $(n+2)$-{\bfi form\/}
$\Omega_\Lambda$ on $\Lambda$ by
\begin{equation}
\Omega_\Lambda = -\mathbf{d} \Theta_\Lambda.  
\label{eqn2B:10}
\end{equation}
Note that if $n = 0$ (i.e., $X$ is one-dimensional), then $\Lambda = 
T^*Y$ and $\Theta_\Lambda$ is the standard canonical one-form.  If
$i_{\Lambda Z}: Z \to \Lambda$  denotes the inclusion, the {\bfi
canonical\/} $(n+1)$-{\bfi form\/} $\Theta$ on $Z$ is defined by
\begin{equation}
\Theta = i^*_{\Lambda Z} \Theta_\Lambda 
\label{eqn2B:11}
\end{equation}
and the {\bfi canonical\/} $(n+2)$-{\bfi form\/} $\Omega$ on
$Z$ is defined by
\begin{equation}
\Omega = -\mathbf{d} \Theta = i^*_{\Lambda Z} \Omega_\Lambda.  
\label{eqn2B:12}
\end{equation}
The pair $(Z, \Omega)$ is called {\bfi multiphase space\/} or {\bfi
covariant phase space\/}. It is an example of a {\bfi multisymplectic
manifold\/}; see Remark~3 below.

From (\ref {eqn2B:3}) and (\ref {eqn2B:9})--(\ref {eqn2B:12}), one
finds that the coordinate expression  for $\Theta$ is 
\begin{equation}
\Theta = p_A{}^\mu dy^A \wedge d^{\ps n}\ns x_\mu + p\,d^{\ps n+1}\ns x, 
\label{eqn2B:13}
\end{equation}
 and so
\begin{equation}
\Omega = dy^A \wedge dp_A{}^\mu \wedge d^{\ps n}\ns x_\mu 
- dp \wedge d^{\ps n+1} \ns x.
\label{eqn2B:14}
\end{equation}

\begin{remarks}[Remarks \ 1.]
There are other possible choices of  multiphase space (cf.
Eche\-verria-Enr\'{\i}ques, Mu\~noz-Lecanda, and Rom\'an-Roy [2000] and
the  references cited at the beginning of this chapter), but this one
is the most relevant and  efficient for our purposes.  There are good
reasons for this choice: it is natural, as shown in  Proposition~2.1,
and it carries a canonical multisymplectic structure.  We refer to
Gotay  [1991a,b]  for more on why one makes this particular choice of
multiphase space.

\paragraph{\bf 2.}	
A notable feature of the covariant formalism is the appearance
of several momenta $p_A{}^\mu$,  spatial in addition to temporal, for
each field component $y^A$.  The Legendre transformation  of Chapter 3 
will provide a physical interpretation of these ``multimomenta'' as
well as of the  ``covariant Hamiltonian'' $p$.  As we shall see, the
affine nature of our constructions provides a unification of the two
parts of the Legendre transformation, namely the definitions of the
momenta and the Hamiltonian, into a single ``covariant'' entity.

\paragraph{\bf 3.}
Whether or not there is a proper abstract definition of
``multisymplectic manifold''  is open to question. As a first try, we
could say that a {\bfi multisymplectic manifold\/} is a manifold $Z$
endowed with a closed
$k$-form  $\Omega$ which is nondegenerate in the sense that
$\mathbf{i}_V\Omega \neq 0$ whenever $V$ is a nonzero tangent vector. 
This would include our case, symplectic manifolds ($k =2$), and
manifolds with a volume form ($k = \dim Z$).  But for all cases
intermediate between symplectic and volume structures  there is no
Darboux theorem. Indeed, the set of germs of such $k$-multisymplectic
forms for $2 <k< \dim Z$ has functional moduli: there are local
differential invariants of such a form, distinguishing it from
arbitrarily nearby multisymplectic forms.   Such a  theory would thus
require  extra structure, perhaps mimicking  the fibration information
contained in our $Z$, or else would proceed in a manner having very
little in common with modern symplectic geometry where the Darboux
theorem is such a central organizing fact.  For attempts at such a
``multisymplectic geometry'' see Martin [1988] or Cantrijn, Ibort, and
de Le\'on [1999]. \end{remarks}

We next recall that there is a characterization of the canonical
one-form $\theta$ on $T^*Q$ by $\beta^*\theta = \beta$  for any
one-form $\beta$ on $Q$ (Abraham and Marsden [1978],
Proposition~3.2.11). Here is an  analogue of that fact for the
canonical $(n + 1)$-form.
\begin{prop}\label{prop2.2} 
If $\sigma$ is a section of $\pi_{X\ns Z}$ and
$\phi = \pi_{Y\ns Z} \circ \sigma$, then
\begin{equation}
\sigma^*\Theta  = \phi^*\sigma, 
\label{eqn2B:15}
\end{equation}
  where $\phi^*\sigma$ means the pull-back by $\phi$ to $X$ of
$\sigma$ regarded as an $(n + 1)$-form on $Y$ along $\phi$.
\end{prop}
\begin{proof} 
Evaluate the left-hand side of equation (\ref{eqn2B:15}) on
tangent vectors $v_1, \dots, v_{n+1}$ to $X$ at $x$: 

\begin{align} 
\left( \sigma^*\Theta)(x)(v_1, \dots, v_{n+1} \right) 
& =
\Theta \left( \sigma(x) \right) \left( T\sigma\cdot v_1, \dots,
T\sigma\cdot v_{n+1} \right) 
\nonumber\\[1.5ex]
& =
\sigma(x) \left( T\pi_{Y\ns Z} \cdot T\sigma \cdot v_1, \dots,
T\pi_{Y\ns Z}\cdot T\sigma\cdot v_{n+1} \right) 
\nonumber\\[1.5ex]
& = \sigma(x) \left( T\phi\cdot v_1, \dots, T\phi\cdot
v_{n+1} \right) 
\nonumber\\[1.5ex] 
& = \left( \phi^*\sigma \right) (x) \left( v_1, \dots,
v_{n+1} \right).\tag*{\qedsymbol}
\end{align}
\renewcommand{\qedsymbol}{}
\end{proof}
\vspace{-12pt}
\startrule
            \addcontentsline{toc}{subsection}{Examples}
\begin{examples}
\mbox{}\\[-18pt]
\paragraph{\bf a\ \; Particle Mechanics.}\enspace 
Let $X = \mathbb R$ and $Y = \mathbb R \times
Q$.  Then $Z = T^*Y = T^*\mathbb R \times T^*Q$ 
has coordinates $(t, p, q^1, \dots,  q^N,  \,p_1, \dots, p_N)$,
\begin{align}
\Theta &= p_A dq^A + p\,dt 
\label{2Bex:a1}
\end{align}
and
\begin{align}
\Omega &= dq^A \wedge dp_A + dt \wedge dp.   
\label{2Bex:a2}
\end{align}
In this case the multisymplectic approach reduces to the extended
state space formulation of  classical mechanics.

\paragraph{\bf b\ \; Electromagnetism.}\enspace 
For the Maxwell theory, coordinates on $Z$ are
\begin{equation*}
 (x^\mu, A_\nu, p, \Ff^{\nu\mu}).   
\label{2Bex:b1}
\end{equation*}
Later we shall see that  $\Ff^{\nu\mu}$ can be identified, via the
Legendre transform, with the electromagnetic field {\it density\/},
so we use this notation in anticipation.  From (\ref {eqn2B:13}) and
(\ref {eqn2B:14}), we get
\begin{align}
\Theta &= \Ff^{\nu\mu} dA_\nu \wedge d^{\ps 3}\ns x_\mu 
+ p \,d^{\ps 4}\ns x 
\label{2Bex:b2}
\end{align}
and
\begin{align}
\Omega &= dA_\nu \wedge d\Ff^{\nu\mu} \wedge d^{\ps 3}\ns x_\mu 
        - dp \wedge d^{\ps 4}\ns x. 
\label{2Bex:b3}
\end{align}
\medskip

When the spacetime metric $g$ is treated parametrically, $Z$ is
replaced by ${\tilde Z} = Z \times_X S_2^{\ps 3,1}(X)$ with coordinates
$(x^{\mu},A_{\nu},p,\mathfrak{F}^{\nu\mu};g_{\sigma\ns\rho}).$ There
are no multimomenta conjugate to the $g_{\sigma\ns\rho}$. The
expressions for the canonical forms remain the same.


\paragraph{\bf c\ \; A Topological Field Theory.}\enspace 
 In the case of the Chern--Simons theory, coordinates on $Z$ are
$(x^{\mu},A_{\nu},p,p^{\nu\mu}).$ The canonical forms are
\begin{equation}
\Theta = p^{\nu\mu}dA_{\nu} \wedge d^{\ps 2}\ns x_{\mu} 
+ p\,d^{\ps 3}\ns x
\label{eq:2Bex:d1}
\end{equation}
\noindent and
\begin{equation}
\Omega = dA_{\nu} \wedge dp^{\nu\mu} \wedge d^{\ps 2}\ns x_{\mu} - dp
\wedge d^{\ps 3}\ns x.
\label{eq:2Bex:d2}
\end{equation}
\paragraph{\bf d\ \; Bosonic Strings.}\enspace 
Corresponding to $Y  = (X \times M) \times_X S^{1,1}_2(X)$, the
coordinates on $Z$ are $(x^\mu, \phi^A, h_{\sigma\ns\rho}, p, p_A{}^\mu,
q^{\sigma\ns\rho \mu})$, and the canonical forms are given by
\begin{equation}
\Theta  =  \left( 
p_A{}^\mu d\phi^A + q^{\sigma\ns\rho \mu} dh_{\sigma\ns\rho}
\right) \wedge d^{\ps 1}\ns x_\mu + p\,d^{\ps 2}\ns x 
\label{2Bex:e1}
\end{equation}
and      
\begin{equation}\label{2Bex:e2}
\qquad\qquad\Omega  =  \left( d\phi^A \wedge dp_A{}^\mu +
dh_{\sigma\ns\rho}
\wedge dq^{\sigma\ns\rho \mu}\right) \wedge d^{\ps 1}\ns x_\mu - dp \wedge 
d^{\ps 2}\ns x. 
\end{equation}
\end{examples}
\vskip-18pt
\startrule
\vskip 12pt

We conclude this section with a few remarks on Poisson brackets.
Poisson brackets are not  essential for the main points in this paper,
but we believe that they are important for linking up the  results of
this paper with those of Marsden, Montgomery, Morrison, and Thompson
[1986] and for the future development of  covariant reduction (see
Marsden [1988]). For more thorough discussions of Poisson brackets in
field theory, we refer the reader to the recent papers by Kanatchikov
[1997], Lawson [2000], Castrillon and Marsden [2002], and Forger,
Paufler, and R\" omer [2002a,b].

Recall that on a symplectic manifold $(P, \Omega)$, the Hamiltonian
vector field $X_{\ns f}$ of a  function $f \in \cf(P)$ is defined by 
$$\Omega(X_{\ns f}, V) = V \hook\, {\mathbf{d}}f$$
for any $V \in\Fx(P)$, and the Poisson bracket is defined by 
$$
\{f, h\} = \Omega(X_{\ns f},X_h).
$$
The multisymplectic analogue of a function is an $n$-form. In
Goldschmidt and Sternberg [1973] and Kijowski and Szczyrba [1975] a
Poisson bracket  is defined on certain types of $n$-forms.  A special
case relevant for later considerations is as  follows.  Let $V$ be a
vector field on $Y$ which is $\pi_{XY}$-projectable;  that is, there
is a vector field $U$ on $X$ such that
\begin{equation*}
 T\pi_{XY} \circ V = U \circ \pi_{XY}. 
\label{2B:16}
\end{equation*}
 Consider an $n$-form on $Z$ of the type 
\begin{equation}
 f(z) = \pi^*_{Y\!Z}( V \hook \, z)
\label{2B:17}
\end{equation}
 for some $\pi_{XY}$-projectable vector field $V$; we call such an
$f$ a {\bfi momentum observable\/}. (As we shall see in \S 4C, special
covariant momentum maps can be regarded as momentum observables.)  The
{\bfi Hamiltonian vector field\/} $X_{\ns f}$ of such a  momentum
observable $f$ is defined by
\begin{equation*}
 {\mathbf{d}}f = X_{\ns f} \hook\, \Omega 
\label{2B:18}
\end{equation*}
where $\Omega$ is the canonical $(n+2)$-form on $Z$.  Since $\Omega$ is
nondegenerate, this uniquely defines $X_{\ns f}$.  The {\bfi Poisson
bracket\/} of two such $n$-forms is the $n$-form defined by
\begin{equation}
\{f, h\} = X_h \hook (X_{\ns f}\hook\,\Omega). 
\label{2B:19}
\end{equation}
For momentum observables $f$ and $h$, $\{f, h\}$ is, up to the
addition of exact terms, another  momentum observable; these constitute
an important class since, as we shall see, it includes
superhamiltonian and supermomentum functions, for instance. Also, the
development in Chapter 5 of Part II shows that when (\ref{2B:19}) is
integrated over a hypersurface, it produces a canonical cotangent
bundle bracket.

\section{
Lagrangian Dynamics}
 
	This chapter summarizes some of the basic facts concerning the
Euler--Lagrange equations,  the Legendre transformation and the Cartan
form that will be needed later.  There are many  formulations available
in the literature, including those of Carath\'eodory  [1935], Weyl
[1935], Lepage [1936,\,1941,\,1942], Cartan [1922], Dedecker
[1957,\,1977], Trautman [1967],  Hermann [1968],  \'Sniatycki [1970a,
\linebreak
1984], Krupka [1971], Ouzilou [1972], Goldschmidt and Sternberg
[1973],  Garc\'\i a [1974], Szczyrba [1976a], Sternberg [1977],
Kupershmidt [1980], Aldaya and de  Azc\'arraga [1980a], Griffiths
[1983], and Anderson [1992]. For second order theories, see Krupka and
\v{S}tep\'ankov\'{a} [1983] and, for the $n^{th}$ order case, Hor\'ak
and Kol\'a\v r [1983], Mu\~noz [1985], and Gotay [1991a,c]. Exactly
which route one takes to end up with the basic results is, to some
extent, a matter of personal preference.  For example, one may
introduce the Cartan form  axiomatically and then define the Legendre
transformation in terms of it; see for instance 
\'Sniatycki [1970a,\,1984] for this point of view. We have chosen to
introduce the Legendre  transformation directly and then use it to
define the Cartan form to most closely parallel the  standard
treatment in classical mechanics (Abraham and Marsden [1978], Section
3.6). We shall continue to confine  ourselves to first order theories.
\medskip   

As before, let $\pi_{XY} : Y \to  X$ be a bundle over $X$ whose
sections will be the fields of the  theory.  In Chapter 2 we studied
the spaces $J^{1}Y$  and $Z \cong J^1 Y^\star$ which are the
field-theoretic analogues of the tangent and cotangent bundles in
particle mechanics.  Here we are concerned with the relationship
between the two and with Lagrangian dynamics on the former.

\subsection{
The Covariant Legendre Transformation}
 
Let
$$
\cl : J^1 Y \to \Lambda^{n+1} X,
$$ 
the {\bfi Lagrangian density\/}, be a given smooth bundle map over
$X$.  In coordinates, we write
\begin{equation}
\cl  =  L \left( x^\mu, y^A, v^A{}_\mu \right) d^{\ps n+1}\ns x. 
\label{eqn3A:1}
\end{equation}
We remark that since $X$ is assumed oriented, we may take $\cl(\gamma)$ to
be an $(n+1)$-form on $X$. If  $X $ were not orientable, then we should
more properly take  $\cl(\gamma)$ to be a density of weight one on  $X$.

Our first goal is to construct the covariant Legendre transformation
for $\cl$.  This is a fiber-preserving  map
$$
\mathbb F\cl : J^1 Y \to  J^1 Y ^\star \cong  Z
$$ 
over $Y$ which has the coordinate expressions
\begin{equation}
 p_A{}^\mu = \frac{\partial L}{\partial v^A{}_\mu}, \quad 
p = L -\frac{\partial L}{\partial v^A{}_\mu} v^A{}_\mu
\label{eqn3A:2}
\end{equation}
for the multimomenta $p_A{}^\mu$ and the covariant Hamiltonian
$p$.  An intrinsic definition follows.

If $\gamma \in J^1_y Y$ then $\mathbb F \cl(\gamma)$ is the affine
approximation to $\cl \! \bigm| \!\ns J^1_yY$ at $\gamma$. Thus
$\mathbb F \cl(\gamma)$ is the affine map from $J^{1}_yY$ to
$\Lambda^{n+1}_x X$ (where $y \in Y_x$) given by 
\begin{equation}
\langle\mathbb F\cl(\gamma),\gamma' \rangle = \cl(\gamma) +\left.
\frac{d}{d\varep} \cl\left(\gamma +
\varep(\gamma' - \gamma)\right )
\right|_{\varep = 0}, 
\label{eqn3A:3}
\end{equation}
 where $\gamma' \in  J^1_y Y$.  If one wishes, the second term in
(\ref {eqn3A:3}) can be interpreted in terms of a standard fiber
derivative. To derive the coordinate expressions (\ref {eqn3A:2}),
suppose $\gamma = v^A{}_\mu$ and $\gamma' = w^A{}_\mu$. Then the
right hand side of (\ref {eqn3A:3}) reads
\begin{equation*}
\left( L(\gamma) + \frac{\partial L}{\partial v^A{}_\mu} (w^A{}_\mu -
v^A{}_\mu) \right) d^{\ps n+1} \ns x, 
\label{eqn3A:4}
\end{equation*}
 which is an affine function of $w^A{}_\mu$ with linear and constant
pieces given by the first and second  equations of (\ref {eqn3A:2})
respectively.  Hence (\ref {eqn3A:2}) is indeed the coordinate
description of $\mathbb F\cl$. (The coordinatization of
$J^{1}Y^\star$ was defined by (\ref{eqn2B:1}).)

\subsection{
The Cartan Form}
 
	The {\bfi Cartan form\/} is the $(n+1)$-form $\Theta_{\cl}$ on
$J^{1}Y$  defined by
\begin{equation}
\Theta_{\cl} =  \mathbb F\cl^*\Theta, 
\label{eqn3B:1}
\end{equation}
 where $\Theta$ is the canonical $(n+1)$-form on $Z$.  We  also
define the $(n+2)$-form $\Omega_\cl$ by
\begin{equation}
\Omega_\cl = -\mathbf{d} \Theta_\cl = \mathbb F \cl^* \Omega,  
\label{eqn3B:2}
\end{equation}
where $\Omega = -\mathbf{d}\Theta$ is the canonical $(n+2)$-form on
$Z$.  

In coordinates, (\ref{eqn2B:13}), (\ref{eqn2B:14}) and (\ref{eqn3A:2})
yield
\begin{equation}
\Theta_\cl = \frac{\partial L}{\partial v^A{}_\mu} dy^A \wedge
d^{\ps n}\ns x{}_\mu + \left(L - \frac{\partial L} {\partial v^A{}_\mu}
v^A{}_\mu \right) d^{\ps n+1}\ns x  
\label{eqn3B:3}
\end{equation}
 and
\begin{equation}
\Omega_\cl = dy^A \wedge d\ns \left(\frac{\partial L}{\partial
v^A{}_\mu}\right) \wedge d^{\ps n}\ns x_\mu - d\ns \left(L -
\frac{\partial L}{\partial v^A{}_\mu}v^A{}_\mu\right) \wedge d^{\ps
n+1}\ns x.
  \label{eqn3B:4}
\end{equation}

One application of the Cartan form is to reexpress the Lagrangian for
holonomic fields as
\begin{equation}
\cl(j^1\ns\phi)  =  (j^1\ns\phi)^*\Theta_\cl.  
\label{eqn3B:5}
\end{equation}
To see this, we use (\ref {eqn3B:3}) to compute
\begin{multline*} 
(j^1\ns\phi)^*\Theta_\cl = 
\frac{\partial L}{\partial v^A{}_\mu}(j^1\ns\phi)\,d\phi^A 
\wedge \,d^{\ps n}\ns x_\mu
+ \left(L(j^1\ns\phi) - \frac{\partial L}{\partial
v^A{}_\mu}(j^1\ns\phi)\,\phi^A{}_{,\mu}\right)d^{\ps n+1}\ns x. 
\end{multline*}
The first and last terms cancel and so we obtain (\ref {eqn3B:5}).  
\medskip

It is possible to obtain $\Theta_\cl$ directly via the action
functional; it is constructed from the boundary terms that appear when
the action integral is varied. The details can be found in, e.g., 
P\^aquet [1941], Krupka [1987], and Marsden, Patrick, and Shkoller
[1998].

\subsection{
The Euler--Lagrange Equations}
 
We now use the formalism developed so far to reexpress, in an
intrinsic way, the {\bfi Euler--Lagrange equations\/}, which in
coordinates take the standard form
\begin{equation}
\frac{\partial L}{\partial y^A}(j^1\ns\phi)  -
\frac{\partial}{\partial x^\mu} \left(\frac{\partial L} {\partial
v^A{}_\mu}(j^1\ns\phi)\right) = 0   
\label{eqn3C:1}
\end{equation}
 for a (local) section $\phi$ of $Y$.  The left hand side of (\ref
{eqn3C:1}) is often denoted ${\delta L}/ {\delta\phi^A}$ and is called
the {\bfi Euler--Lagrange derivative\/} of $L$.

\begin{thm}\label{thm3.1}
The following assertions regarding a section
$\phi$ of the bundle $\pi_{XY} : Y \to X$ are equivalent:
\begin{enumerate}
\renewcommand{\labelenumi}{\em (\roman{enumi})}
\item 
 $\phi$ is a stationary point of $\displaystyle \int_X
\cl(j^1\ns\phi)$;
\item 
 the Euler--Lagrange equations {\rm (\ref {eqn3C:1})} hold in
coordinates;
\item
 for any vector field\/ $W$ \!on $J^{1}Y$,
\end{enumerate}
\begin{equation}
(j^1\ns\phi)^{*}\!\left( W \hook \, \Omega_\cl \right)  = 0.
\label{eqn3C:2}
\end{equation}
\end{thm}

\begin{remarks}[Remarks \ 1.]
The theorem still holds if the section $\phi$ is only a local
section: $\phi : U \to Y_U$ where $U$ is an open subset of $X$.  The
only change is that the integral in (i) is now over
$U$.  The case that will be of importance to us is where $U$ is a
neighborhood of a Cauchy surface.

\paragraph{\bf 2.} The meaning of ``stationary point" is in the formal
sense of the calculus of variations.   Namely, a {\bfi variation\/} of
$\phi$ is provided by a curve $\phi_\lambda = \eta_\lambda \circ
\phi$,  where $\eta_\lambda$ is the flow of a vertical vector field $V$
on $Y$ which is compactly supported in $X$.  One says that $\phi$ is a
{\bfi stationary point\/} if
\begin{equation*}
\left.\frac{d}{d\lambda} \left[ \int_X \cl(j^1\ns\phi_\lambda)
\right]\right|_{\lambda=0} = 0   
\label{eqn3C:3}
\end{equation*}
 for all variations $\phi_\lambda$ of $\phi$.

\paragraph{\bf 3.}
This theorem appears more or less as we present it in, for
example, Dedecker [1953],  Trautman [1967],  \'Sniatycki [1970a,\,1984],
Ouzilou [1972], Goldschmidt and Sternberg [1973], and Guillemin  and
Sternberg [1977].

\paragraph{\bf 4.}
If in (\ref {eqn3C:2}) one requires that $s^*(\mathbf{i}_W\Omega_\cl)
= 0$ for any section $s$ of $J^{1}Y \to X$, not necessarily
holonomic---that is, not necessarily of the form $j^{1}\ns\phi$---then
condition (iii) will be equivalent to the Euler--Lagrange equations
{\it and\/} the assertion that $s$ is holonomic under the assumption
that the Lagrangian is ``regular."  A Lagrangian is called {\bfi
regular\/} if the Legendre transformation is of maximal rank or,
equivalently, if $(J^1 Y,\Omega_\cl)$ is a multisymplectic manifold.
In the formulation as we have given it,  {\it no assumption of
regularity is needed\/}; in fact such an assumption would be very
much  inappropriate, since it is not valid in any of our examples.
See, however, Garc\'\i a and Mu\~noz [1985,\,1991] and Saunders
[1992].    

\paragraph{\bf 5.}
Because the Euler--Lagrange equations are equivalent to the intrinsic
conditions (i) and  (iii), they too must be intrinsic. One can in fact
write the Euler--Lagrange derivative intrinsically in a direct way.
See Trautman [1967] and Marsden, Patrick, and Shkoller [1998]. 

\paragraph{\bf 6.}
In the above we have restricted attention to vertical variations. But
it is occasionally necessary to take ``horizontal
variations,'' as in continuum mechanics where they produce
configurational forces which are important in dislocation theory. See
Marsden, Patrick, and Shkoller [1998] for details, and Lew,
Marsden, Ortiz, and West [2002] for further applications of the split
between horizontal and vertical variations.
\end{remarks}

In the course of the proof, we will examine three cases for the vector
field $W$ in (\ref {eqn3C:2}).   The following lemma shows that two of
the cases are trivial.

\begin{lem}\label{lem3.2} 
If $\phi$ is a section of $\pi_{XY}$ and if either $W$ is tangent to
the image of $j^{1}\ns\phi$ or $W$ is $\pi_{Y, J^1 Y}$-vertical,
then     
$$ 
(j^1\ns\phi)^*\! \left( W \hook \,\Omega_\cl \right)  = 0.	
$$
\end{lem}

\begin{proof} 
First assume that $W$ is tangent to the image of $j^1\ns\phi$ in
$J^{1}Y$; that is, $W = T(j^1\ns\phi) \cdot w$ for some vector field
$w$ on $X$.  Then 
\begin{align*} 
(j^1\ns\phi)^* \!\left( W \hook \,\Omega_\cl \right)  
&= 
(j^1\ns\phi)^* \!\left( \left( T(j^1\ns\phi)
\cdot w \right) \hook \,\Omega_\cl \right) 
\\[1.5ex]
&=  w  \hook \,(j^1\ns\phi)^* \Omega_\cl,
\end{align*} 
which vanishes since $(j^1\ns\phi)^*\Omega_\cl$ is an $(n + 2)$-form on
the $(n + 1)$-manifold $X$.

Secondly, if $W$ is $\pi_{Y, J^1 Y}$-vertical, then $W$ has the
coordinate form 
$$
W = (0, 0, W^A{}_\mu ).
$$
A calculation using (\ref{eqn3B:4}) shows that
$$
W \hook \,\Omega_\cl = -W^B{}_\nu \frac{ \partial^2 L}{\partial
v^A{}_\mu \partial v^B{}_\nu } \left( dy^A \wedge d^{\ps n}\ns x_\mu -
v^A{}_\mu d^{\ps n+1}\ns x \right) ,
$$  
which vanishes when pulled back by the jet prolongation of a section of
$Y$. 
\end{proof}

\begin{proof}[Proof of Theorem~\ref{thm3.1}] 
The direct proof of equivalence of (i) and (ii) is a standard argument
in the calculus of variations that we shall omit.  We shall work
towards a proof of the equivalence of (i) and (iii) and then of (iii)
and (ii).
 
To show that (i) is equivalent to (iii), consider a variation 
$\phi_\lambda = \eta_\lambda \circ \phi$ of $\phi$ (as in Remark 2)
corresponding to a $\pi_{XY}$-vertical vector field $V$ on $Y$ with
compact support in $X$.  Using  (\ref {eqn3B:5}), we compute
\begin{align} \label{eqn3C:4}
\left.\frac{d}{d\lambda} 
\left[ \int_X \cl(j^1\ns\phi_\lambda)  
\right]
\right|_{\lambda=0} 
& = \frac{d}{d\lambda}
\left.
\left[
\int_X (j^1\ns\phi_\lambda) ^* \Theta_\cl
\right] 
\right|_{\lambda=0} 
\nonumber\\[2ex] 
& = \frac{d}{d\lambda}\left.
\left[ \int_X (j^1\ns\phi)^* (j^1 \ns\eta_\lambda) ^*\Theta_\cl
\right]
\right|_{\lambda=0} 
\nonumber\\[2ex] 
& = \int_X(j^1\ns\phi)^*\pounds_{j^1\ns V}
\Theta_\cl, 
\end{align}
where 
$$  j^{1}\ns V =  \left.\frac{d}{d\lambda} \, j^{1}\ns
\eta_\lambda \right|_{\lambda=0}$$   
is the jet prolongation of $V$ to $J^{1}Y$, given in coordinates by 
$$ j^{1}\ns V = \left( 0, V^A, \frac{\partial V^A}{\partial x^\mu}
+\frac{\partial V^A}{\partial y^B} v^B{}_\mu \right)$$  
and where $\pounds_{j^1\ns V}$ is the Lie derivative along $j^1\ns V$.
(See \S 4A for an intrinsic definition of jet prolongations and for
more information; in particular, see Remark~1 of that section). Thus,
from (\ref{eqn3C:4}) and 
$$
\pounds{}_{j^1\ns V}\Theta_\cl  = -j^1\ns V \hook \,\Omega_\cl
+ \mathbf{d} (j^1\ns V \hook \,\Theta_\cl),
$$
we get  
\begin{align} \label{eqn3C:5}
\left.\frac{d}{d\lambda} \left[ \int_X \cl(j^1\ns\phi_\lambda) 
\right] \right|_{\lambda=0} 
       & =  -\int_X(j^1\ns \phi) ^{*}( j^1\ns V\,\hook
\,\Omega_\cl )
\nonumber\\[2ex] 
       & \qquad  \qquad + \int_X \mathbf{d}\ps
(j^1\ns \phi)^{*}( j^1\ns V\,\hook
\,\Theta_\cl ) 
\nonumber\\[2ex] 
       & = -\int_X(j^1\ns \phi)^{*}( j^1\ns V\,\hook \,\Omega_\cl )
\end{align}
by Stokes' theorem and the fact that $V$, and hence $j^1\ns V$ is
compactly supported in $X$. From (\ref {eqn3C:5})  it follows that 
(iii) implies (i).

To show the converse, we first observe that any tangent vector field
$W$ on $J^{1}Y$  can be  decomposed into a component tangent to the
image of $j^{1}\ns\phi$ and a $\pi_{X,J^1 Y}$-vertical vector field.
Similarly, any $\pi_{X,J^1 Y}$-vertical vector field can be
decomposed into a jet extension of some $\pi_{XY}$-vertical vector
field $V$ on $Y$ and a $\pi_{Y,J^1 Y}$-vertical vector field.  

Thus, if (i) holds, (\ref {eqn3C:5}) and Lemma~\ref{lem3.2} show that
$$
\int_X(j^1 \ns \phi)^* (W \hook\, \Omega_\cl) = 0
$$ 
for all vector fields $W$ on $J^{1}Y$  with compact support in $X$. 
Since $W$ can be multiplied by an  arbitrary scalar function on $X$, 
an argument like that in the Fundamental Lemma of the Calculus of 
Variations shows that the integrand must vanish for all vector fields
$W$ with compact support in $X$.  A partition of unity argument then
shows that the integrand vanishes for any vector field $W$  whatsoever,
which proves (iii).

To see directly that (iii) is equivalent to the Euler--Lagrange
equations, one readily computes in coordinates that along $j^1\ns\phi$,
\begin{equation}
\left(j^1\ns \phi\right)^* \ns\left(j^1\ns V \hook \,
\Omega_\cl \right) = V^A \left[
\frac{\partial}{\partial x^\mu} \left(
\frac{\partial L}{\partial v^A{}_\mu} (j^1\ns \phi)
\right) -\frac{\partial L}{\partial y^A} (j^1\ns \phi)
\right] d^{\ps n+1} \ns x   
\label{eqn3C:6}
\end{equation}
for all $\pi_{XY}$-vertical vector fields $V$ on $Y$.  Thus (iii)
implies (ii).  On the other hand, (\ref {eqn3C:6}) combined with
the above remarks on decompositions of vector fields and
Lemma~\ref{lem3.2} shows that (ii) implies (iii). 
\end{proof}
\newpage
\startrule
\vskip-12pt
            \addcontentsline{toc}{subsection}{Examples}
\begin{examples}
\mbox{}\\[-18pt]
\paragraph{\bf a\ \; Particle Mechanics.}\enspace 
With the Lagrangian density $\cl = L(t, q, v)dt$,  expressions (\ref
{eqn3A:2}) become $p_A = {\partial L}/{\partial v^A}$ and $p = L -
v^A{\partial L}/{\partial v^A} = -E$,  the Cartan form becomes the
usual one-form for  describing time-dependent mechanics
\begin{equation}\label{3Cex:a0}
\Theta_\cl = \frac{\partial L}{\partial v^A} dq^A - E \,dt,
\end{equation}
and the Euler--Lagrange equations take the standard form 
\begin{equation}
\frac{d}{dt} \frac{\partial L}{\partial v^A} = \frac{\partial
L}{\partial q^A}.
\label{3Cex:a1}
\end{equation}
\medskip

Let us specifically work out the details for a relativistic free
particle of mass $m \neq 0$.  The  Lagrangian density is
\begin{equation}
\cl  =  - m \| \bold v \| \ps dt,   
\label{3Cex:a2}
\end{equation}
where $\bold v = (v^0, v^1, v^2, v^3)$ is the velocity of the
particle (with respect to parameter time $t$) and $\| \bold v \| = (-
g_{AB} v^A v^B)^{1/2}$, $g$ being the spacetime metric.\footnote{\
Sometimes in this context one encounters the Lagrangian density
$$ \cl'  =  - \frac12 m \| \bold v \|^2 \ps dt.$$ 
However, unlike $\cl$ given by (\ref{3Cex:a2}) $\cl'$ is not time
reparametrization-invariant.  In \S11C we shall see how to `repair'
this defect in $\cl'$. Of course, if one does not require time
reparametrization-invariance, then the choice $\cl'$ of Lagrangian
density is certainly permissible. Misner, Thorne, and Wheeler [1973]
give a detailed discussion of these two Lagrangian densities in
\S13.4.} The Cartan form (\ref{3Cex:a0}) is thus
\begin{equation}
\Theta_\cl  =  \frac{mg_{AB} v^B}{\| \bold v \|} dq^A, 
\label{3Cex:a3}
\end{equation}
 whence
\begin{equation}
 p_A  =   \frac{mg_{AB} v^B}{\| \bold v \|} \qquad \text{and} \qquad p 
=  0. 
\label{3Cex:a4}
\end{equation}
 The equations of motion (\ref {3Cex:a1}) are just the geodesic
equations for the spacetime metric $g$:
\begin{equation}
\frac{d}{dt} \left(\frac{g_{AB} v^B}{\| \bold v \|} \right) =
\frac{1}{2\| \bold v \|} g_{BC,A}v^B v^C    
\label{3Cex:a5}
\end{equation}
where $t$ is used as the curve parameter.

\paragraph{\bf b\ \; Electromagnetism.}\enspace 
 For electromagnetism on a fixed background spacetime $X$  with metric
$g$,  the Lagrangian density is\footnote{\ For  convenience, in this
example we use (\ref {3Cex:b1}), with the coefficient $-1/4$, instead
of the Misner, Thorne, and Wheeler [1973] Lagrangian, with
coefficient $-1/{16\pi}$.  One may pass from one convention to the
other by scaling the electromagnetic fields.}
\begin{equation}
\cl = -\frac14 F_{\mu\nu} F^{\mu\nu}\ \root\of{-g} \,d^{\ps 4}\ns x,   
\label{3Cex:b1}
\end{equation}
 where $\root\of{-g}:= \ \root\of{-\text{det } g_{\mu\nu}}$ and,
viewed as a function on $J^1(\Lambda^{1}X)$, the field tensor
$F_{\mu\nu}$ is defined by 
\begin{equation}
F_{\mu\nu}  =  v_{\nu\mu} - v_{\mu\nu}.   
\label{3Cex:b2}
\end{equation}
The Legendre transformation gives the relations
\begin{equation}
\Ff^{\mu\nu} =  F^{\mu\nu}\ \root\of{-g}  \qquad \text{and} \qquad p =
\frac14 F_{\mu\nu} F^{\mu\nu} \ \root\of{-g}.   
\label{3Cex:b3}
\end{equation}
 The  Cartan form $\Theta_\cl$ is
\begin{equation}
\Theta_\cl =\ \root\of{-g}\, F^{\nu\mu} dA_\nu \wedge 
d^{\ps 3}\ns x_\mu +
\frac14 F_{\mu\nu} F^{\mu\nu} \ \root\of{-g} \,d^{\ps 4}\ns x   
\label{3Cex:b4}
\end{equation}
 and the 5-form $\Omega_\cl = -\mathbf{d}\Theta_\cl$ is 
\begin{multline}
\Omega_\cl  =   \  \root\of{-g} \,(g^{\nu\sigma} g^{\mu\rho} -
g^{\nu\rho} g^{\mu\sigma})\, dA_\nu \wedge dv_{\rho \sigma} \wedge
d^{\ps 3} \ns x_\mu \\[1.5ex] - \
\root\of{-g} \,F^{\mu\nu} dv_{\nu\mu} \wedge d^{\ps 4}\ns x.  
\label{3Cex:b5} 
\end{multline}

For a section $A$ of $Y = \Lambda^1 X$, (\ref {3Cex:b2}) gives
$(j^1\! A)^*F = \mathbf{d}A$ and the Euler--Lagrange equations reduce to
\begin{equation}
 (A^{\nu,\mu} - A^{\mu,\nu})_{;\nu}=  0.   
\label{3Cex:b6} 
\end{equation}
Here the semicolon denotes a covariant derivative with respect to
the background metric, so (\ref {3Cex:b6}) is a covariant divergence.
\medskip

When $g$ is treated parametrically, none of the formulas above
change. In particular, since $g$ is not variational, it has no
Euler--Lagrange equations. Even so, varying the Lagrangian density
with respect to $g$ produces the stress-energy-momentum tensor of the
electromagnetic field, as shown in Example {\bf b} of \S4D and
Interlude II.

\paragraph{\bf c\ \; A Topological Field Theory.}\enspace 
The Chern--Simons Lagrangian density is
\begin{equation} {\mathcal L} =
\frac{1}{2}\epsilon^{\mu\nu\sigma}F_{\mu\nu}A_{\sigma}\, d^{\ps 3}\ns x,
\label{eq:3Cex:d1}
\end{equation}
where the curvature $F$ of $A$ is defined by (\ref{3Cex:b2}). 
Intrinsically, $\cl= F \wedge A$. Applying the Legendre
transformation, we get
\begin{equation} p^{\nu\mu} = \epsilon^{\mu\nu\sigma}A_{\sigma}\qquad
\mbox{ and }\qquad p = 0.
\label{eq:3Cex:d2}
\end{equation}
The Cartan form is thus
\begin{equation}
\Theta_{\mathcal L} =
\epsilon^{\mu\nu\sigma}A_{\sigma}\, dA_{\nu} 
\wedge d^{\ps 2}\ns x_{\mu},
\label{eq:3Cex:d3}
\end{equation}
whence
\begin{equation}
\Omega_{\mathcal L} = -
\epsilon^{\mu\nu\sigma}dA_{\sigma} \wedge dA_{\nu} \wedge
d^{\ps 2}\ns x_{\mu}.
\label{eq:3Cex:d4}
\end{equation}

The Euler--Lagrange equations are
$\epsilon^{\mu\nu\sigma}F_{\mu\nu} = 0$, which are equivalent to
the connection $A$ being flat:
\begin{equation} 
(j^{1}\! A )^{*} F = \mathbf{d}A = 0.
\label{eq:3Cex:d5}
\end{equation}
That these equations are only of the first order indicates
that the Chern--Simons theory is highly degenerate.

\paragraph{\bf d\ \; Bosonic Strings.}\enspace 
In the Polyakov model, with $\phi$ and $h$ as independent variables, 
the action is taken as the negative of the ``energy" functional.  Thus
as a function on $J^{1}Y$ , the Lagrangian density is
\begin{equation}
\cl = -\frac12\ \root\of{| h |}\, h^{\sigma\ns\rho}
g_{AB} v^A{}_\sigma v^B{}_\rho\, d^{\ps 2}\ns x.   
\label{3Cex:e1} 
\end{equation}
The Legendre transformation is
\begin{gather}
p_A{}^\mu = -\ \root\of{| h |}\, h^{\mu\nu} g_{AB} v^B{}_\nu  
\label{3Cex:e2} 
\\[2ex]
\rule{0ex}{3ex} q^{\sigma\ns\rho \mu}  =  0   
\label{3Cex:e3} 
\\[2ex] 
p = \frac12 \
\root\of{| h |}\, h^{\mu\nu} g_{AB} v^A{}_\mu v^B{}_\nu,   
\label{3Cex:e4} 
\end{gather}
so that the Cartan form is
\begin{equation}
\Theta_\cl = \ \root\of{| h |} \left( -h^{\mu\nu} g_{AB} v^B{}_\nu
\,d\phi^A \wedge d^{\ps 1}\ns x_\mu + \frac12 h^{\mu\nu} 
g_{AB}v^A{}_\mu v^B{}_\nu \, d^{\ps 2}\ns x \right)   
\label{3Cex:e5} 
\end{equation}
 and
\begin{multline}\label{3Cex:e6} 
\Omega_\cl = d\phi^A   \wedge d\left(-\ \root\of{| h |}\, h^{\mu\nu}
g_{AB} v^B{}_\nu \right) \wedge d^{\ps 1}\ns x_\mu
\\[1.5ex] 
- d\left( \frac12 \ \root\of{| h|} \,
  h^{\mu\nu} g_{AB}v^A{}_\mu v^B{}_\nu \right) \wedge d^{\ps 2}\ns x .
\end{multline}

The Euler--Lagrange equations reduce to
\begin{gather}
(h^{\mu\nu} g_{AB} (\phi) \phi^B{}_{,\nu})_{;\mu} = 0  
\label{3Cex:e7} 
\\[1.5ex]
\left(\frac12 h^{\mu\nu} g_{AB}(\phi)\phi^A{}_{,\mu}\phi^B{}_{,\nu}
\right)
h_{\alpha\beta}=g_{CD}(\phi)\phi^C{}_{,\alpha}\phi^D{}_{,\beta}.  
\label{3Cex:e8} 
\end{gather}
Equation (\ref {3Cex:e7}) says that $\phi$ satisfies the harmonic map
equation. Equation (\ref {3Cex:e8}) does two things: 
\renewcommand{\labelenumi}{(\roman{enumi})}
\begin{enumerate}
\item 
it says $h$ is conformally related to the induced metric $\phi^*g$ on
$X$: $\Lambda^2h_{\alpha\beta}=(\phi^* g)_{\alpha \beta}$,  and   
 \item 
 it determines the conformal factor: $\Lambda^2=\frac12 h^{\mu\nu}
g_{AB}(\phi)\phi^A{}_{,\mu}\phi^B{}_{,\nu}.$ 
\end{enumerate}
At first glance, it might appear that (ii) is significant while (i) is
insignificant, since in two dimensions ``any two metrics are
conformally related."  However, the content of (\ref {3Cex:e8}) is, in
fact, just the reverse.  The actual theorem states that in two
dimensions, given two metrics, there is a diffeomorphism which
transforms one metric into a multiple of the other.  Equation (\ref
{3Cex:e8}) says that in {\it any\/} coordinate system $h$ is
proportional to $\phi^*g$.  There is no need for a diffeomorphism.

We summarize the content of (\ref {3Cex:e8}): There are three
equations.  Two say that the  components of $h$ and $\phi^*g$ are
proportional while the third determines the proportionality  constant. 
The first two have content while the third is an identity. 
\vskip -18pt
\end{examples}
\startrule
\vskip -24pt

\section[Covariant Momentum Maps and Noether's Theorem]{
Covariant Momentum Maps and Noether's \\Theorem}

A key feature of relativistic field theories is that not all of the
Euler--Lagrange equations  necessarily describe the temporal evolution
of fields; some of the equations may impose constraints  on the choice
of initial data. Those constraints which are first class in the sense
of Dirac [1950,\,1964] reflect the gauge symmetry of the theory and
play a central role in our development.  In Part III we will see that
these first class constraints are related to the vanishing of various
momentum maps.

To lay the groundwork for this, we now study symmetries in the
Lagrangian and  multisymplectic formalisms. First, we show how
automorphisms of $Y$ lift to $J^{1}Y$  in a way  compatible, via the
Legendre transform, with their canonical lifts to $Z$.  Second, we
introduce the  notion of a covariant momentum map for a group $\cg$
acting on $Z$ by covariant canonical  transformations. Third, we prove
that the lifted action of a group $\cg$ on $J^{1}Y$  has a covariant 
momentum map (with respect to the Cartan form) which is given by the
pull-back to $J^{1}Y$  of the  covariant momentum map on $Z$ by
$\mathbb F \cl$.  Then we prove  Noether's theorem, here formulated in 
terms of covariant momentum maps.

\subsection{
Jet Prolongations}
 
Here we show how to naturally lift automorphisms of $Y$ to
automorphisms of $J^{1}Y$.  In  effect, we shall construct the
covariant analogue of the tangent map.

Let $\eta_Y : Y \to Y$ be a $\pi_{XY}$-bundle automorphism covering a
diffeomorphism $\eta_X : X \to X$.  If $\gamma : T_xX \to T_yY$ is an
element of $J^{1}Y$, let $\eta_{J^1 Y}(\gamma) : T_{\eta_X\ns (x)}X \to
T_{\eta_Y\ns (y)}Y$ be defined by
\begin{equation}
\eta_{J^{1}Y} (\gamma)  =  T\eta_Y \circ \gamma \circ T\eta^{-1}_X. 
\label{eqn4A:1}
\end{equation}
The $\pi_{Y,J^1 Y}$-bundle automorphism $j^1\ns\eta_Y :=
\eta_{J^1 Y}$ so constructed is called the {\bfi first jet extension\/}
or {\bfi prolongation\/} of $\eta_Y$ to $J^{1}Y$.  In coordinates, 
if $\gamma = (x^\mu, y^A, v^A{}_\mu)$, then 
\begin{equation}
\eta_{J^1 Y}(\gamma) = 
\left( \eta^\mu_X(x)
, \eta_Y^A(x,y) 
,  \left[\partial_\nu \eta_Y^A + 
      \left(\partial_B\eta_Y^A
      \right)  v^B{}_\nu
 \right] 
 \partial_\mu \!
 \left(\eta^{-1}_X\right)^\nu 
\right),  
\label{eqn4A:2}
\end{equation}
where $\partial_B  = \partial/\partial y^B$. Occasionally we are
interested in the case when $\eta_Y$ is vertical; that is, $\eta_X$ is
the identity. Equations (\ref {eqn4A:1}) and (\ref {eqn4A:2}) simplify
accordingly in this case.

If $V$ is a vector field on $Y$ whose flow is $\eta_\lambda$, so that
$$ 
V \circ \eta_\lambda = \frac{d\eta_\lambda}{d\lambda}, 
$$ 
then its {\bfi first jet extension\/} or {\bfi prolongation\/} $j^1\ns
V := V_{J^1 Y}$ is the vector field on $J^{1}Y$  whose flow is
$j^1\ns\eta_\lambda $:
\begin{equation}
 j^{1}\ns V \circ j^{1}\ns \eta_\lambda 
=  \frac{d}{d\lambda}\, j^1\ns \eta_\lambda .  
\label{eqn4A:3}
\end{equation}
In coordinates, by differentiating (\ref {eqn4A:2}), we obtain
\begin{equation}
 j^1\ns V  = \left( V^\mu, V^A, \frac{\partial V^A}{\partial x^\mu} +
\frac{\partial V^A}{\partial y^B} v^B{}_\mu - v^A{}_\nu \frac{\partial
V^\nu}{\partial x^\mu} \right).  
\label{eqn4A:4}
\end{equation}

It is important to note that if $\phi$ is a section of $\pi_{XY}$ and
$\eta_Y$ is a $\pi_{XY}$-bundle  automorphism covering $\eta_X$,  then
$\eta_Y\cdot \phi := \eta_Y \circ \phi \circ \eta^{-1}_X$ is another
section and
\begin{equation}
 j^{1}\!\left( \eta_Y\cdot \phi \right)  =  j^{1}\ns \eta_Y
\circ j^{1}\ns\phi \circ \eta^{-1}_X.  
\label{eqn4A:5}
\end{equation}
See Garc\'\i a [1974] and Aldaya and de Azc\'arraga [1980a] for more
information.

\begin{remarks}[Remarks \ 1.]
If $\phi_\lambda =  \eta_\lambda \circ \phi$ is a variation of $\phi$
(as in Remark 2 of \S 3C), then $\eta_X$ is the identity and (\ref
{eqn4A:5}) becomes
\begin{equation*}
 j^{1}\ns \phi_\lambda =  j^{1}\ns\eta_\lambda  \circ j^{1}\ns\phi.  
\label{eqn4A:6}
\end{equation*}
In this case, the generating vector field of the flow
$\eta_\lambda$ is vertical and (\ref {eqn4A:4}) reduces to
\begin{equation*}
 j^1\ns V  =  \left( 0, V^A, \frac{\partial V^A}{\partial x^\mu} +
\frac{\partial V^A}{\partial y^B} v^B{}_\mu \right).  
\label{eqn4A:7}
\end{equation*}

\paragraph{\bf 2.} In the above mentioned references, an additional
term $v^B{}_\mu v^A{}_\lambda {\partial V^\lambda}/{\partial y^B}$ is
present in the third  component of (\ref {eqn4A:4}); this is absent
here since we consider only transformations of $Y$ which  are bundle
automorphisms so $V^\lambda$ is independent of $y^B$. 

\paragraph{\bf 3.} {\sl A word of caution\/}:\enspace One can also
view $V$ as a section of the bundle $TY \to Y$ and take its  first jet
in the sense of equation (\ref {eqn2A:4}).  Then one gets the section
of $J^1(TY) \to Y$ given  in coordinates by  
\begin{equation*}
 (x^\mu, y^A) \mapsto  \left( x^\mu, y^A, V^\mu, V^A, \frac{\partial
V^\mu}{\partial x^\nu},
\frac{\partial V^\mu}{\partial y^B}, \frac{\partial V^A}{\partial
x^\nu}, \frac{\partial V^A} {\partial y^B} \right).  
\label{eqn4A:8}
\end{equation*}
This section is not to be confused with $j^1\ns V$ as  defined by (\ref
{eqn4A:3}) and (\ref {eqn4A:4}); they are two different objects.

\paragraph{\bf 4.}
From the definitions one finds that 
$ j^{1}\!\left( \left[ V, W \right] \right) = \left[ j^1\ns V, j^1\ns W
\right]$ .  
\end{remarks}

\subsection{
Covariant Canonical Transformations}
 
A {\bfi covariant canonical transformation\/} is a $\pi_{X\ns
Z}$-bundle map $\eta_Z : Z \to Z$ covering a  diffeomorphism $\eta_X :
X \to X$ such that
\begin{equation}
\eta^*_Z\Omega =  \Omega. 
\label{eqn4B:1}
\end{equation}
We say $\eta_Z$ is a {\bfi special covariant canonical
transformation\/} if
\begin{equation}
\eta^*_Z\Theta =  \Theta. 
 \label{eqn4B:2}
\end{equation}

If $\eta_Y :  Y \to Y$ is a $\pi_{XY}$-bundle automorphism (also
covering a diffeomorphism $\eta_X : X 
\to X$), its {\bfi canonical lift\/} $\eta_Z : Z \to Z$ is defined by  
\begin{equation}
\eta_Z(z)  =  (\eta_Y^{-1})^*z;  
\label{eqn4B:3}
\end{equation}
that is,
\begin{equation*}
\eta_Z(z) \left( v_1, \dots, v_{n+1} \right)  =  z \left( T\eta^{-1}_Y
\cdot v_1,\dots, T\eta^{-1}_Y \cdot v_{n+1} \right), 
\label{eqn4B:4}
\end{equation*}
where  $z \in Z_y$ and $v_1, \dots, v_{n+1} \in T_{\eta_Y\ns (y)}Y.$
Since $\eta_Y$ is a $\pi_{XY}$-bundle map, $T\eta_Y$ maps vertical
vectors to vertical vectors, so $\eta_Z$ does indeed map $Z$ to $Z$. 
In coordinates, if we write $z = p \,d^{\ps n+1}\ns x + p_A{}^\mu dy^A
\wedge d^{\ps n}\ns x_\mu$, and notice that
$$ 
\left( \eta^{-1}_Y \right) ^* dx^\mu  =  \left( \eta^{-1}_X \right) ^*
dx^\mu   
$$ 
and
\begin{equation*}
\left( \eta^{-1}_Y \right) ^*dy^A  =  \partial_\nu 
\!\left(
\eta^{-1}_Y \right) ^A dx^\nu + \partial_B \!\left( \eta^{-1}_Y \right)
^A dy^B, 
\label{eqn4B:5}
\end{equation*}
we get
\begin{align}\label{eqn4B:6}
\eta_Z(z) = \left( \eta^{-1}_Y \right) ^*\!z  = &\left( p  +
\partial_\nu\! \left( \eta^{-1}_Y \right) ^A p_A{}^\mu \partial_\mu
\eta^\nu_X \right) J^{-1}d^{\ps n+1}\ns x \nonumber \\[1.5ex]
&\quad \ \ns + \left( \partial_A \!\left( \eta^{-1}_Y \right) ^B 
p_B{}^\nu \partial_\nu \eta^\mu_X \right) J^{-1}dy^A 
\wedge d^{\ps n}\ns x_\mu,
\end{align}
where $J$ is the Jacobian determinant of $\eta_X$ and where the
quantities in (\ref {eqn4B:6}) are evaluated at the  appropriate
points.  

\begin{prop}\label{prop4.1}
Lifts are special covariant canonical transformations.
\end{prop}

This is readily proved using a method similar to the
cotangent bundle case  (Abraham and Marsden [1978], Theorem~3.2.12).

\medskip Here is a second way to understand the lift of a
$\pi_{XY}$-bundle automorphism $\eta_Y : Y \to Y$  covering a
diffeomorphism $\eta_X : X \to X$,  which proceeds via $J^{1}Y^\star$
rather than directly to $Z$.  We define the {\bfi lift\/} of $\eta_Y$
to $J^{1}Y^\star$ to be the affine dual of the prolongation
$\eta_{J^{1} Y}$; that is, if $z : J^{1}_y Y \to \Lambda^{n+1}_x X$ is
an element of $J^{1}_yY^\star$, let
$$
\eta_{J^{1}  Y  ^\star}(z) : J^{1}_{\eta_Y\ns (y)} Y \to  \Lambda^{n+1}
_{\eta_X\ns (x)}  X
$$  
be the affine map defined by
\begin{equation}
\langle \eta_{J^1 Y^\star}(z),\gamma\rangle  = 
\left( \eta^{-1}_X \right) ^* \!\left\langle z,
 \eta_{J^1 Y}^{-1}\left( \gamma 
\right) \right\rangle.  
\label{eqn4B:7}
\end{equation}
In coordinates, if $z = (p, p_A ^\mu)$ acts on $\gamma = (x^\mu, y^A,
v^A{}_\mu)$  as in (\ref{eqn2B:1}), then from (\ref{eqn4A:2}),
\begin{align} \label{eqn4B:8}
&\langle \eta_{J^1 Y^\star}(z),\gamma\rangle \\[2ex]
&\quad = \left(\eta^{-1}_X\right)^*
    \left[
    \left(p + p_A{^\mu} \partial_\mu\eta_X^\nu
    \left[\partial_\nu \!
    \left(\eta^{-1}_Y\right)^A 
              + v^B{_\nu} \partial_B \!
              \left(\eta^{-1}_Y
              \right)^A
    \right]
    \right)d^{\ps n+1}\ns x 
    \right]
 \nonumber   \\[2ex]
&\quad = \left[\left(p +\partial_\nu \!
 \left(\eta^{-1}_Y \right)^{A} p_A{^\mu}\partial_\mu
            \eta_X^\nu \right)       
    + {}\left( \partial_A \! \left( \eta^{-1}_Y \right)^B p_B{^\nu}
\partial_\nu
          \eta_X^\mu \right) v^A{}_\mu\right]J^{-1} d^{\ps n+1}\ns x
          \nonumber
\end{align}
 so that the terms in (\ref {eqn4B:8}) correspond to the components in
(\ref {eqn4B:6}) via (\ref {eqn2B:8}).  Consequently, we have  given a
coordinate proof of

\begin{prop}\label{prop4.2}
The canonical isomorphism $\Phi : Z \to J^1 Y^\star$ is equivariant
with respect to the lifts $\eta_Z$ and $\eta_{J^1 Y^\star}$; that is,
\begin{equation}
\Phi \circ \eta_Z  =  \eta_{J^1 Y^\star} \circ \Phi  
\label{eqn4B:9}
\end{equation}
\end{prop}

Here is an invariant proof of Proposition~4.2 using (\ref {eqn2B:6}),
(\ref {eqn4A:1}), (\ref {eqn4B:3}) and (\ref {eqn4B:7}) above.  For $z
\in Z$  and $\gamma \in J^1 Y$,  we have
$$
\begin{aligned}
\left\langle\eta_{J^1 Y^\star}\!\left( \Phi(z) \right) ,
\gamma\right\rangle
& =
\left( \eta^{-1}_X \right) ^* \!\left\langle\Phi(z),
\eta_{J^1 Y}^{-1}(\gamma) \right\rangle
\\[1.5ex]
& =
\left( \eta^{-1}_X \right) ^*\!\left\langle \Phi(z),  T\eta^{-1}_Y
\circ
\gamma \circ T\eta_X \right\rangle
\\[1.5ex]
& =
\left( \eta^{-1}_X \right) ^* \left[ \left( T\eta^{-1}_Y \circ \gamma 
\circ T\eta_X \right) ^* \ns z \right] 
\\[1.5ex]  & =
\left( \eta^{-1}_X \right) ^*\!
\eta_X^*\gamma^*\!\left(\eta^{-1}_Y\right)^*z  = \gamma^* 
\left( \eta_Z(z)
\right) = \left\langle\Phi
\left ( \eta_Z(z) \right),\gamma\right\rangle
\end{aligned}
$$
and so (\ref {eqn4B:9}) holds. Thus, the identification of $Z$ and
$J^{1}Y^\star$ respects the action of a lifted covariant canonical
transformation.  
\medskip

Let $V$ be a vector field on $Y$ whose flow $\eta_\lambda$ maps fibers
of $\pi_{XY}$ to fibers. The  {\bfi canonical lift\/} of $V$ to $Z$ is
the vector field $V_Z$ that generates the canonical lift of this flow
to $Z$.  By differentiating (\ref{eqn4B:6}) we obtain
\begin{equation}
V_Z = (V^\mu, V^A, - pV^\nu{}_{,\nu} - p_B{}^\nu V^B{}_{,\nu},
p_A{}^\nu V^\mu{}_{,\nu} - p_B{}^\mu V^B{}_{,A} - p_A{}^\mu
V^\nu{}_{,\nu})  
\label{eqn4B:10}
\end{equation}
in coordinates, where $V = (V^\mu, V^A)$.  By either direct
calculation or by Proposition~\ref{prop4.1}, we get
$\pounds_{V_Z}\Theta =  0.  $

\subsection{
Covariant Momentum Maps}
 
We now introduce the covariant analogue of momentum maps in symplectic
and Poisson  geometry (see for example, Abraham and Marsden [1978],
Marsden, Weinstein, Ratiu,  Schmid, and Spencer [1983], Guillemin and
Sternberg [1984],  Marsden [1992], and Marsden and Ratiu [1999].

Let $\cg$ denote a Lie group (perhaps infinite-dimensional) with Lie
algebra $\fg$  that acts on $X$ by diffeomorphisms and acts on $Z$ (or
$Y$) as $\pi_{X\ns Z}$- (or $\pi_{XY}\mbox{-})$ bundle automorphisms. 
For $\eta \in \cg$, let $\eta_X, \eta_Y$, and $\eta_Z$ denote the
corresponding transformations of $X, Y$, and $Z$, and for $\xi \in
\fg$,  let $\xi_X, \xi_Y$, and $\xi_Z$ denote the corresponding
infinitesimal generators.  If $\cg$ acts on $Z$ by covariant canonical
transformations, then the Lie derivative of $\Omega$  along $\xi_Z$ is
zero:
\begin{equation}
\pounds_{\xi_Z} \Omega = 0  
\label{eqn4C:1}
\end{equation}
and if $\cg$ acts by {\it special\/} covariant canonical transformations,
then 
\begin{equation}
\pounds_{\xi_Z} \Theta = 0  .
\label{eqn4C:2}
\end{equation}

If $\cg$ acts on $Z$ by covariant canonical transformations, then a
{\bfi covariant momentum  map\/} (or a {\bfi multimomentum map\/}) for
this action is a map
$$ 
J : Z \to \fg^* \otimes \Lambda^{n} Z  =  L(\fg, \Lambda^{n} Z)	 
$$ 
covering the identity on $Z$ such that
\begin{equation}
\mathbf{d}J(\xi)  = \mathbf{i}_{\xi_Z} \Omega,  
\label{eqn4C:3}
\end{equation}
where $J(\xi)$ is the $n$-form on $Z$ whose value at $z \in Z$ is $
\langle J(z), \xi\rangle$. A covariant momentum map is said to be
$\operatorname{Ad}^*$-{\bfi equivariant\/} if
\begin{equation}
 J \left( \text{Ad}^{-1}_\eta \xi \right)  =  \eta^*_Z \left[
J(\xi) \right] .  
\label{eqn4C:4}
\end{equation}
If $\cg$ acts by {\it special\/} covariant canonical transformations,
there is an explicit formula for a {\bfi special covariant momentum
map\/}.  We claim that
\begin{equation}
J(\xi)  = \mathbf{i}_{\xi_Z} \Theta 
\label{eqn4C:5}
\end{equation}
is a covariant momentum map.  Indeed, 
$$ 
\mathbf{d}J(\xi)  
=  \mathbf{d}\mathbf{i}_{\xi_Z} \Theta 
=  (\pounds_{\xi_Z}
- \mathbf{i}_{\xi_Z}\mathbf{d})\Theta =
\mathbf{i}_{\xi_Z}\Omega.
$$

\begin{prop}\label{prop4.3}
Special covariant momentum maps are 
$\operatorname{Ad}^*$-equivariant.
\end{prop}

The proof is analogous to that for the cotangent bundle case (Abraham
and Marsden  [1978], Theorem~4.2.10).  For the particular case of
lifted actions, we have:

\begin{prop}\label{prop4.4}
If the action of $\cg$ on $Z$ is the lift of an action of $\cg$ on $Y$, 
then the special covariant momentum map is given by
\begin{equation}
 J(\xi)(z) = \pi^*_{Y\!Z} \mathbf{i}_{\xi_Y} z. 
\label{eqn4C:6}
\end{equation}
This momentum map is  $\operatorname{Ad}^*$-equivariant.
\end{prop}

This follows from the recognition that $\xi_Y = T\pi_{Y\!Z} \cdot
\xi_Z$, the definitions of $J(\xi)$ and $\Theta$, and Proposition~\ref
{prop4.3}.

In coordinates, if $\xi_Y$ has components $(\xi^\mu, \xi^A)$,  then
equation (\ref {eqn4C:6}) reads
\begin{equation}
 J(\xi)(z) = (p_A{}^\mu \xi^A + p\, \xi^\mu)\, d^{\ps n}\ns x_\mu -
p_A{}^\mu
\xi^\nu dy^A \wedge d^{\ps n-1}\ns x_{\mu\nu},  
\label{eqn4C:7}
\end{equation}
where
\begin{equation*}
 d^{\ps n-1}\ns x_{\mu\nu}  = \mathbf{i}_{\partial_\nu}
\mathbf{i}_{\partial_\mu} d^{\ps n+1}\ns x.  
\label{eqn4C:8}
\end{equation*}
\medskip

In passing, we reformulate equivariance for lifted actions in terms of
Poisson brackets, just  as one does in the usual context of symplectic
and Poisson manifolds. Observe that according to (\ref{eqn4C:6}),
special covariant momentum maps are ``momentum observables'' in the
sense of (\ref{2B:17}).

\begin{prop}\label{prop4.5}
For lifted actions, 
\begin{equation}
\{J(\xi), J(\zeta)\} = \mathbf{d}(\mathbf{i}_{\xi_Z}\mathbf{i}_{\zeta_Z}
\Theta) + J([\xi,\zeta]).  
\label{eqn4C:9}
\end{equation}
\end{prop}

\begin{proof} 
Set $\eta_\lambda = \exp(\lambda\zeta)$. Differentiating the
equivariance relation (\ref {eqn4C:4}) with respect to $\lambda$ at
$\lambda = 0$ gives
$$  J([\xi, \zeta])  =  \pounds_{\zeta_Z}J(\xi)  = 
\mathbf{d}\mathbf{i}_{\zeta_Z}J(\xi) + \mathbf{i}_{\zeta_Z}
\mathbf{d}J(\xi)  =  \mathbf{d}\mathbf{i}_{\zeta_Z}J(\xi) +
\mathbf{i}_{\zeta_Z}\mathbf{i}_{\xi_Z}\Omega $$  
by (\ref{eqn4C:3}). This, (\ref{eqn4C:5}), and (\ref {2B:19}) give
(\ref {eqn4C:9}). 
\end{proof}

\startrule
\vskip-12pt
            \addcontentsline{toc}{subsection}{Examples}
\begin{examples}
\mbox{}\\

Since $X$ is assumed oriented, $\operatorname{Diff}(X) $ will henceforth
denote the group of \emph{orientation-preserving} diffeomorphisms of $X$.

\paragraph{\bf a\ \; Particle Mechanics.}\enspace 
For classical mechanics with time reparametrization freedom  (see, for
example, Kucha\v r [1973]), we let $\cg = \operatorname{Diff}(\mathbb
R)$, the diffeomorphism group of $X = \mathbb R$. Then $\cg$ acts on
$Y = \mathbb R \times Q$ by time reparametrizations and this action
may be lifted to  $Z = T^*\mathbb R \times T^*Q$.  We identify $\fg$
with the space $\cf(\mathbb R)$ of smooth functions on $\mathbb R$ and
identify $\cf(\mathbb R)^*$, the space of densities on $\mathbb R$,
with $\cf(\mathbb R)$. In terms of the coordinates of Example {\bf a}
of \S 2B, we have 
\[
\chi_Y = \chi(t)\frac{\partial}{\partial t}
\]
for $\chi \in \cf(\mathbb R)$. The momentum map
$$ 
J : Z  =  T^*\mathbb R \times T^*Q \to \cf(\mathbb R)^* \approx
\cf(\mathbb R)
$$ 
is defined by 
\begin{equation}
\langle J(t, p, q^1, \dots, q^N, p_1, \dots, p_N),\chi\rangle  = 
p\,\chi(t).   
\label{ex4C:a1} 
\end{equation}

If we consider a classical mechanical system with a fixed time
parametrization and  with a group $G$ acting on $Q$, then the covariant
momentum map induced on $Z$ is the  usual momentum map on
$T^*Q$ regarded as a map on $Z$.

In this example, note that {\it covariant canonical maps include
time-dependent canonical  transformations\/}.

\paragraph{\bf b\ \; Electromagnetism.}\enspace 
The appropriate invariance group for electromagnetism on a fixed 
spacetime background $X$ is $\cg = \cf(X)$, the additive group of
smooth functions on spacetime.   Let $\cg$ act on $Y = \Lambda^1 X$ as
follows:  For $f \in \cf(X)$ and $A \in \Lambda^1_x X$,
\begin{equation}
 f \cdot A  = A + \mathbf{d}f(x) \in \Lambda^1_x X.   
\label{ex4C:b1} 
\end{equation}
For $\chi \in \fg \cong \cf(X)$, the infinitesimal generator is
given by
\begin{equation}\label{ex4C:b2}
\chi_Y(A) = \chi_{,\nu} \frac{\partial }{\partial A_{\nu}}. 
\end{equation}
 From (\ref {eqn4C:7}), the corresponding covariant momentum map is
\begin{equation}\label{ex4C:b3}
\langle J(x,A,p,\Ff),\chi\rangle  =  \Ff^{\nu\mu} \chi_{,\nu} \, d^{\ps
3}\ns x_\mu. 
\end{equation}
\medskip

If we treat the metric $g$ on $X$ parametrically, we may enlarge the
group $\mathcal G$ to be the semi-direct product
$\mbox{Diff}(X) \; \circledS \; {\mathcal F}(X)$, with $\mbox{Diff}(X)$
acting on ${\mathcal F}(X)$ by push-forward; that is, $\eta \cdot f = f
\circ \eta^{-1}$. Then $\mathcal G$ acts on ${\tilde Y} = Y \times_X
S_2^{\ps 3,1}(X)$ by
\begin{equation} 
(\eta,f) \cdot (A;g) = \left( (\eta^{-1})^*(A + \mathbf{d}f(x)); \left(
\eta^{-1}
\right) ^*g\right).
\label{ex4C:b4}
\end{equation}
The infinitesimal generator $(\xi,\chi)_{\tilde Y}$ at the point
$(A;g) \in {\tilde Y}_x$ is
\begin{equation}
(\xi_{S_2^{3,1}(X)}(g))_{\sigma\ns\rho}\frac{\partial}{\partial
g_{\sigma\ns\rho}} + (\xi_Y(A))_{\nu}\frac{\partial}{\partial
A_{\nu}} + (\chi_Y(A))_{\nu}\frac{\partial}{\partial A_{\nu}} +
\xi^{\mu}(x)\frac{\partial}{\partial x^{\mu}},
\label{ex4C:b5}
\end{equation}
where
\begin{equation} (\xi_{S_2^{3,1}(X)}(g))_{\sigma\ns\rho} = -
(g_{\sigma\mu}\xi^{\mu}_{\:\:\: ,\rho} +
g_{\rho\mu}\xi^{\mu}_{\:\:\: ,\sigma}),
\label{ex4C:b6}
\end{equation}
\begin{equation} (\xi_Y(A))_{\nu} = -
A_{\mu}\xi^{\mu}_{\:\:\:,\nu},
\label{ex4C:b7}
\end{equation}
and $\chi_Y(A)$ is given by (\ref{ex4C:b2}). From
(\ref {eqn4C:7}), the multimomentum map ${\tilde J}: {\tilde Z}
\rightarrow \mathfrak{g}^* \otimes \Lambda^3 Z $ for this action is
\begin{multline}
\langle {\tilde J}(x,A,p,\Ff;g),(\xi,\chi)\rangle 
= \\[1.5ex] \left(- {\mathfrak{F}}^{\nu\mu}A_{\tau}
\xi^{\tau}_{\:\:\: ,\nu} 
+ {\mathfrak{F}}^{\nu\mu}\chi_{,\nu} + p\,\xi^{\mu}\right)d^{\ps
3}\ns x_{\mu} - {\mathfrak{F}}^{\tau\mu}\xi^{\nu}dA_{\tau} \wedge
d^{\ps 2}\ns x_{\mu\nu}.
\label{ex4C:b8}
\end{multline}

\paragraph{\bf c\ \; A Topological Field Theory.}\enspace  Just as in
Example {\bf b} above, we take ${\mathcal G} = \mbox{Diff}(X)
\; \circledS \; {\mathcal F}(X)$ acting on $Y$ according to
\begin{equation} 
(\eta,f) \cdot A = (\eta^{-1})^*(A + \mathbf{d}f(x)).
\label{ex4C:d1}
\end{equation}
The corresponding infinitesimal generator is
\begin{equation}
(\xi,\chi)_Y(A) =  (\xi_Y(A))_{\nu}\frac{\partial}{\partial A_{\nu}} +
(\chi_Y(A))_{\nu}\frac{\partial}{\partial A_{\nu}} +
\xi^{\mu}(x)\frac{\partial}{\partial x^{\mu}},
\label{ex4C:d2}
\end{equation}
where $(\xi_Y(A))_{\nu}$ and $\chi_Y(A)$ are given by (\ref{ex4C:b7})
and (\ref{ex4C:b2}), respectively. Then we compute
\begin{multline}
\langle J(x^\mu,A_\nu,p,p^{\nu\mu}),(\xi,\chi)\rangle 
= \\[1.5ex] \left( -
p^{\nu\mu}A_{\tau}\xi^{\tau}_{\:\:\: ,\nu} +
p^{\nu\mu}\chi_{,\nu} + p\,\xi^{\mu}\right) d^{\ps 2}\ns x_{\mu} -
p^{\tau\mu}\xi^{\nu}dA_{\tau}
\wedge d^{\ps 1}\ns x_{\mu\nu}.
\label{ex4C:d3}
\end{multline}

\paragraph{\bf d\ \; Bosonic Strings.}\enspace  
Corresponding to $Y = (X \times M) \times_X S^{1,1}_2(X)$, we take the
invariance group for bosonic strings to be the semi-direct product
\begin{equation*}
\cg = \operatorname{Diff}(X) \,\circledS\; 
{\rm Con}^{1,1}_2 ( X),
\label{ex4C:e1} 
\end{equation*}
where ${\rm Con}^{1,1}_2(X)$ is the group of conformal transformations
of $S^{1,1}_2(X)$ (which may be identified with the group $\cf(X,
\mathbb R^+)$ of positive real-valued functions on $X$) and $\eta \in 
\operatorname{Diff}(X)$ acts on $\Lambda \in {\rm Con}^{1,1}_2(X)$ by
push-forward:
\begin{equation}
\eta \cdot \Lambda = \Lambda \circ \eta^{-1}.   
\label{ex4C:e2} 
\end{equation}
 Then $(\eta,\Lambda) \in \cg$ acts on $(x,\phi, h) \in Y_x$ by
\begin{equation}
\left( \eta,\Lambda \right) \cdot \left(\phi, h \right) = \left(\phi
\circ \eta, (\eta^{-1})^*(\Lambda^2(x)h) \right) .  
\label{ex4C:e3} 
\end{equation}
Hence, for $(\xi,\lambda)  \in \fg \cong \mathfrak  X (X)\, \circledS
\; \mathcal F (X)$, the corresponding infinitesimal generator
$(\xi,\lambda) _Y$ at $(\phi, h) \in Y$ is
\begin{equation}
 (\lambda_Y(h))_{\sigma\rho} \frac{\partial}{\partial h_{\sigma\rho}} +
(\xi_Y(h))_{\sigma\rho}\frac{\partial}{\partial h_{\sigma\rho}} 
+ \xi^\mu \frac{\partial}{\partial x^\mu},   
\label{ex4C:e4} 
\end{equation}
where
\begin{equation}
 (\lambda_Y(h))_{\sigma\rho} = 2\lambda h_{\sigma\rho}   
\label{ex4C:e5} 
\end{equation}
 and
\begin{equation}
 (\xi_Y(h))_{\sigma\rho}  =  -(h_{\sigma\mu} \xi^\mu{}_{,\rho} + 
h_{\rho\mu} \xi^\mu{}_{,\sigma}).   
\label{ex4C:e6} 
\end{equation}
Note that there is no component in the direction $\partial/\partial
\phi^A$ since $\phi$ is a scalar field on $X$. From (\ref {eqn4C:7}),
we get the covariant momentum map
\begin{align} \label{ex4C:e7} 
\langle
J(x^\mu,\phi^A,h_{\sig\rho},
p,p_A{}^{\mu},& \, q^{\sig\rho\mu}),(\xi,\chi)\rangle \nonumber
 \\[1.5ex] = \mbox{} &
\big[ q^{\sigma\rho \mu}
\left( 2\lambda h_{\sigma\rho}- h_{\sigma\nu} \xi^\nu{}_{,\rho}-
h_{\rho \nu} \xi^\nu{}_{,\sigma} \right) + p\,\xi^\mu \big] d^{\ps
1}\ns x_\mu \nonumber
\\[1.5ex] & -  (p_A{}^\mu \xi^\nu d\phi^A + q^{\sigma\rho \mu} \xi^\nu
dh_{\sigma\rho}) \eps_{\mu\nu}
\end{align}  
where we have written $\eps_{\mu\nu} = d^{\ps 2}\ns x_{\mu\nu}$.

\begin{remark}[Remark]
As we shall see in Part III the covariant momentum maps (\ref{ex4C:b3}),
(\ref{ex4C:d3}), and (\ref{ex4C:e7}) are closely related to the
divergence, spatial flatness, and superhamiltonian and supermomentum
constraints of electromagnetism, Chern--Simons theory, and string theory,
respectively.
\end{remark}
\end{examples}
\vskip -18pt
\startrule
\vskip 12pt

The covariant momentum maps discussed and illustrated in this section
do not involve the  Lagrangian at all. In the next section, however, we
will show that when a theory possesses  symmetries, an important link
between the Lagrangian and the covariant momentum map may be 
established. 
 
\subsection{
Symmetries and Noether's Theorem}
 
Having assembled the necessary background in \S \S 4A--C, we are now
able to state and  prove Noether's Theorem. This fundamental result
will be used in Part III to show that the energy-momentum map in the
Lagrangian representation vanishes identically on solutions of the
Euler--Lagrange equations.  As we shall see in Chapters 10 and 11,
the vanishing of this energy-momentum  map on a single Cauchy surface
is often equivalent to the totality of the first class constraints of
the  theory.  However, its vanishing on a whole family of
hypersurfaces is equivalent to the entire  Euler--Lagrange system under
a suitable hypothesis on the gauge group; see \S 9C.
\medskip

Suppose that a group $\cg$ acts on $Y$ by bundle automorphisms. 
Prolong this action to $J^{1}Y$  by setting $\eta \cdot \gamma =
\eta_{J^1 Y} (\gamma)$ as in \S 4A. We say that the Lagrangian density
$\cl$ is {\bfi equivariant\/} with respect to $\cg$ if for all $\eta
\in \cg$ and $\gamma \in J^1_x Y$, 
\begin{equation}
\cl \left( \eta_{J^1 Y}(\gamma) \right) = \left(
\eta^{-1}_X \right) ^* \cl(\gamma),  
\label{eqn4D:1} 
\end{equation}
where $\left( \eta^{-1}_X \right)^* \cl(\gamma) $ means that the
$(n+1)$-form $\cl(\gamma)$ at $x \in X$ is pushed forward to an $(n +
1)$-form at $\eta_X(x)$; equality in (\ref {eqn4D:1}) thus means
equality of $(n+1)$-forms at the point $\eta(x)$. The infinitesimal
version of this is the following equation obtained from (\ref
{eqn3A:1}) and (\ref {eqn4A:4}):
\begin{equation}
\frac{\partial L}{\partial x^\mu} \xi^\mu + \frac{\partial L}{\partial
y^A} \xi^A +
\frac{\partial L}{\partial v^A{}_\mu} \left(\xi^A{}_{,\mu} - v^A{}_\nu
\xi^\nu{}_{,\mu} +  v^B{}_\mu 
\frac{\partial \xi^A}{\partial y^B} \right) + L\,\xi^\mu{}_{,\mu} = 0.
\label{eqn4D:2} 
\end{equation}
The quantity $\delta_\xi L$ defined by the left hand side of this
equation is called the {\bfi variation\/} of $L$ in the direction
$\xi$.

One of the fundamental assumptions that we shall make in this work is
the following:
\paragraph{\bf A1 \ Covariance.}\enspace
{\it  
The group $\cg$ acts on $Y$ by $\pi_{XY}$-bundle automorphisms and the
Lagrangian density $\cl$ is equivariant with respect to $\cg$.
}\bigskip

This is a strong assumption but, amazingly, the gauge groups of
most field theories satisfy it. One general class of exceptions
are the topological field theories, whose gauge groups tend to
leave $\Omega_{\mathcal L}$, but not $\mathcal L$, invariant. We will
discuss how to deal with one such case in Example {\bf c} below.
This assumption has a number of important consequences, some of which
are stated in the  following proposition.

\begin{prop}\label{prop4.6} 
Let $\cl$ be equivariant with respect to the lifted action of $\cg$ on
$J^{1}Y$.  Then

\begin{enumerate}
\renewcommand{\labelenumi}{\em (\roman{enumi})}
\item 
 $\mathbb F\cl$ is also equivariant; that is, $\eta_Z \circ
\mathbb F\cl = \mathbb F\cl \circ \eta_{J^1 Y}$. Thus, the diagram
$$
\begin{CD}                             J^1 Y @>{\mathbb F\cl}>> Z \\
@V{\eta_{J^1 Y}}VV @VV{\eta_Z}V \\ J^1 Y @>{\mathbb F\cl}>> Z
\end{CD}
$$ 
commutes for every $\eta \in \cg$, 
\item 
 the Cartan form $\Theta_\cl$ is invariant\textup{;} that is,
$\eta^*_{J^1 Y}\Theta_\cl =  \Theta_\cl\,$ for all $\eta \in \cg$, and
\item 
 the map $J^\cl(\xi) := \mathbb F\cl^*J(\xi):J^1Y \rightarrow
\Lambda^n(J^1Y)$ is a momentum map for the lifted action of $\cg$ on
$J^{1}Y$  relative to $\Omega_\cl$;  that is, for all $\xi \in \fg$,
\begin{equation}\label{eqn4D:3} 
\xi_{J^1 Y} \hook \,\Omega_\cl = \mathbf{d}J^\cl(\xi), 
\end{equation}
where $\xi_{J^1 Y} = j^1\ns\xi_Y$ is the infinitesimal generator
corresponding to $\xi$. Moreover,
\begin{equation} J^{\mathcal L}(\xi) = \xi_{J^1 Y} \hook \,
\Theta_{\mathcal L}.
\label{eqn4D:3.1}
\end{equation}
\end{enumerate}
\end{prop}

\begin{proof} 
\mbox{}\\[-12pt] 
\begin{enumerate}
\renewcommand{\labelenumi}{(\roman{enumi})}
\item    
It is convenient to identify $Z$ with $ J^{1}Y^\star$ and use
Proposition~\ref{prop4.2}. From (\ref {eqn4B:7}) and (\ref {eqn3A:3}),
we have
$$
\begin{aligned}
\left\langle\eta_{J^1 Y^\star}\!\left(\mathbb
F\cl(\gamma)\right),\gamma'\right\rangle &= \left( \eta^{-1}_X \right)
^*\!\left\langle\mathbb F\cl(\gamma), \eta^{-1}_{J^1
Y}(\gamma') \right\rangle
\\[2ex]
&= \left( \eta^{-1}_X \right) ^* \left[\cl(\gamma) + \left.
\frac{d}{d\varep}
\cl
\left(
\gamma +
\varep[\eta^{-1}_{J^1 Y}(\gamma') - \gamma] \right) \right
|_{\varep=0}\right]. 
\end{aligned}
$$ 
On the other hand, using (\ref {eqn3A:3}), we have
$$
\left\langle\mathbb F\cl(\eta_{J^1 Y}(\gamma)),\gamma'\right\rangle = 
\cl(\eta_{J^1 Y}(\gamma)) + \left.
\frac{d}{d\varep} \cl \big( \eta_{J^1 Y}(\gamma)) + \varep[\gamma' -
\eta_{J^1 Y}(\gamma)] \big)\right |_{\varep=0}.
$$
These are equal by the covariance condition {\bf A1}.

\item  
The proof is analogous to that for Corollary~{4.2.14}
of Abraham and Marsden [1978].

\item  
The infinitesimal version of (i) states
\begin{equation}\label{eqn4D:4} 
\xi_Z \circ \mathbb{F}\cl  =  T\mathbb{F}\cl \circ \xi_{J^1 Y}.  
\end{equation}
Pulling (\ref{eqn4C:3}) back from $Z$ to $J^{1}Y$  along the Legendre
transformation and using (\ref{eqn4D:4})  yields (\ref{eqn4D:3}).
Similarly, pulling (\ref{eqn4C:5}) back by $\mathbb{F}{\mathcal
L}$ yields (\ref{eqn4D:3.1}).\hfill\qedsymbol 
\end{enumerate}
\renewcommand{\qedsymbol}{}
\end{proof}

Because of this theorem, we call $J^\cl$ the {\bfi covariant momentum
map\/} in the  {\bfi Lagrangian representation\/}.
From (\ref{eqn4D:3.1}) and (\ref{eqn3B:3}) or, equivalently,
from (\ref{eqn4C:7}) and (\ref{eqn3A:2}), we have the explicit
formula
\begin{multline} J^{\mathcal L}(\xi) = 
\\[2ex]  \left(\frac{\partial
L}{\partial v^A_{\:\:\:\mu}}\xi^A + \left[L - \frac{\partial
L}{\partial
v^A_{\:\:\:\nu}}v^A_{\:\:\:\nu}\right]\xi^{\mu}\right)
d^{\ps n}\ns x_{\mu} -
\frac{\partial L}{\partial v^A_{\:\:\:\mu}}\xi^{\nu}dy^A \wedge
d^{\ps n-1}\ns x_{\mu\nu}.
\label{eqn4D:4.1}
\end{multline}

If $\phi$ is a solution of the Euler--Lagrange equations, then from
Theorem~{\ref{thm3.1},
$$ 
(j^1\ns \phi)^*(W \hook \,\Omega_\cl) = 0
$$ 
for any vector field $W$ on $J^{1}Y$.  In particular, setting $W =
\xi_{J^1 Y}$ and applying $(j^1\ns \phi)^*$ to (\ref{eqn4D:3}), we
obtain the following basic {\bfi Noether conservation law\/}:

\begin{thm}[Divergence Form of Noether's Theorem]\label{thm4.7}
If the covariance assumption {\bf A1} is  satisfied, then for each $\xi
\in \fg$,
\begin{equation}\label{eqn4D:5} 
\mathbf{d}\ns \left[(j^1\ns \phi)^*J^\cl(\xi)\right] = 0 
\end{equation}
for any section $\phi$ of $\pi_{XY}$ satisfying the Euler--Lagrange
equations.
\end{thm}

The quantity $(j^1\ns \phi)^*J^\cl(\xi)$ is called the {\bfi Noether
current}, and Theorem 4.7 is often referred to as the {\bfi first
Noether theorem}.\footnote{\ The ``second Noether theorem,'' which we
call the ``Vanishing Theorem,'' will be discussed in Chapter 9.}  We
now proceed to compute  coordinate expressions for the Noether current
and its divergence without assuming that the field  equations are
satisfied.  In the process, we will derive a criteria for when a
theory has a converse to Noether's theorem; that is, when the Noether
conservation law implies the Euler--Lagrange equations.

From (\ref{eqn4D:4.1}) we get
$$
\begin{aligned} 
(j^1\ns \phi)^* & J^\cl(\xi)    \\[2ex]
& = \left( \frac{\partial
L}{\partial v^A{}_\mu} (j^1\ns \phi) ( \xi^A \circ
\phi ) + L (j^1\ns \phi) \xi^\mu -
\frac{\partial L}{\partial v^A{}_\nu} (j^1\ns \phi) \,\phi^A{}_{,\nu}
\xi^\mu
\right) d^{\ps n}\ns x_\mu \\[2ex] & \ \quad 
-\frac{\partial L}{\partial v^A{}_\mu} (j^1\ns \phi)\,
\phi^A{}_{,\lambda} \xi^\nu dx^\lambda \wedge d^{\ps n-1}\ns
x_{\mu\nu} 
\end{aligned}
$$ 
which, upon using the identity
$$ 
dx^\lambda \wedge d^{\ps n-1}\ns x_{\mu\nu}  =  \delta^\lambda_\nu
\, d^{\ps n}\ns x_\mu -
\delta^\lambda_\mu\, d^{\ps n}\ns x_\nu,	
$$ 
simplifies to
\begin{equation}\label{eqn4D:6} 
(j^1\ns \phi)^*\! J^\cl(\xi)  =  \left[ \frac{\partial
L}{\partial v^A{}_\mu}(j^1\ns \phi) ( \xi^A \circ \phi
-\phi^A{}_{,\nu}\xi^\nu ) + L(j^1\ns \phi) \xi^\mu\right] d^{\ps n}\ns
x_\mu. 
\end{equation}
Defining the {\bfi Lie derivative\/} of $\phi$  along $\xi$ by
\begin{equation}\label{eqn4D:7}
\pounds_\xi\phi = T\phi \circ \xi_X - \xi_Y \circ \phi; \quad
\text{i.e.,} \quad \left( \pounds_\xi\phi \right) ^A  = 
\phi^A{}_{,\nu} \xi^\nu - \xi^A \circ \phi,   
\end{equation}
(\ref{eqn4D:6}) may be rewritten as
\begin{equation}
(j^1\ns \phi)^*J^\cl \left( \xi \right)  =
\left[-\frac{\partial L}{\partial v^A{}_\mu} (j^1\ns \phi)
(\pounds_\xi\phi)^A + L(j^1\ns \phi)\xi^\mu\right] d^{\ps n}\ns x_\mu    
\label{eqn4D:8} 
\end{equation}
where $\phi$ is any section of $\pi_{XY}$, not necessarily
satisfying the Euler--Lagrange equations.

Next, taking the exterior derivative of the Noether current
(\ref{eqn4D:6}) and using the identity
$$
\mathbf{d}(V^\mu d^{\ps n}\ns x_\mu)=\partial_\rho V^\mu dx^\rho \wedge
d^{\ps n}\ns x_\mu=\partial_\mu V^\mu d^{\ps n+1}\ns x	
$$ 
(which justifies the name ``divergence'' for the resulting
quantity), we obtain after some  rearrangement and cancellation,

\begin{align} \label{eqn4D:9}
\mathbf{d}\left[(j^1\ns \phi)^*J^\cl(\xi)\right] 
& = \partial_\mu \left[ \frac{\partial
L}{\partial v^A{}_\mu} (j^1\ns \phi) (\xi^A \circ \phi -
\phi^A{}_{,\nu}\xi^\nu) + L(j^1\ns \phi)\xi^\mu\right] d^{\ps n+1}\ns x 
\nonumber\\[2ex]
& =
\left\{ \left[\frac{\partial L}{\partial y^A} (j^1\ns \phi) -
\frac{\partial}{\partial x^\mu}
\left( \frac{\partial L}{\partial v^A{}_\mu} (j^1\ns
\phi)\right)\right] (-\xi^A \circ \phi +
\phi^A{}_{,\nu} \xi^\nu)\right. 
\nonumber\\[2ex]
& \qquad + \frac{\partial
L}{\partial x^\mu} (j^1\ns \phi)\xi^\mu + \frac{\partial L}
{\partial y^A}(j^1\ns \phi) (\xi^A \circ \phi) 
\nonumber\\[2ex]
& \qquad + \frac{\partial L}{\partial
v^A{}_\mu} (j^1\ns \phi) \left[ \frac{\partial \xi^A}{\partial x^\mu} +
\frac{\partial \xi^A}{\partial y^B} v^B{}_\mu - \frac{\partial
\xi^\nu}{\partial x^\mu} v^A{}_\nu
\right] (j^1\ns \phi) 
\nonumber\\[2ex]
& \left. \qquad + \: L(j^1\ns \phi) \frac{\partial
\xi^\mu}{\partial x^\mu}\right\} d^{\ps n+1}\ns x 
\nonumber\\[2ex]
& = \left\{ \frac{\delta
L}{\delta\phi^A} (\pounds_\xi\phi)^A + \delta_\xi L \right\}
(j^1\ns \phi)\, d^{\ps n+1}\ns x. \\ \nonumber
\end{align}   
Here $\delta L/\delta\phi^A$ is the Euler--Lagrange derivative of
$\cl$ defined in (\ref{eqn3C:1}), which vanishes when the field
equations are satisfied, and $\delta_\xi L$ is the variation of $\cl$ 
defined in (\ref{eqn4D:2}), which vanishes when the Lagrangian is
equivariant.  Thus equation (\ref{eqn4D:9}) again makes it obvious
that the Noether conservation laws hold when the Lagrangian is
equivariant and the Euler--Lagrange equations are satisfied.
\medskip

To obtain a converse, we say that the $\cg$-action is {\bfi
vertically transitive\/} if for each $x\in X$ and each
$y \in Y_x$ and each (local) section $\phi$ through $y$,  
\begin{equation}\label{eqn4D:10} 
\left\{(\pounds_\xi \phi)(x) \mid \xi \in \fg\right\} = V_yY.   
\end{equation}
Looking forward to Part II, \S 7A, we note that this is the
infinitesimal version of the statement that at each point $x \in X$,
there exists a group element that transforms any given value of the
{\it fields\/} at $x$ into any other specified value.  This is true
for metrics and connections under the diffeomorphism group and for
connections under bundle automorphisms, but is rarely true for other
fields.

Equation (\ref{eqn4D:9}) now implies

 \begin{thm}\label{thm4.8}
Assume that the action of $\cg$ on $Y$ is vertically transitive and
that the Lagrangian density is $\cg$-equivariant.  Then a local
section $\phi$ of $Y$ is a solution to the Euler--Lagrange equations 
{\rm  (3C.1)} if and only if it satisfies the Noether conservation
law  {\rm  (\ref{eqn4D:5})} for all $\xi \in \fg$.
\end{thm}

To check whether the $\cg$-action for a given field theory is
vertically transitive, we may use  the equivalent formulations given
in:

\begin{prop}\label{prop4.9}
Assume $\cg$ acts on $Y$ and let
\begin{equation*}
\fg(y)  = \{\xi_Y(y)\mid \xi \in \fg\}	
\end{equation*}
denote the span of the infinitesimal generators of the action at $y
\in Y$; that is, the tangent space to the orbit through $y$.  Vertical
transitivity of the action of $\cg$ on $Y$ is equivalent to either of
the following conditions:
\begin{enumerate}
\renewcommand{\labelenumi}{\em (\roman{enumi})}
\item 
 For each $y \in Y$ and each section $\phi$ with $\phi(x) = y$, the
image of $\phi$ is trans\-verse to $\fg(y)$;  that is,
\begin{equation}\label{eqn4D:11}
\hspace{-5ex}\fg(y) + \hbox {\rm  im}\;T_x\phi = T_yY.  
\end{equation}

\item 
 For each $y \in Y$ and every choice of bundle
coordinates $(x^\mu, y^A)$, we have 
\begin{equation}\label{eqn4D:12} 
\hspace{-6ex}\left\{\xi^A(y)\partial_A \mid \xi \in \fg\right\} = V_yY, 
\end{equation}
where $\xi_Y = \xi^\mu \partial_\mu + \xi^A\partial_A$.
\end{enumerate}
\end{prop}

Note that an action can be vertically transitive even if none of the
$\xi_Y$ are vertical; note also from (\ref{eqn4D:7}) that
$-(\pounds_\xi\phi)(x)$ is exactly the vertical component of
$\xi_Y(\phi(x))$.

\startrule
\vskip -12pt
            \addcontentsline{toc}{subsection}{Examples}
\begin{examples}
\mbox{}\\

In each of the following examples, we compute the Noether current and
its divergence  without assuming the fields satisfy the Euler--Lagrange
equations, and we then check for vertical  transitivity.

\paragraph{\bf a\ \; Particle Mechanics.}\enspace  From the
infinitesimal covariance condition (\ref{eqn4D:2}), we see that a
particle Lagrangian is $\operatorname{Diff}(\mathbb R)$-equivariant
(that is to say, time reparametrization-invariant) if and only if it is
time-independent and the energy $E =v^A({\partial L}/{\partial v^A}) -
L$ vanishes. Suppose also that there is a group $G$ acting on $Q$
whose  prolongation to $TQ$ leaves the Lagrangian invariant. Take
$\cg  = \operatorname{Diff}(\mathbb R) \times G$ and write elements
of $\fg$ as pairs $(\chi, \xi)$. Then (\ref{eqn4D:6}) reads
\begin{align} \label{ex4D:a1} 
(j^1\ns\phi)^*J^\cl(\chi,\xi) &= \left[\frac{\partial L}{\partial v^A}
(\xi^A - v^A\chi) + L\chi\right] \nonumber \\[1.5ex] 
&= (J^L(\xi) - E\chi)\nonumber
\\[1.5ex] &= J^L(\xi),   
\end{align}
where $J^L(\xi) = (\partial L/\partial v^A) \xi^A$ is the usual
momentum map for $G$ acting on $Q$ in Lagrangian mechanics. Then
(\ref{eqn4D:5}) asserts that on solutions of the Euler--Lagrange
equations,
\begin{equation}
\frac{d}{dt} J^L(\xi)  = 0.   
\label{ex4D:a2}  
\end{equation}

This group $\cg$ is vertically transitive iff $G$ acts transitively on
$Q$.  
\medskip

In the special case of the relativistic free particle, the Lagrangian
density (\ref{3Cex:a2})  is $\operatorname{Diff}(\mathbb
R)$-equivariant. Then the corresponding momentum map vanishes
identically, consistent with  the fact that $p=0$, cf. (\ref{3Cex:a4})
and (\ref{ex4C:a1}).  On the other hand, suppose that $(Q, g)$ is 
Minkowski space and that $G$ is the Poincar\'e group.  Then $\cl$ is
$G$-invariant and (\ref{ex4D:a2}) expresses the fact that relativistic
energy-momentum and angular momentum are constant along dynamical
trajectories. This is equivalent to the statement that the dynamical
trajectories are geodesics of $g$.

\paragraph{\bf b\ \; Electromagnetism.}\enspace 
 Let $\cg = \cf(X)$ act on $Y = \Lambda^1 X$ as
in Example {\bf b} of \S 4C.  The Maxwell Lagrangian density
(\ref{3Cex:b1}) is invariant under the prolongation of $\cf(X)$ to 
$J^1(\Lambda^1 X)$.  In this case (\ref{eqn4D:5}) reads
\begin{align}\label{ex4D:b1}  
\mathbf{d} \!\left[ (j^1\! A)^*J^\cl(\chi) \right] 
& = \mathbf{d} \left[ (j^1\!A)^* \!\left( F^{\nu\mu}\
\root\,\of{-g}\,\chi_{,\nu} \,d^{\ps 3}\ns x_\mu \right) \right]
\nonumber \\[1.5ex]
& = \mathbf{d} \left[ \left( A^{\mu,\nu} - A^{\nu,\mu}\right) \
\root\,\of{-g} \,\chi_{,\nu} \,d^{\ps 3}\ns x_\mu \right] 
\nonumber \\[1.5ex] 
& = \left[ \left( A^{\mu,\nu} - A^{\nu,\mu} \right)
\ \root\,\of{-g}  \,\,\right]_{,\mu} \,\chi_{,\nu}\,d^{\ps 4}\ns x = 0	
\end{align}   
where we have used $dx^\lambda \wedge d^{\ps 3}\ns x_\mu =
\delta^\lambda_\mu \, d^{\ps 4}\ns x$ and noted that the terms
$\chi_{,\mu\nu}$ have dropped out by symmetry.  Since $\chi$ is
arbitrary, (\ref{ex4D:b1}) is equivalent to
\begin{equation} 
(A^{\mu,\nu} - A^{\nu,\mu}){}_{;\mu} = 0   
\label{ex4D:b2}  
\end{equation}
which are Maxwell's equations.  From (\ref{ex4C:b2}) it is clear that
the group $\cf(X)$ is vertically transitive.

Observe that when the spacetime metric $g$ is fixed, the Maxwell
Lagrangian density (\ref{3Cex:b1}) is not
$\mbox{Diff}(X)$-equivariant: ${\mathcal L}((\eta,0)_{J^1 Y}
(\gamma)) = {\mathcal L}(\gamma)$, whereas
\[ \left( \eta_X^{-1} \right) ^* {\mathcal L}(\gamma) = {\mathcal
L} \left( \gamma \right) J^{-1},\]
$J$ being the Jacobian determinant of $\eta_X$. We can remedy this by
allowing $g$ to vary parametrically, in which case the $\sqrt{-g}$
factor in $\mathcal L$ will give rise to a factor of
$J$ which will cancel the offending factor $J^{-1}$ in the above.
Combined with our previous results, we see that parametrized
electromagnetism is $(\mbox{Diff}(X) \; \circledS \; {\mathcal
F}(X))$-equivariant: $\delta_{(\xi,\chi)}L = 0$. But there is a price
to be paid for this extended equivariance: Noether's theorem no longer
holds for ${\tilde J}^{\mathcal L} = \mathbb{F}{\mathcal L}^*{\tilde
J}$. Indeed, from (\ref{eqn4D:9}) we see that even if the section
$A$ of $\Lambda^1 X$ satisfies Maxwell's equations,
\begin{align*} 
\mathbf{d} \!\left[ \left((j^1\! A) \times g
\right) ^* \!{\tilde J}^{\mathcal L}(\xi,\chi) \right] 
& = 
\frac{\partial L}{\partial g_{\sigma\ns\rho}}(\pounds_{\xi}
g)_{\sigma\ns\rho}\, d^{\ps 4}\ns x \\[2ex]
& =  \mathfrak{T}^{\sigma\ns\rho} \left( \pounds_{\xi} g
\right) _{\sigma\ns\rho}d^{\ps 4}\ns x
\end{align*}
need not vanish, where
\begin{equation} 
\mathfrak{T}^{\sigma\ns\rho} =
-\left(\frac{1}{4}g^{\sigma\ns\rho}F_{\alpha\beta}F^{\alpha\beta} +
g^{\rho\beta}F^{\alpha\sigma}F_{\beta\alpha}\right)\sqrt{-g}
\label{ex4D:b3}
\end{equation}
is the stress-energy-momentum tensor of the electromagnetic field (cf.
Gotay and Marsden [1992] and Interlude II). The reason, of course, is
that
$g$ is not variational, whence ${\delta L}/{\delta g_{\sigma\ns\rho}} =
{\partial L}/{\partial g_{\sigma\ns\rho}} \neq 0$ necessarily.

\paragraph{\bf c\ \; A Topological Field Theory.}\enspace  The
behavior of the Chern--Simons field is exactly opposite to that of
parametrized electromagnetism: the Lagrangian density
(\ref{eq:3Cex:d1}) is equivariant under the action of
$\mbox{Diff}(X)$, but not under that of ${\mathcal F}(X)$. Indeed, we
compute
\begin{equation}
\delta_{(\xi,\chi)}L =
\frac{1}{2}\epsilon^{\mu\nu\sigma}F_{\mu\nu}\chi_{,\sigma}.
\label{ex4D:d1}
\end{equation}
\noindent Consequently the Legendre transformation is not
equivariant, and the Cartan form is not invariant. 

Nonetheless, we can verify by direct computation that the Noether
current 
\begin{multline} (j^1\! A)^*J^{\mathcal L}(\xi,\chi) = \\[2ex] \left(
\epsilon^{\mu\nu\sigma}(-A_{\tau}\xi^\tau{}_{,\nu} -
A_{\nu,\tau}\xi^{\tau} + \chi_{,\nu}) + \frac{1}{2}
\epsilon^{\nu\tau\sigma}F_{\nu\tau}\xi^{\mu}\right)\!
A_{\sigma}\, d^{\ps 2}\ns x_{\mu}
\label{ex4D:d10}
\end{multline}
is conserved; we obtain
$${\bf d}\left[(j^1\! A)^*J^{\mathcal L}(\xi,\chi)\right] =
\eps^{\mu\nu\sig}F_{\mu\nu}\left(A_{\sig ,\tau}\xi^\tau +
A_\tau\xi^\tau{}_{,\sig} + \frac{1}{2}\chi_{,\sig}\right) $$
which is zero when the Chern--Simons equations $F_{\mu\nu} = 0$ hold.
Alternately, we could appeal to (\ref{eqn4D:9}) and observe from
(\ref{ex4D:d1}) that $\delta_{(\xi,\chi)}L$ vanishes on shell. 

As in the case of electromagnetism, the action of ${\rm
Diff}(X)\,\circledS \, {\mathcal F}(X)$ on $\Lambda^1(X)$ is vertically
transitive. Since in particular the value of $\chi_{,\sig}(x)$ is
arbitrary, the conservation of the Noether current implies that
$F_{\mu\nu}(x) = A_{\nu,\mu}(x)-A_{\mu,\nu}(x)= 0$ at each $x\in X$.

\paragraph{\bf d\ \; Bosonic Strings.}\enspace 
The harmonic map Lagrangian density defined by equation (\ref{3Cex:e1})
is  equivariant under the action of $
\operatorname{Diff}(X) \,\circledS \, {\rm Con}^{1,1}_2(X)$ on 
$Y = (X\times M)\times_X S^{1,1}_2(X)$. Using
(\ref{ex4C:e4})--(\ref{ex4C:e6}) to compute the Lie derivatives
$\pounds_\xi\phi, \pounds_\xi h$, and $\pounds_\lambda h$ via
(\ref{eqn4D:7}), (\ref{eqn4D:9}) gives
\begin{multline}  \label{ex4D:e1}  
\mathbf{d}\!\left[j^1(\phi, h)^*J^\cl(\xi,\lambda) \right]  
= \\[2ex] \left(\frac{\delta L}{\delta \phi^A} \phi^A{}_{,\nu}\xi^\nu
\right. 
+ \left. 
\frac{\delta L}{\delta h_{\sigma\rho}} \left(h_{\sigma\rho,\nu}\xi^\nu
- 2\lambda h_{\sigma\rho} + h_{\sigma\nu}\xi^\nu{}_{,\rho} 
+ h_{\rho \nu} \xi^\nu{}_{,\sigma}\right)\right) \!d^{\ps 2}\ns x.
\end{multline}
At a given point $x \in X$, the quantities $\lambda$, $\xi^\nu$, and
$\xi^\nu{}_{,\rho}$ can be specified arbitrarily. The arbitrariness of
$\xi^\nu{}_{,\rho}$ implies that $\delta L/\delta h_{\sigma\rho} = 0$,
and so we obtain the ``conformal" Euler--Lagrange equation
(\ref{3Cex:e8}) as a conserved Noether  current. The arbitrariness of
$\xi^\nu$ then gives
$$ \frac{\delta L}{\delta \phi^A} \phi^A{}_{,\nu} = 0 $$
which is {\it not\/} equivalent to the harmonic map equation
(\ref{3Cex:e7}). This is consistent with the fact that   $\cg$ is not
vertically transitive; indeed (\ref{eqn4D:10}) is not satisfied since,
by (\ref{ex4C:e4}), $(\xi,\lambda) _Y$ has no component in the
$\partial/\partial\phi^A$ direction. 
\end{examples}
\vskip -18pt
\startrule

\newpage
\startrule

\section*{\large\bf\sc
Interlude I---On Classical Field Theory}
\addcontentsline{toc}{section}{Interlude I---On Classical Field Theory}
\markboth{Interlude I}{On Classical Field Theory}

Classical field theories come in many forms and varieties, and it is
useful to attempt a broad  classification  of such theories, based upon
certain features which play an important role in this  work. The first
of these features devolves upon the notion of {\bfi gauge group\/}
which is, roughly  speaking, the largest   localizable group of
transformations of the theory under which the Lagrangian density is 
equivariant (so that the group maps solutions of the Euler--Lagrange
equations to solutions).  The  concept of gauge group is explored more
fully in Part III, Chapter 8; for this discussion, we use the term
loosely.   It will be helpful in what follows to keep in mind the
examples in Chapter 4, since the groups studied  there are in fact the
gauge groups of their respective theories.

Consider a classical field theory with gauge group $\cg$.  Suppose that
$\cg \subset \operatorname{Aut}(Y)$, the  automorphism group of the
covariant configuration bundle $Y$. We may distinguish two basic types
of field theories based upon the relationship between the gauge group
$\cg$ and the (spacetime) diffeomorphism  group
$\operatorname{Diff}(X)$.  The first consists of those which are {\bfi
param\-etr\-ized\/} in the sense that the natural homomorphism
$\operatorname{Aut}(Y) \to \operatorname{Diff}(X)$ given by $\eta_Y
\mapsto \eta_X$ maps $\cg$ onto $\operatorname{Diff}(X)$ (or at least
onto a ``sufficiently large" subgroup thereof, such as the compactly
supported diffeomorphisms).  This terminology reflects the fact that
such a theory is invariant under (essentially) arbitrary relabeling of
the points of the parameter ``spacetime" $X$.  (In relativity theory,
cf. Anderson [1967], one would say that $X$ is a {\it relative\/}
object in the theory.) The param\-etr\-ized theory \emph{par
excellence} is of course general relativity, in which case $\cg$
equals the spacetime diffeomorphism group.  Another example is
provided by the relativistic free particle, whose gauge group consists
of (oriented) time reparametrizations.  The Polyakov and Nambu strings
are also param\-etr\-ized theories, as are most topological field
theories (see Horowitz [1989]).

At the other extreme, there are the {\bfi background\/} theories in
which $\cg \subset \operatorname{Aut}_{\operatorname{Id}}(Y)$, the 
automorphisms of $Y$ which cover the identity on $X$ (or at least in
which the image of $\cg$ in  $\operatorname{Diff}(X)$ is ``sufficiently
small"---the Poincar\'e group, for instance).  In this case $X$ is
called an {\it absolute\/} element of the theory, much like time in
Newtonian mechanics, or Minkowski spacetime in special relativity. 
The prototypical background theory is electromagnetism treated on a
{\it fixed\/}, non-interacting, background spacetime; here $\cg$
consists of shifts of the vector potential by exact 1-forms and so is
completely ``internal."  Yang--Mills--Higgs theory is similar. 
Traditionally,  Newtonian mechanical systems are treated as background
theories; usually, such systems do not  even have gauge groups.

In many ways, param\-etr\-ized theories are preferred over background
theories.  According to  the principle of general covariance, one
should model physical systems so that their descriptions  contain no
absolute objects.  From a more prosaic standpoint, param\-etr\-ized
theories are easier to  handle since, for example, they readily admit
compatible slicings (necessary for the space + time  decomposition of
the theory, cf. Part II, \S 6A and \S 6D). Only in a parametrized
theory can one properly correlate the dynamics and the initial value
constraints of a theory with its gauge freedom, cf. Part II \S 7F, and
Part III Chapter 13; in fact, the system must effectively be
param\-etr\-ized to do Hamiltonian dynamics ``correctly," as our 
development will show.  For more information on param\-etr\-ized
theories compared to background theories, see  Anderson [1967] and
Kucha\v r [1973].

Fortunately, most background theories can readily be converted into
param\-etr\-ized theories.  The conversion usually involves the
promotion of previously fixed objects, notably a metric on $X$,   into
the ranks of the field variables. In this way one can often make the
Lagrangian density $\operatorname{Diff}(X)$-equivariant, and thereby
enlarge the gauge group to ``include" the diffeomorphism group  of
$X$.  Thus, for example, when one encounters a background theory, such
as Yang--Mills--Higgs  on Minkowski spacetime, one can parametrize it
by simply ``turning the metric on." This is precisely how we have
parametrized electromagnetism in Example {\bf b}. As long as no  new
terms are added to the Lagrangian and the previously fixed fields are
{\it not\/} varied to give Euler--Lagrange equations, the converted
theory, while now param\-etr\-ized, remains physically equivalent  to
the original background theory. Kucha\v r [1973] finds an alternate way
to parametrize a background theory, by introducing certain slicing
parameters, treating them as fields, and letting them evolve
dynamically. Further details of this approach can be found in Kucha\v
r [1976].

Among either class of field theories, we may further distinguish {\bfi
metric\/} and {\bfi non-metric\/}  theories.  Metric theories are
those which carry a (Lorentzian, Riemannian or general signature) 
metric {\it on the parameter space\/} $X$.  Most familiar classical
field theories are metric, with topological field theories and
theories whose action functionals measure length, area, volume, etc.,
being important exceptions. Most mechanical systems (relativistic or
nonrelativistic), on the other hand, are non-metric.  (In this regard,
we emphasize that our terminology refers to the metric structure of
$X$, which in mechanics is usually taken to be the time line; the
kinetic energy metrics familiar in mechanics live instead on the fibers
of $Y \to X$.\footnote{\  Our terminology here is also at odds with the
standard usage of ``metric" and ``non-metric" as applied to theories of
gravity.})

In metric theories the metric may be an absolute object, in that it
does not ``evolve" or  dynamically interact with any of the other
elements of the theory (as in electromagnetism on a fixed  spacetime
background), or it may be treated as one of the field variables. 
Again, the principle of  general covariance suggests that the metric
should be treated as a relative object.  When this is the  case, the
metric may be {\bfi variational\/} in the sense that it is varied to
give field equations (as in theories of gravity), or it may appear
essentially as a parameter in that it may take different values, but is
{\it not\/} varied to give Euler--Lagrange equations. Electromagnetism
on a variable but non-dynamic spacetime is an example of a theory in
which the metric is treated {\bfi parametrically\/} in this fashion. 
Theories which contain dynamic or parametric metrics are typically
param\-etr\-ized whereas theories with fixed metrics are usually
background theories.

A word of caution: If a theory has some fields which are parametric
(but not variational) then  the Noether conservation laws may not hold
because the parametric field is not varied and so does  not satisfy any
field equations. We have already encountered this in Example {\bf b}
of \S 4D.

As noted above, our formalism is best applied to param\-etr\-ized
theories. Likewise, it is  often advantageous to work in the context
of metric theories, with the metric being treated  dynamically.  This
can sometimes be achieved using the ``Polyakov trick," which
transforms the Nambu  string into the Polyakov string.  We show how
this goes for the relativistic free particle in Part III, \S 11C.   We
suspect that this procedure could be useful in treating some of the
``nonstandard" vector field  theories of Isenberg and Nester [1977].

We summarize our classification with a chart, indicating in which
categories various field  theories fall.
\vskip 24pt

\begin{figure}[ht] \label{Gimmsy5-1}
\begin{center}
\includegraphics[scale=1,angle=0]{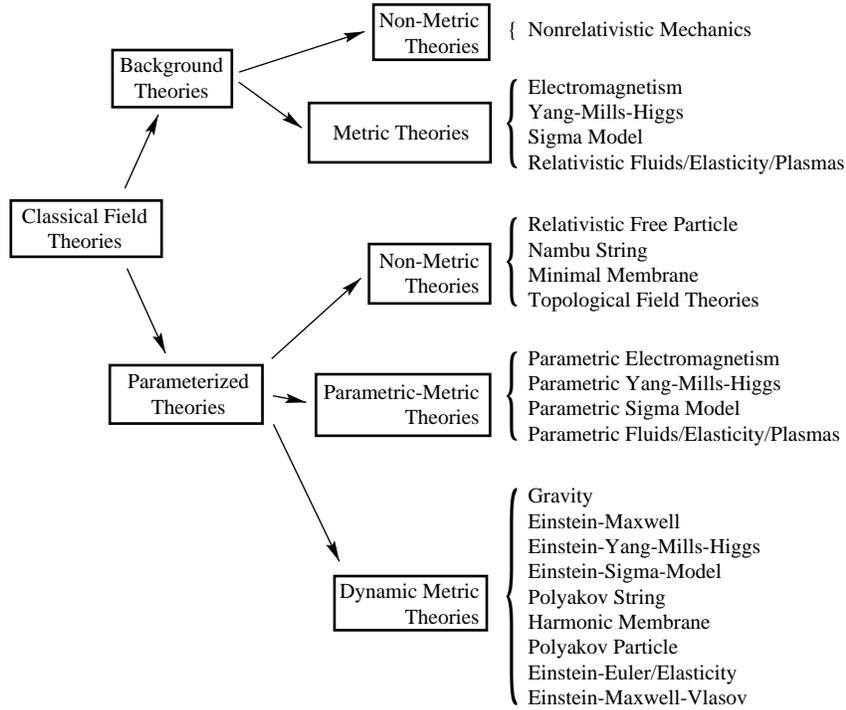}
\caption{Classification of Field Theories}
\end{center}
\end{figure}

In this work, we discuss example theories from each category. But there
are many other  theories, not touched upon here, to which our formalism
may be applied. Two especially important  ones are fluids and
elasticity (cf. Bao, Marsden, and Walton [1985]).

Let us specifically consider the case of a fluid in the {\bfi material
representation\/}.  The basic  variables are the {\bfi particle
placement fields\/}  $\phi : \Sigma \times \mathbb R \to M$ which have
the physical interpretation that $\phi(\{x\} \times \mathbb R)$ is the
world line of the particle with material label $x \in \Sigma$,  where
$\Sigma$ is the fluid ``reference manifold" and $M$ is spacetime. 
Thus $Y$ is the trivial bundle $(\Sigma \times \mathbb R) \times M$. 
Consistent with standard practice in continuum mechanics, $M$ carries
the spacetime metric while $\Sigma \times \mathbb R$ carries a
reference  metric $h$ out of which things like the internal energy are
built (cf. Marsden and Hughes [1983]). Observe that
$\operatorname{Diff}(\Sigma \times \mathbb R)$ encodes the material
symmetries (including particle relabeling and time reparametrization
symmetries).  In this formulation, the theory is already
param\-etr\-ized; note also that the metric $h$ on
$\Sigma \times \mathbb R$ is treated parametrically.

This approach to fluids is in keeping with our treatment of strings. 
But it is not particularly  suited to a discussion of general
relativistic fluids.  In this context it is more convenient to use the
{\bfi inverse material representation\/}.  The fields then include the
gravitational field (treated variationally or parametrically) with the
fluid described by {\bfi particle labeling fields\/} $\psi : X \to
\Sigma \times \mathbb R$, where $\psi = \phi^{-1}$.  In addition to
the material symmetries given by
$\operatorname{Diff}(\Sigma \times \mathbb R)$ we have the
spacetime-based  $\operatorname{Diff}(X)$ gauge symmetries associated
with the gravitational field.  The same ``switch" to the inverse
material representation obviously can be made for strings, and similar
remarks apply to elasticity (with some qualifications).

\newpage



\section*{References}
\markboth{References}{References}
\addcontentsline{toc}{section}{References}

\begin{description}

\item[]
 R. Abraham and J. Marsden [1978],
{\it Foundations of Mechanics\/},
Second Edition,
Addison-Wesley,
Menlo Park, California.

\item[]
 V. Aldaya and J. A. de Azc\'arraga  [1980a],
Geometric formulation of classical mechanics and
field theory,
{\em Rivista del Nuovo Cimento\/},
{\bf  3},
1--66.

\item[]
 V. Aldaya and J. A. de Azc\'arraga [1980b],
Higher-order Hamiltonian formalism in field theory,
{\em J.~Phys. A: Math. Gen.\/},
{\bf  13},
2545--2551.

\item[]
S. C. Anco and R. S. Tung [2002],
Covariant Hamiltonian boundary conditions in General
Relativity for spatially bounded space-time regions,
{\it J. Math. Phys.,}
{\bf 43},
5531--5566.

\item[]
 I. Anderson [1992],
Introduction to the variational bicomplex,
{\em Contemp. Math.\/},
{\bf 132},
51--73.

\item[]
 J. L. Anderson [1967],
{\it Principles of Relativity Physics\/},
Academic Press,
New York.

\item[]
 J. M. Arms  [1977],
Linearization stability of the Einstein-Maxwell system,
{\em J.~Math. Phys.\/},
{\bf  18},
830--833.

\item[]
 J. M. Arms [1981],
The structure of the solution set for the Yang-Mills equations,
{\em Math. Proc. Camb. Phil. Soc.\/},
{\bf  90},
361--372.

\item[]
 J. M. Arms, R. Cushman, and M. J. Gotay [1991],
A universal reduction procedure for Hamiltonian
group actions,
{\it The Geometry	of Hamiltonian Systems\/}
(T. Ratiu, ed.),
M.S.R.I. Publ. {\bf 22},
33--51,
Springer-Verlag, New York.

\item[]
 J. M. Arms, M. J. Gotay, and G. Jennings [1990],
Geometric and algebraic reduction for singular
momentum mappings,
{\em Adv. in Math.\/},
{\bf  79},
43--103.

\item[]
 J. M. Arms, J. E. Marsden, and V. Moncrief [1981],
Symmetry and bifurcations of momentum
mappings,
{\em Commun. Math. Phys.\/},
{\bf  78},
455--478.

\item[]
 J. Arms, J. Marsden, and V. Moncrief [1982],
The structure of the space of solutions of Einstein's
equations:  II. Several Killings Fields and the Einstein-Yang-Mills 
equations, 
{\em Ann. Phys.\/}
{\bf  144},
81--106.

\item[]
 R. Arnowitt, S. Deser, and C. W.  Misner [1962],
The dynamics of general relativity,
{\it  Gravitation, an Introduction to Current Research\/}
(L. Witten, ed.),
227--265,
Wiley, New York.

\item[]
 A. Ashtekar [1986],
New variables for classical and quantum gravity,
{\em Phys. Rev. Lett.\/},
{\bf  57},
2244--2247.

\item[]
 A. Ashtekar [1987],
New Hamiltonian formulation of general relativity,
{\em Phys. Rev. D\/},
{\bf  36},
1587--1602.

\item[]
 A. Ashtekar, L. Bombelli, and O. Reula [1991],
The covariant phase space of asymptotically flat gravitational fields,
 {\it Mechanics, Analysis and Geometry: 200 Years After Lagrange\/}
(M. Francaviglia ed.),
417--450,
North-Holland,  Amsterdam.

\item[]
 D. Bao [1984],
A sufficient condition for the linearization stability of $N = 1$
supergravity: a
preliminary report,
{\em Ann. Phys.\/},
{\bf  158},
211--278.

\item[]
 D. Bao, J. Isenberg, and P. B. Yasskin [1985],
The dynamics of the Einstein Dirac system. I. A
principal bundle formulation of the theory and its canonical analysis,
{\em Ann. Phys.\/},
{\bf  164},
103--171.

\item[]
 D. Bao, J. Marsden, and R. Walton [1985],
The Hamiltonian structure of general relativistic perfect
fluids,
{\em Commun. Math. Phys.\/},
{\bf  99},
319--345.

\item[]
 C.  Batlle, J. Gomis, and J. M. Pons [1986],
Hamiltonian and Lagrangian constraints of the bosonic
string,
{\em Phys. Rev. D\/},
{\bf  34},
2430--2432.

\item[]
 R. Beig [1989],
On the classical geometry of bosonic string dynamics,
{\em Int.~J. Theor. Phys.,}
{\bf 30},
211--224.

\item[]
 F. J.  Belinfante [1940],
On the current and the density of the electric charge, the energy, 
the linear momentum and the angular momentum of arbitrary fields,
{\em Physica\/},
{\bf  vii},
449--474.

\item[]
 E. Binz, J. \'Sniatycki, and H. Fischer [1988],
{\it The Geometry of Classical Fields\/},
North Holland,
Amsterdam.

\item[]
T. J.  Bridges [1997],
Multi-symplectic structures and wave propagation,
{\it Math. Proc. Camb. Phil. Soc.\/},
{\bf 121},
147--190.

\item[]
F. Cantrijn, L. A. Ibort, and M. de Le\'on [1999],
On the geometry of multisymplectic manifolds,
{\em J. Aust. Math. Soc. A},
{\bf 66},
303--330.

\item[]
 C. Carath\'eodory [1935],
{\it Variationsrechnung und Partielle Differentialgleichungen erster
Ordnung\/},
Teubner,
Leipzig  (Reprinted by Chelsea, New York, 1982).

\item[]
 \'E. Cartan [1922],
{\it Le\c cons sur les Invariants Int\'egraux\/},
Hermann,
Paris.

\item[]
M. Castrillon Lopez and J. E. Marsden [2002],
Some remarks on Lagrangian and Poisson reduction for
field theories, 
{\it Preprint}.

\item[]
D. Christodoulou [2000],
{\it The Action Principle and Partial Differential Equations},
Princeton University Press, Princeton.

\item[]
\v{C}. Crnkovi\'{c} and E. Witten [1987],
Covariant description of
canonical formalism in geometrical theories,
{\it Newton's Tercentenary Volume\/}
(S.W. Hawking and W. Israel, eds.),
666--684,
Cambridge University Press, Cambridge.

\item[]
 P. Dedecker [1953],
Calcul des variations, formes diff\'erentielles et champs 
g\'eod\'esiques, 
{\em Colloq. Int. de G\'eom\'etrie
Diff\'erentielle\/} (Strasbourg),
17--34.

\item[]
 P. Dedecker [1957],
Calcul des variations et topologie alg\'ebrique,
{\em Mem. Soc. Roy. Sc. Li\`ege\/},
{\bf  19},
1--216.

\item[]
 P. Dedecker [1977],
On the generalization of symplectic geometry to multiple integrals in
the calculus of variations, {\em Lecture Notes in Math.\/},
{\bf  570},
395--456,
Springer-Verlag,
New York.

\item[]
 Th. De Donder [1930],
{\it Th\'eorie Invariantive du Calcul des Variations\/},
Gau\-thier-Villars,
Paris.

\item[]
 P. A. M. Dirac [1950],
Generalized Hamiltonian dynamics,
{\em Can.~J.~Math.\/},
{\bf  2},
129--148.

\item[]
 P. A. M. Dirac [1964],
{\it Lectures on Quantum Mechanics\/},
Academic Press,
New York.

\item[]
A. Eche\-verria-Enr\'{\i}ques, M. C. Mu\~noz-Lecanda, and N.
Rom\'an-Roy [2000], 
Geometry of multisymplectic Hamiltonian first-order field theories,
{\it J. Math. Phys.,}
{\bf 41},
7402--7444.

\item[]
G. Esposito, G. Gionti, and C. Stornaiolo [1995],
Spacetime covariant form of Ashtekar's constraints,
{\em Nuovo Cimento},
{\bf B110},
1137--1152.

\item[]
 M. Ferraris and M. Francaviglia [1983],
On the global structure of Lagrangian and Hamiltonian
formalisms in higher order calculus of variations,
{\it  Proc. of the Meeting ``Geometry and Physics''\/}
(M. Modugno, ed.),
43--70,
Pita\-gora Editrice,
Bologna.

\item[]
A. E.  Fischer and J. E. Marsden [1979a],
Topics in the dynamics of general relativity,
{\it  Isolated Gravitating Systems in General Relativity\/} 
(J.~Ehlers, ed.), Italian Physical Society,
322--395.

\item[]
A. E.  Fischer and J. E. Marsden [1979b],
The initial value problem and the dynamical formulation
of general relativity,
{\it  General Relativity\/}
(S.W. Hawking and W. Israel, eds.),
138--211,
Cambridge Univ. Press, Cambridge.

\item[]
A. E.  Fischer, J. E. Marsden, and V. Moncrief [1980],
The structure of the space of solutions of
Einstein's equations I:  One Killing field,
{\em Ann. Inst. H. Poincar\'e\/},
{\bf 33},
147--194.

\item[]
M. Forger, C. Paufler and H. R\" omer [2002a],
The Poisson bracket for Poisson forms in multisymplectic field theory,
{\em Preprint\/} math-ph/0202043.

\item[]
M. Forger, C. Paufler and H. R\" omer [2002b],
A general construction of Poisson brackets on exact multisymplectic
manifolds,
{\em Preprint\/} math-ph/0208037.

\item[]
 P. L. Garc\'\i a [1974],
The Poincar\'e-Cartan invariant in the calculus of variations,
{\em Symp. Math.\/},
{\bf 14},
219--246.

\item[]
 P. L. Garc\'\i a and J. Mu\~noz [1991],
Higher order regular variational problems,
{\it  Symplectic Geometry and Mathematical Physics\/}
(P. Donato, C. Duval, J.~Elhadad, and G.~M. Tuynman, eds.),
Birkh\"auser, Boston.

\item[]
 K. Gaw\c edzki  [1972],
On the geometrization of the canonical formalism in the classical field
theory,
{\em Rep. Math. Phys.\/},
{\bf  3},
307--326.

\item[]
 H. Goldschmidt and S. Sternberg [1973],
The Hamilton-Cartan formalism in the calculus of
variations,
{\em Ann. Inst. Fourier\/},
{\bf  23},
203--267.

\item[]
 M.J. Gotay [1988],
A multisymplectic approach to the KdV equation,
{\it  Differential Geometric Methods in Theoretical Physics\/}
(K. Bleuler and M. Werner eds.),
NATO Advanced Science Institutes Series C: Mathematical and 
Physical Sciences,
{\bf 250},
295--305,
Kluwer,
Dordrecht.

\item[]
 M. J. Gotay [1991a],
A multisymplectic framework for classical field theory and the
calculus of
variations I. Covariant Hamiltonian formalism,
{\it  Mechanics, Analysis, and
Geometry: 200 Years After Lagrange\/} 
(M. Francaviglia, ed.),
203--235,
North Holland,
Amsterdam.

\item[]
 M. J. Gotay [1991b],
A multisymplectic framework for classical field theory and the
calculus  of variations II. Space + time decomposition,
{\em Diff. Geom. Appl.\/},
{\bf  1},
375--390.

\item[]
 M. J. Gotay [1991c],
An exterior differential systems approach to the Cartan form,
{\it  Symplectic Geometry and Mathematical 			Physics\/}
(P. Donato, C.~Duval, J.~Elhadad, and G.M. Tuynman, eds.),
160--188,
Birkh\"auser, Boston.

\item[]
 M. J. Gotay and J. E. Marsden [1992],
Stress-energy-momentum tensors and the Belinfante--Rosenfeld formula,
{\em Contemp. Math.\/},
{\bf 132},
367--391.

\item[]
 M. J. Gotay, J. M. Nester, and G. Hinds [1978],
Presymplectic manifolds and the Dirac--Bergmann
theory of constraints,
{\em J.~Math. Phys.\/},
{\bf  19},
2388--2399.

\item[]
 P.~A.~Griffiths [1983],
{\it Exterior Differential Systems and the Calculus of Variations\/},
Birkh\"auser,
Boston.

\item[]
 V. Guillemin and S. Sternberg [1977],
Geometric asymptotics,
{\em Amer. Math. Soc. Surveys\/},
{\bf  14}.

\item[]
 V. Guillemin and S. Sternberg [1984],
{\it Symplectic Techniques in Physics\/},
Cambridge University Press, Cambridge.

\item[]
 C. G\"unther [1987],
The polysymplectic Hamiltonian formalism in field theory and the
calculus of variations, 
{\em J.~Diff. Geom.\/},
{\bf  25},
23--53.

\item[]
A.~J.~Hanson, T. Regge, and C. Teitelboim [1976],
Constrained Hamiltonian systems,
{\em Accademia Nazionale dei Lincei}, {\bf 22}, Rome.

\item[]
 R. Hermann  [1968],
{\it Differential Geometry and the Calculus of Variations\/},
Academic Press, Second Edition,
Math. Sci. Press,
Brookline, Massachusetts.

\item[]
 R. Hermann  [1975],
{\it Gauge Fields and Cartan-Ehresmann Connections Part A\/},
Math Sci Press,
Brookline, Massachusetts.

\item[]
 D. Holm [1985],
Hamiltonian formalism of general relativistic adiabatic fluids,
{\em Physica\/},
{\bf 17D},
1--36.

\item[]
 M. Hor\'ak and I. Kol\'a\v r [1983],
On the higher order Poincar\'e-Cartan forms,
{\em Cz. Math. J.\/},
{\bf  33},
467--475.

\item[]
 G. Horowitz [1989],
Exactly soluble diffeomorphism invariant theories,
{\em Commun. Math. Phys.\/},
{\bf  129},
417--437.

\item[]
 J. Isenberg and J. Nester [1977],
The effect of gravitational interaction on classical fields: a
Hamilton--Dirac analysis,
{\em Ann. Phys.\/},
{\bf  107},
56--81.

\item[]
 J. Isenberg and J. Nester [1980],
Canonical gravity,
{\it  General Relativity and Gravitation, Vol. 1\/}
(A. Held, ed.),
23--97,
Plenum Press,
New York.

\item[]
I.~V.~ Kanatchikov [1997],
On field theoretic generalizations of a Poisson algebra,
{\em Rep. Math. Phys.\/},
{\bf 40},
225--234.

\item[]
I.~V.~Kanatchikov [1998],
Canonical structure of classical
field theory in the polymomentum phase space,
{\em Rep. Math. Phys.\/},
{\bf 41},
49--90.

\item[]
 H.~A.~Kastrup [1983],
Canonical theories of Lagrangian dynamical systems in physics,
{\em Phys. Rep.\/},
{\bf   101},
1--167.

\item[]
 J. Kijowski [1973],
A finite-dimensional canonical formalism in the classical field theory,
{\em Commun. Math. Phys.\/},
{\bf  30},
99--128.

\item[]
 J. Kijowski [1974],
Multiphase spaces and gauge in the calculus of variations,
{\em Bull. Acad. Sc. Polon.\/},
{\bf  22},
1219--1225.

\item[]
 J. Kijowski and W. Szczyrba [1975],
Multisymplectic manifolds and the geometrical construction
of the Poisson brackets in the classical field theory,
{\it  G\'eom\'etrie Symplectique et Physique Math\'ematique\/} 
(J.-M. Souriau, ed.),
347--379,
C.N.R.S.,
Paris.

\item[]
 J. Kijowski and W. Szczyrba [1976],
A canonical structure for classical field theories,
{\em Commun. Math. Phys.\/},
{\bf  46},
183--206.

\item[]
 J. Kijowski and W. Tulczyjew [1979],
A symplectic framework for field theories,
{\em Lecture Notes in Physics\/},
{\bf  107},
Springer-Verlag,
New York.

\item[]
 I. Kol\'a\v r [1984],
A geometric version of the higher order Hamilton formalism in fibered
manifolds,
{\em J. Geom.  Phys.\/},
{\bf  1},
127--137.

\item[]
 Y. Kosmann-Schwarzbach [1981],
Hamiltonian systems on fibered manifolds,
{\em Lett. Math. Phys.\/},
{\bf   5},
229--237.

\item[]
S. Kouranbaeva and S. Shkoller [2000],
A variational approach to
second-order multisymplectic field theory,
{\em J. Geom. Phys.\/},
{\bf 35},
333--366.

\item[]
 D. Krupka [1971],
Lagrange theory in fibered manifolds,
{\em Rep. Math. Phys.\/},
{\bf  2},
121--133.

\item[]
 D. Krupka [1987],
Geometry of Lagrangean structures. 3.,
{\em Supp. Rend. Circ. Mat.\/} (Palermo), ser II,
{\bf 14},
187--224.

\item[]
 D. Krupka and O. \v{S}t\v{e}p\'ankov\'a  [1983],
On the Hamilton form in second order calculus of
variations,
{\it  Proc. of the Meeting ``Geometry and Physics''\/} 
(M.~Modugno, ed.),
85--101,
Pitagora Editrice,
Bologna.

\item[]
 K. Kucha\v r [1973],
Canonical quantization of gravity,
{\it  Relativity, Astrophysics and Cosmology\/} 
(W. Israel, ed.),
237--288,
Reidel,
Dordrecht.

\item[]
 K. Kucha\v r [1974],
Geometrodynamics regained: a Lagrangian approach,
{\em J. Math. Phys.\/},
{\bf  15},
708--15.

\item[]
 K. Kucha\v r [1976],
Geometry of hyperspace I--IV,
{\em J.~Math. Phys.\/},
{\bf  17},
777--820;
{\bf 18},
1589--1597.

\item[]
 M. Kummer [1981],
On the construction of the reduced space of a Hamiltonian system 
with symmetry,
{\em Indiana Univ. Math.~J.\/},
{\bf  30},
281--291.

\item[]
H.~P.~K\"unzle and J. M. Nester [1984],
Hamiltonian formulation of gravitating perfect fluids and the Newtonian
limit,
{\em J.~Math. Phys.},
{\bf 25},
1009--1018.

\item[]
 B. A. Kupershmidt [1980],
Geometry of jet bundles and the structure of Lagrangian and Hamiltonian
formalisms,
{\em Lecture Notes in Math.\/},
{\bf  775},
162--218,
Springer-Verlag,
New York, N.Y.

\item[]
J. K. Lawson [2000],
A frame bundle generalization of multisymplectic field theories,
{\em Rep. Math. Phys.},
{\bf 45},
183--205.

\item[]
 M. de Le\'on and P. R.  Rodrigues [1985],
{\it Generalized Classical Mechanics and Field Theory\/},
North Holland,
Amsterdam.

\item[]
 Th. H. J.  Lepage [1936],
Sur les champs g\'eod\'esiques du calcul des variations,
{\em Bull. Acad. Roy. Belg., Cl. Sci.\/},
{\bf  22},
716--729, 1036--1046.

\item[]
 Th. H. J.  Lepage [1941],
Sur les champs g\'eod\'esique des int\'egrales multiples,
{\em Bull. Acad. Roy. Belg., Cl. Sci.\/},
{\bf  27},
27--46.

\item[]
 Th. H. J.  Lepage [1942],
Champs stationnaires, champs g\'eod\'esiques et formes int\'e\-grables, 
{\em Bull. Acad. Roy. Belg., Cl. Sci.\/}, 
{\bf 28}, 
73--92, 247--265.

\item[]
A. Lew, J. E. Marsden, M. Ortiz, and M. West [2002], 
Asynchronous variational integrators, 
{\em Arch. Rat. Mech. Anal.} 
(in press).

\item[]
 L. Lusanna [1991],
The second Noether theorem as the basis of the theory of singular
Lagrangians and Hamiltonian constraints,
{\em Riv. Nuovo Cimento},
{\bf 14}(3),
1--75.

\item[]
 L. Lusanna [1993],
The Shanmugadhasan canonical transformation, function groups and the
extended second Noether theorem, {\em Int.~J.~Mod. Phys. A\/},
{\bf 8},
4193--4233.

\item[]
 J. E. Marsden [1981],
Lectures on geometric methods in mathematical physics,
{\it SIAM--CBMS},
{\bf 37}.

\item[]
 J. E. Marsden [1988],
The Hamiltonian formulation of classical field theory,
{\em Contemp. Math.\/},
{\bf  71},
221--235.

\item[]
 J. E. Marsden [1992],
Lectures in Mechanics,
{\em London Math. Soc. Lecture Notes\/},
{\bf  174},
Cambridge University Press, Cambridge.

\item[]
 J. E. Marsden and T. J. R. Hughes [1983],
{\it Mathematical Foundations of Elasticity\/},
Prentice-Hall,
Redwood City, California (Reprinted by Dover, 1994).

\item[]
J. E. Marsden, R. Montgomery, P. J. Morrison, and W.B. Thompson [1986],
Covariant Poisson brackets for classical fields,
{\em Ann. Phys.\/},
{\bf  169},
29--48.

\item[]
 J. E. Marsden, R. Montgomery, and T. Ratiu [1990],
Symmetry, Reduction, and Phases in Mechanics,
{\em Mem. Amer. Math. Soc.\/},
{\bf  436}.

\item[]
J. E. Marsden, G. W. Patrick, and S. Shkoller [1998],
Multisymplectic geometry, variational integrators and nonlinear
PDEs,
{\em Comm. Math. Phys.} 
\textbf{199}, 
351--395.

\item[]
J. E. Marsden,  S. Pekarsky, S. Shkoller, and M. West [2001],
Variational methods, multisymplectic geometry and continuum mechanics, 
{\em J. Geom. Phys.}, 
\textbf{38}, 
253--284.

\item{}
J. E. Marsden and T. S. Ratiu [1999], 
Introduction to mechanics and symmetry, Second Edition, 
{\em Texts in Applied Mathematics,} 
{\bf 17},
Springer-Verlag, New York.

\item[]
J. E. Marsden and S. Shkoller [1999],
Multisymplectic geometry,
covariant Hamiltonians and water waves, 
{\em Math. Proc. Camb. Phil. Soc.},
\textbf{125}, 553--575.

\item[]
 J. E. Marsden, A. Weinstein, T. Ratiu, R. Schmid, and R. G.  
Spencer [1983],
Hamiltonian systems with symmetry, coadjoint orbits and plasma
physics,
{\it Proc. IUTAM-
ISIMM Symposium on Modern Developments in Analytical Mechanics\/}
(S. Benenti, M. Francaviglia and A.
Lichnerowicz, eds.),
{\em Atti della Accademia delle Scienze di Torino\/},
{\bf  117},
289--340.

\item[]
 G. Martin [1988],
A Darboux theorem for multi-symplectic manifolds,
{\em Lett. Math. Phys.\/},
{\bf  16},
133--138.

\item[]
K. Mikami and A.~Weinstein [1988], 
Moments and reduction for symplectic groupoid actions, 
{\em Publ. RIMS Kyoto Univ.},
\textbf{24},
121--140.

\item[]
 C. W.  Misner, K. Thorne, and J. A. Wheeler [1973],
{\it Gravitation\/},
W.H. Freeman,
San Francisco.

\item[]
J.-J. Moreau [1982], 
Fluid dynamics and the calculus of horizontal variations,
{\em Internat. J. Engrg. Sci.},
\textbf{20}, 
389--411.

\item[]
 J. Mu\~noz [1985],
Poincar\'e-Cartan forms in higher order variational calculus on fibered
manifolds,
{\em Rev. Mat. Iberoamericana\/},
{\bf  1},
85--126.

\item[]
Y. Nambu [1970],
Duality and Hydrodynamics,
{\em Lectures for the Copenhagen Summer Symposium\/},
 unpublished.

\item[]
 P. J. Olver [1993],
Applications of Lie groups in differential equations, Second Edition, 
{\em Graduate Texts in Mathematics\/},
{\bf  107},
Springer-Verlag,
New York.

\item[]
 R. Ouzilou [1972],
Expression symplectique des probl\`emes variationnels,
{\em Symp. Math.\/},
{\bf  14},
85--98.

\item[]
P. V. P\^aquet [1941], 
Les Formes differenti\'elles ext\'erieurs $\Omega_n$
dans le calcul des variations, 
\emph{Bull. Acad. Roy. Belg., Cl. des Sci.}, 
{\bf 27}, 
65--84.

\item[]
A. M. Polyakov [1981],
Quantum geometry of bosonic strings,
{\em Phys. Lett.,}
{\bf 103B},
207--210.

\item[]
 R. Ragionieri and R. Ricci [1981],
Hamiltonian formalism in the calculus of variations,
{\em Bollettino U.M.I.\/},
 18-B,
119--130.

\item[]
 L. Rosenfeld [1940],
Sur le tenseur d'impulsion---\'energie,
{\em M\'em. Acad. Roy. Belg. Sci.\/},
{\bf  18},
1--30.

\item[]
G.A. Sardanashvily [1993],
{\em Gauge Theory in Jet Manifolds},
Hadronic Press, Palm Harbor, Florida.

\item[]
D. J. Saunders [1989], 
The geometry of jet bundles,
{\em London Mathematical Society Lecture Note Series},
{\bf 142},
Cambridge University Press, Cambridge.

\item[]
D. J. Saunders [1992],
The regularity of variational problems,
{\em Contemp. Math.},
{\bf 132},
573--593.

\item[]
 J. Scherk [1975],
An introduction to the theory of dual models and strings,
{\em Rev. Mod. Phys.\/},
{\bf  47},
123--164.

\item[]
 W. F. Shadwick [1982],
The Hamiltonian formulation of regular $r$th-order Lagrangian field
theories,
{\em Lett. Math. Phys.\/},
{\bf  6},
409--416.


\item[]
 R. Sjamaar and E. Lerman [1991],
Stratified symplectic spaces and reduction,
{\em Ann. Math.\/}
{\bf 134},
375--422.

\item[]
 J. \'Sniatycki [1970a],
On the geometric structure of classical field theory in Lagrangian
formulation,
{\em Proc. Camb. Phil. Soc.\/},
{\bf  68},
475--484.

\item[]
 J. \'Sniatycki  [1970b],
On the canonical formulation of general relativity,
{\it  Proc. Journ\'ees Relativistes 1970\/},
127--135,
Facult\'e des Sciences,
Caen.

\item[]
 J. \'Sniatycki [1974],
Dirac brackets in geometric dynamics,
{\em Ann. Inst. H. Poin\-car\'e\/},
{\bf  A20},
365--372.

\item[]
 J. \'Sniatycki [1984],
The Cauchy data space formulation of classical field theory,
{\em Rep. Math. Phys.\/},
{\bf  19},
407--422.

\item[]
 S. Sternberg [1977],
Some preliminary remarks on the formal variational calculus of 
Gel'fand and Dikki,
{\em Lecture Notes in Math.\/},
{\bf  676},
399--407,
Springer-Verlag,
New York.

\item[]
 K. Sundermeyer [1982],
Constrained dynamics,
{\em Lecture Notes in Physics\/},
{\bf  169},
Springer-Verlag,
New York.

\item[]
 W. Szczyrba [1976a],
Lagrangian formalism in the classical field theory,
{\em Ann. Polon. Math.\/},
{\bf   32},
145--185.

\item[]
 W. Szczyrba [1976b],
A symplectic structure of the set of Einstein metrics:  a canonical
formalism for general relativity, 
{\em Commun. Math. Phys.\/},
{\bf  51},
163--182.

\item[]
 A. Trautman [1967],
Noether equations and conservation laws,
{\em Commun. $\!$Math. Phys.\/},
{\bf  6},
248--261.

\item[]
 H. Weyl [1935],
Geodesic fields in the calculus of variation for multiple integrals,
{\em Ann. Math.\/},
{\bf  36},
607--629.

\item[]
G. J. Zuckerman [1987],
Action principles and global geometry,
{\it Mathematical Aspects of String Theory,}
(S. T. Yau, ed.),
{\em Adv. Ser. Math. Phys.}, 
{\bf 1},
259--284,
World Scientific, Singapore.

\end{description}



\newcommand{\indsp}{\null\phantom{2\enspace}}

\newpage
\section*{\Large\bfseries Table of Contents
for Parts II--V}
\addcontentsline{toc}{section}{Table of Contents for Parts II--V}
\markboth{Table of Contents}{ Parts II--V}
\bigskip

\section*{\large\bfseries II---Canonical Analysis of Field Theories}

\begin{enumerate}
\medskip
\item[\bf 5\quad] {\bf Symplectic Structures on Cauchy Surfaces}\\	
      5A\quad  Cauchy Surfaces and Spaces of Fields\\	
      5B\quad  Canonical Forms on $T^*\mathcal{Y}_\tau$\\	
      5C\quad  Presymplectic Structure on $\mathcal{Z}_\tau$	\\	
      5D\quad  Reduction of $\mathcal{Z}_\tau$ to $T^*\mathcal{Y}_\tau$

\medskip
\item[\bf 6\quad] {\bf Initial Value Analysis of Field Theories}
\\	
      6A\quad  Slicings\\	
      6B\quad  Space + Time Decomposition of the Jet Bundle\\	
      6C\quad  The Instantaneous Legendre Transform\\	
      6D\quad  Hamiltonian Dynamics\\	
      6E\quad  Constraint Theory

\medskip
\item[\bf 7\quad] {\bf The Energy-Momentum Map}
\\	
      7A\quad  Induced Actions on Fields\\	
      7B\quad  The Energy-Momentum Map\\	
      7C\quad  Induced Momentum Maps on $\mathcal{Z}_\tau$\\	
      7D\quad  Induced Momentum Maps on $T^*\mathcal{Y}_\tau$\\	
      7E\quad  Momentum Maps for Lifted Actions\\	
      7F\quad  The Hamiltonian and the Energy-Momentum Map
\end{enumerate}

\medskip
\noindent{\bf Interlude II---The Stress-Energy-Momentum Tensor}

\section*{\large\bfseries 
III---Gauge Symmetries and Initial Value Constraints}

\begin{enumerate}
\medskip
\item[\bf 8\quad] {\bf The Gauge Group}
\\	
      8A\quad  Principal Bundle Construction of the Gauge Group\\	
      8B\quad  Covariance, Localizability, and Gauge Groups\\	
      8C\quad  Gauge Transformations

\medskip
\item[\bf 9\quad] {\bf The Vanishing Theorem and Its Converse}
\\	
      9A\quad  Flexibility\\	
      9B\quad  The Vanishing Theorem\\	
      9C\quad  The Converse of the Vanishing Theorem

\medskip
\item[\bf 10\quad] {\bf Secondary Constraints and the Energy-Momentum
Map}
\\	
      10A\quad  The Final Constraint Set Lies in the Zero Level of
the\newline 
                \phantom{\null 11A\quad }Energy-Momentum Map\\	
      10B\quad  First Class Secondary Constraints

\medskip
\item[\bf 11\quad] {\bf Primary Constraints and the Momentum Map}
\\	
      11A\quad  The Foliation $\dot\cg_\tau$\\	
      11B\quad  The Primary Constraint Set Lies in the Zero Level of the
\newline 
                \phantom{\null 10B\quad }Momentum Map\\	
      11C\quad  First Class Primary Constraints

\end{enumerate}

\medskip
\noindent{\bf Interlude III---Singularities in Solution Spaces of 
\newline 
               \phantom{Interlude III---}Classical Relativistic Field
Theories}


\section*{\large\bfseries IV---The Adjoint Formalism  }

\begin{enumerate}
\medskip
\item[\bf 12\quad] {\bf The Dynamic and Atlas Fields}
\\	
      12A\quad  The Dynamic Bundle\\	
      12B\quad  Bundle Considerations\\	
      12C\quad  The Atlas Bundle

\medskip
\item[\bf 13\quad] {\bf The Adjoint Formalism}
\\	
      13A\quad  Linearity of the Hamiltonian\\	
      13B\quad  Model Bundles \\	
      13C\quad  The Adjoint Form, Reconstruction and Decomposition
\end{enumerate}

\medskip
\noindent{\bf Conclusions}


\section*{\large\bfseries V--- Palatini Gravity}

\begin{enumerate}
\medskip
\item[\bf 14\quad] {\bf  Application to Palatini Gravity}
\\	
      14A\quad  Covariant Analysis\\	
      14B\quad  Canonical Analysis\\	
      14C\quad  Energy-Momentum Map Analysis\\
      14D\quad  The Adjoint Formalism

\end{enumerate}


\end{document}